\documentclass[prb,twocolumn,superscriptaddress,longbibliography,floatfix]{revtex4-2}
\usepackage{amsmath,amssymb}
\usepackage{graphicx}
 \pdfoutput=1

\date{\today}

\begin{document}
\title{
Nonlinear Fermi-liquid transport through a quantum dot \\ 
in asymmetric tunnel junctions
}

\author{Kazuhiko Tsutsumi}
\affiliation{Department of Physics, Osaka City University, 
Sumiyoshi-ku, Osaka 558-8585, Japan}

\author{Yoshimichi Teratani}
\affiliation{Department of Physics, Osaka City University, 
Sumiyoshi-ku, Osaka 558-8585, Japan}
\affiliation{Nambu Yoichiro Institute of Theoretical 
and Experimental Physics, Sumiyoshi-ku, Osaka 558-8585, Japan}

\author{Rui Sakano}
\affiliation{Institute for Solid State Physics, 
the University of Tokyo, 5-1-5 Kashiwanoha, Kashiwa, Chiba 277-8581, Japan}

\author{Akira Oguri}
\affiliation{Department of Physics, Osaka City University, 
Sumiyoshi-ku, Osaka 558-8585, Japan}
\affiliation{Nambu Yoichiro Institute of Theoretical 
and Experimental Physics, Sumiyoshi-ku, Osaka 558-8585, Japan}

\begin{abstract}

We study the nonlinear conductance through a quantum dot, 
specifically its dependence on the asymmetries 
in the tunnel couplings and bias voltages $V$, at low energies. 
Extending the microscopic Fermi-liquid theory for 
the Anderson impurity model,
we obtain an exact formula for the steady current $I$ 
up to terms of order $V^3$ in the presence of these asymmetries.
The coefficients for the nonlinear terms are described in terms of  
a set of the Fermi-liquid parameters: 
the phase shift, static susceptibilities, and three-body correlation functions 
 of electrons in the quantum dots,  defined with respect to the equilibrium ground state.  
 We calculate  these correlation functions, 
using the numerical renormalization group approach (NRG), 
over a wide range of impurity-electron filling that can be controlled 
by a gate voltage in real systems.  
The NRG results 
show that the order $V^2$ nonlinear current
 is enhanced significantly in the valence fluctuation regime. 
 It is caused by the order $V$ energy shift of the impurity level, 
induced in the presence of the tunneling or bias asymmetry. 
Furthermore,  in the valence fluctuation regime,  
we also find that the order $V^3$ nonlinear current exhibits a shoulder structure,  
for which the three-body correlations that evolve for large asymmetries play 
an essential role. 
\end{abstract}
\maketitle

\section{Introduction}
\label{Introduction}
The Kondo effect is one of the most interesting  phenomena 
occurring in strongly correlated fermion systems \cite{Kondo,hewson_1993} 
and has been explored originally for dilute magnetic alloys \cite{Anderson1961,Nozieres1974,Wilson1975,Yamada1975II,10.1143/PTP.54.316,Shiba1975,Yoshimori1976}. 
It occurs also in quantum dots coupled to electron reservoirs \cite{Ng-Lee1988,Glazman1988,M-W-L1991,Kawabata1991},  
and has intensively been studied for three decades 
\cite{Goldhaber-Gordon1998nature,Goldhaber-Goldon1998PRL,Cronenwett-Oosterkamp-Kouwenhoven1998,vanderWiel2000}.
One of the advantages of the quantum dots for studying the Kondo effect is  
that 
the information of the many-body quantum states    
can be probed in a highly tunable way. For instance, 
recent development makes it possible to observe directly 
 the Kondo screening cloud \cite{KondoCloud2020}.
Furthermore, 
quantum dots have various variations, such as     
a single dot \cite{Schmid1998,Simmel1999,Sasaki2004,GrobisGoldhaber-Gordon,ScottNatelson,Heiblum,Delattre2009,KobayashiKondoShot,Izumida1998,AO2001,Anders2008,WeichselbaumVonDelft}, multiple dots \cite{Izumida2000,Borda2003}, 
and the carbon-nanotube (CNT) dots  
\cite{RMP-Kouwenhoven,Teratani2016,Ferrier2016,Ferrier2017,Teratani2020PRB}.  
The orbital degrees of freedom of the CNT dots 
also provide an interesting variety in the Kondo singlet state 
with the SU(4) symmetry 
\cite{Sasaki2000,Pablo2005,Choi2005,Eto2005,Sakano2006,Sakano2007,Makarovski2007,Anders-SU4,SU4_Kondo_ferro_Weymann,MoraEtal2009,Mora2009,Cleuziou2013,MantelliMocaZarandGrifoni,Teratani2020PRL}.

Here we focus on the effects of the asymmetries in the tunnel junctions,     
which real quantum dots inevitably have more or less, 
on the nonlinear current $I$ 
 in the low-energy Fermi-liquid regime \cite{Nozieres1974,Wilson1975,Yamada1975II,10.1143/PTP.54.316,Shiba1975,Yoshimori1976}. Specifically, 
we examine the asymmetry in the chemical potentials $\mu_L$ and $\mu_R$
of the source and drain electrodes,  applied such that  $eV\equiv \mu_L-\mu_R$,  
as well as the asymmetry in the tunnel couplings $\Gamma_L$ and $\Gamma_R$. 

These asymmetries vary the occupation number of quantum dots,  
and  deform the Kondo cloud of conduction electrons 
which screens the local moment at temperatures $T \ll T^\ast$, 
lower than the Kondo energy scale $T^\ast$.
The tunnel-coupling and bias asymmetries also affect the differential conductance 
$dI/dV$, which takes the following form at $T=0$ up  to terms of order $(eV)^2$, 
\begin{align}
\!\!
\frac{dI}{dV} \,=\, g_0^{} 
\left[\,\sin^2\delta\,+C_V^{(2)}\,\frac{eV}{T^\ast} \, -C_V^{(3)}
\biggl(\frac{eV}{T^\ast}\biggr)^2 +\cdots \,\right].
 \nonumber
\end{align}
Here, 
$g_0^{}=(2e^2/h)\, 4\Gamma_L\Gamma_R/(\Gamma_L+\Gamma_R)^2$. 
The first term on the right-hand side with  $\sin^2 \delta$ represents  
the linear conductance, with  $\delta$ the phase shift which is related 
to the average number of electrons in quantum dots. 
The nonlinear terms  $C_V^{(2)}$ and $C_V^{(3)}$ include additional  
information about two-body and three-body correlations 
of electrons passing through the quantum dots, 
and this is the main topic of this work.  
 These two terms depend strongly on  the tunneling and bias asymmetries;
especially $C_V^{(2)}$ emerges only in the presence of these asymmetries.

Effects of these asymmetries on the nonlinear currents
in the Fermi-liquid regime have been developed for over a decade.  
Sela and Malecki \cite{Sela-Malecki2009} have derived $C_V^{(3)}$ for   
the particle-hole symmetric Anderson impurity model, 
for which the phase shift takes the value $\delta=\pi/2$ 
and $C_V^{(2)}$ identically vanishes 
even in the presence of the junction asymmetries.
They have also shown using an effective Hamiltonian   
 that $C_V^{(3)}$ for this case becomes independent of 
 the junction-asymmetry parameters, $\Gamma_L-\Gamma_R$ 
and $(\mu_L+\mu_R)/2 -E_F$, 
in the limit of strong Coulomb interaction $U\to \infty$. 
Here, $E_F$ is the Fermi level at equilibrium $eV=0$. 
Aligia \cite{Aligia2011}  has derived a  more generalized formula 
 which is applicable to arbitrary electron fillings,  
using the renormalized perturbation theory \cite{HewsonRPT2001}. 
The formula for  $C_V^{(2)}$ has successfully  been derived 
in terms of the renormalized parameters of the Fermi-liquid 
and the junction-asymmetry parameters.
However, the formula for  $C_V^{(3)}$ away from half filling was described in terms of   
 the second derivative of the real part of self-energy 
$\Sigma_{\sigma}^{r}(\omega)$ with respect to the frequencies $\omega$
whose behavior has been less understood until very recently. 
The lack of the knowledge about 
the second derivative of  $\mathrm{Re}\, \Sigma_{\sigma}^{r}(\omega)$, 
which determines the higher-order energy shift of quasiparticles, 
makes the application range of their formula narrow.

Mora {\it et al\/} \cite{MoraEtal2009} 
have made an important step toward a deeper understanding 
of the higher-order contributions away from half filling 
by extending Nozi$\acute{\mathrm{e}}$res' phenomenological Fermi-liquid theory.
They have derived the formulas 
for $C_V^{(2)}$ and $C_V^{(3)}$ for the SU($N$) Kondo model,  
for which the number of  impurity electrons $m$  
takes only integer values $m =1,2,\ldots, N-1$,
taking also into account the tunnel asymmetry. 
More recently, Mora, Filippone 
{\it et al\/} \cite{MMvDZ2015,FMvDM2018} 
have extended the Nozi$\acute{\mathrm{e}}$res' description  
further to treat the Anderson impurity model for arbitrary electron fillings. 
The corresponding microscopic approach of  Yamada-Yosida 
\cite{Yamada1975II,10.1143/PTP.54.316,Shiba1975,Yoshimori1976}  
has recently been expanded in the Keldysh formalism to explore higher-order Fermi-liquid corrections \cite{AO2017_I,AO2017_II,AO2017_III}.  
In particular, it has been clarified that the second derivative 
of $\mathrm{Re}\, \Sigma_{\sigma}^{r}(\omega)$ is determined by 
the three-body correlation functions of the impurity electrons, 
defined with respect to the equilibrium ground state.  
However, despite this recent progress, 
effects of junction-asymmetries on the nonlinear transport 
still have not  been explored in detail, so far, 
for quantum dots at arbitrary electron fillings 
\cite{MMvDZ2015,FMvDM2018,AO2017_I,AO2017_III}.

The purpose of this paper is to study how these  
 asymmetries of the  tunnel couplings and the bias voltages 
affect the nonlinear transport of correlated electrons away from half filling. 
To this end, we derive the exact formulas for 
the coefficients $C_V^{(2)}$ and  $C_V^{(3)}$ 
applicable to the asymmetric tunnel couplings and bias voltages. 
These coefficients are expressed in terms of the Fermi-liquid (FL) parameters  
defined with respect to the equilibrium ground state, i.e.,\  
the phase shift $\delta$, 
the linear susceptibilities $\chi_{\sigma\sigma'}^{}$, and 
the nonlinear susceptibilities $\chi_{\sigma_1\sigma_2\sigma_3}^{[3]}$. 
While $C_V^{(2)}$ is determined by $\delta$ and $\chi_{\sigma\sigma'}^{}$, 
the coefficient  $C_V^{(3)}$ depends also on the three-body 
correlations $\chi_{\sigma_1\sigma_2\sigma_3}^{[3]}$.  
Furthermore, we calculate  $C_V^{(2)}$ and  $C_V^{(3)}$ 
over a wide range of electron fillings, using the numerical renormalization group (NRG) \cite{Wilson1975,KWW1980}. 
We have reported some preliminary results previously in conference proceedings \cite{Tsutsumi2020vol}. 
In the present paper,  we describe further developments, 
including the complete formulation,  
and explore the behavior of  $C_V^{(2)}$ and  $C_V^{(3)}$ 
for various junction asymmetries occurring   
in a wide parameter space.

The formula for  $C_V^{(2)}$  agrees with the previous result of 
Aligia \cite{Aligia2011}, and this term reflects  
the shift of impurity level  induced at finite bias voltages $eV$ 
in the presence of the tunnel-coupling and bias asymmetries.  
We find using also  NRG  that  $C_V^{(2)}$ is enhanced significantly 
in the valence fluctuation regime.

The coefficient $C_V^{(3)}$  
depends  on two  different  three-body correlations 
 $\chi_{\uparrow\uparrow\uparrow}^{[3]}$ 
and $\chi_{\uparrow\downarrow\downarrow}^{[3]}$ 
for the SU(2) Anderson impurity model away from half filling.  
The three-body correlations are enhanced in the valence fluctuation regime,  
outside the Kondo ridge.   
We find  that $C_V^{(3)}$ has a shoulder-type weak plateau  
in the gate-voltage dependence  for large tunnel asymmetries. 
This structure is caused by the three-body correlations, 
especially by an enhancement  of  $\chi_{\uparrow\downarrow\downarrow}^{[3]}$ 
in the valence fluctuation regime. 
Recent experiments, which were carried out  for highly symmetric tunnel junctions,  
have successfully demonstrated that the three-body contributions 
can be deduced from the nonlinear current \cite{Hata2021}.  
Out results  open the possibility to determine the contributions of 
 $\chi_{\uparrow\uparrow\uparrow}^{[3]}$ 
and $\chi_{\uparrow\downarrow\downarrow}^{[3]}$  
separately through measurements that can tune tunnel asymmetries.

This paper is organized as follows. 
In Sec.\ \ref{Formulation}, 
we describe the formulation and Fermi-liquid parameters. 
Section  \ref{Differential conductance} is devoted to  
the derivation of  the nonlinear transport coefficients 
 through  quantum dots.  
In Sec.\ \ref{Gate voltage dependence of FL parameters}, 
we describe the NRG results for quasiparticle parameters, 
 including the three-body correlations.  
The NRG results for $C_V^{(2)}$ and  $C_V^{(3)}$  
are presented for junctions with tunnel and bias asymmetries   
 in Secs.\ \ref{Cv2Results} and \ref{Cv3Results}, respectively. 
Our summary is given in Sec.\ \ref{Conclusion}.

\section{Formulation}
\label{Formulation}

\subsection{Anderson impurity model for quantum dots}

\begin{figure}[t]
	\begin{minipage}[r]{\linewidth}
	\centering
	\includegraphics[keepaspectratio,scale=0.5]{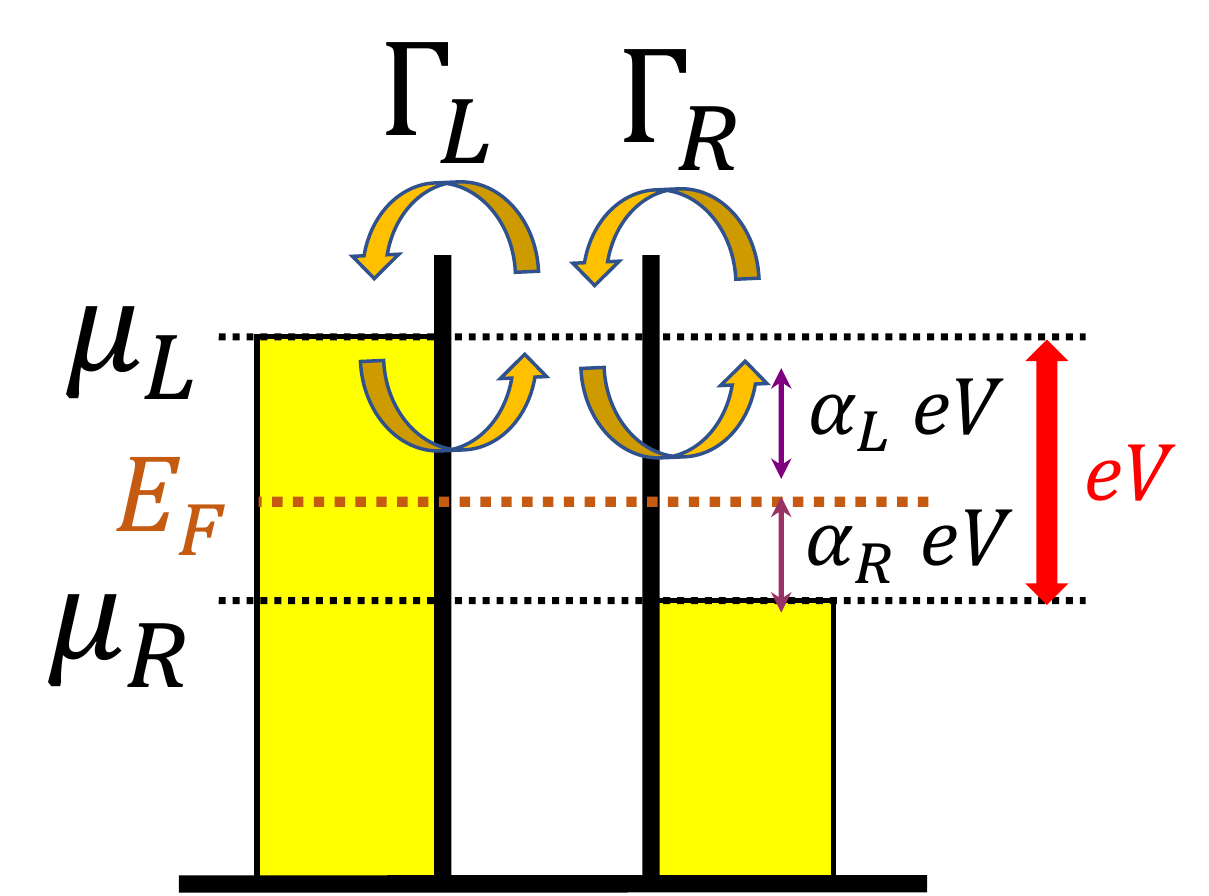}
	\end{minipage}
	\caption{
Schematic diagram of a quantum dot coupled 
to two leads via the tunnel couplings $\Gamma_L$ and $\Gamma_R$. 
The  bias voltage $eV  = \mu_L -\mu_R$ is applied in  a way such that  
 $\mu_L = E_F + \alpha_L eV$ and $\mu_R= E_F -\alpha_R eV$, 
with $\alpha_L+\alpha_R=1$. The chemical potentials of 
the left and right leads are measured from the Fermi energy 
at equilibrium $E_F=0$. 
The tunnel coupling is smaller for the  barrier with thicker width. 
}
  \label{schematic}
\end{figure}

We consider a single quantum dot coupled to two noninteracting leads,    
which is illustrated in Fig.\ \ref{schematic} and is  
described by the Anderson impurity model \cite{Anderson1961}: 
 $H= H_d + H_c + H_T$ with  
\begin{align}
H_d\, =& \ \sum_{\sigma=\uparrow, \downarrow}\epsilon_{d\sigma}n_{d\sigma}
+ U\, n_{d\uparrow}n_{d\downarrow}\,,
\qquad 
\\ 
H_c\, =&\sum_{\nu=L, \,R}\sum_{\sigma=\uparrow, \downarrow}\int_{-D}^{D}d\epsilon\,\,\epsilon \,c_{\epsilon\nu\sigma}^{\dagger}c_{\epsilon\nu\sigma}^{\,}\,,
\,\, \ \ 
\\
H_T\,=  & \  \sum_{\nu=L,\,R}\sum_{\sigma=\uparrow, \downarrow}v_\nu\,
(\psi_{\nu,\,\sigma}^{\dagger}d_\sigma^{}+d_\sigma^{\dagger}\,\psi_{\nu,\,\sigma}^{\,})\,,
\\
\psi_{\nu,\,\sigma}\equiv&\int_{-D}^{D}d\epsilon\,\sqrt{\rho_c}\,c_{\epsilon\nu\sigma}\,.
\end{align}
Here, 
$d_\sigma^\dagger$ creates an impurity electron with spin $\sigma$ 
and energy $\epsilon_{d\sigma} \equiv \epsilon_d -\sigma b$. 
The parameter $b$ represents a magnetic field,  
 $n_{d\sigma}=d_\sigma^\dagger d_\sigma^{}$, and    
$U$ is the Coulomb interaction between electrons in the quantum dot. 
The operator $c_{\epsilon\nu\sigma}^\dagger$ creates a conduction electron 
with spin $\sigma$ and continuous energy $\epsilon$ 
in the left ($L$) or right ($R$) lead.  
It satisfies the anticommutation relation
 $\bigl\{c_{\epsilon\nu\sigma}\,,\,c_{\epsilon'\nu'\sigma'}^\dagger\bigr\}=\delta(\epsilon-\epsilon')\,
\delta_{\nu\nu'}\delta_{\sigma\sigma'}$ with $\delta(\epsilon-\epsilon')$ 
the Dirac $\delta$ function. 
We assume that the conduction band is flat 
 over the range $-D<\epsilon<D$ 
with the constant density of states $\rho_c=1/(2D)$.
The linear combination of the conduction electrons $\psi_{\nu,\sigma}$  
couples to the electrons in the quantum dot 
via tunneling matrix element $v_\nu$.  
The resonance width of the impurity level is given by $\Delta=\Gamma_L+\Gamma_R$ 
for $U=0$,  with $\Gamma_\nu=\pi\rho_c \nu^2$  
the hybridization energy scale due to each of the leads ($\nu =L,R$).

\subsection{Fermi-liquid parameters}
\label{Fermi-liquid parameters}

We introduce here a set of the renormalized parameters 
which determine behavior of the transport coefficients, 
such as the conductance, current noise, and thermal conductivities, 
in the low-energy Fermi-liquid regime \cite{Teratani2020PRL}.  
Properties of the quasiparticles can be described in terms of 
the retarded Green's function:
\begin{align}
G_\sigma^r(\omega,T,eV)&\equiv-i\int_{0}^{\infty}dt\,e^{i(\omega+i0^+) t}\,\Bigl\langle\,\Bigl\{d_\sigma^{}(t),\,d_\sigma^\dagger\Bigr\}\Bigr\rangle 
\nonumber
\\ 
&=\frac{1}{\omega-\epsilon_{d\sigma}+i\Delta-\Sigma_\sigma^r(\omega,T,eV)}
\,,
\label{eG} 
\\ 
A_{\sigma}(\omega,T,eV)
&=\,-\frac{1}{\pi}\,\mathrm{Im}\, G_{\sigma}^r(\omega,\,T,\,eV)\,.
\label{spectral_green}
\end{align}
Here, $\Sigma_\sigma^r$    
is the self-energy due to the Coulomb interaction 
and $A_{\sigma}$ is the spectral function of impurity electrons.
The  nonequilibrium steady-state 
average $\langle \cdots \rangle$ is taken with 
 the statistical density matrix, 
 which is constructed at  finite bias voltages $eV$ and temperatures $T$,   
using the Keldysh formalism \cite{M_W1992,Hershfield1992}. 

The low-energy properties can be deduced specifically 
from the behavior of $G_\sigma^r$ at small $\omega$, $T$, and $eV$ 
in the vicinity of  the equilibrium ground state. 
In the following discussion, we use the notation
\begin{align}
\rho_{d\sigma}(\omega) \equiv A_{\sigma}(\omega,0,0), 
\qquad 
\Sigma_\sigma^\mathrm{eq}(\omega) \equiv  \Sigma_\sigma^r(\omega,0,0).
\end{align}
For small frequencies $\omega$, 
the Green's function takes the form  
\begin{align}
G_\sigma^\mathrm{eq}(\omega) \,\equiv \,
  G_\sigma^r(\omega,0,0)
\simeq\, 
\frac{z_\sigma}{\omega-\widetilde{\epsilon}_{d\sigma}+i\widetilde{\Delta}_{\sigma}}\,.
\end{align}
Here, $z_\sigma$ is the wave function renormalization factor, 
and $\widetilde{\Delta}_{\sigma}$ 
and $\widetilde{\epsilon}_{d\sigma}$ 
are the parameters that describe the renormalized resonance level, defined by 
\begin{align}
z_\sigma \, \equiv& \ \left\{\, 
1-\left.\frac{\partial 
\Sigma_\sigma^\mathrm{eq}(\omega)}{\partial \omega}\right|_{\omega=0}\, \right\}^{-1},
\label{zdef} \\
\widetilde{\Delta}_{\sigma}\, \equiv & \ z_\sigma\Delta,
\qquad 
\widetilde{\epsilon}_{d\sigma} \,\equiv\, z_\sigma\Bigl[\,
\epsilon_{d\sigma}+\Sigma_\sigma^\mathrm{eq}(0 )\Bigr]\,.
\label{eddef}
\end{align}

The occupation number  $\langle n_{d\sigma}\rangle$  
of the impurity level  determines ground-state properties. 
It can be obtained  from the free energy 
$\Omega\equiv-T\ln \left[\mathrm{Tr}\,e^{-H/T}\right]$ 
or  the phase shift $\delta_\sigma$ using the Friedel sum rule,    
\begin{align}
\langle n_{d\sigma}\rangle 
\,=& \ 
\frac{\partial \Omega}{\partial \epsilon_{d\sigma}}\xrightarrow{T\to0}\frac{\delta_\sigma}{\pi}\,,  
\label{ndFriedel}
\\
\delta_\sigma\, \equiv  & \ \cot^{-1}
\left[\frac{\epsilon_{d\sigma}
+\Sigma_{\sigma}^\mathrm{eq}(0)}{\Delta}\right]\,.
\end{align}
It also determines the density of states  $\rho_{d\sigma}$, 
at the Fermi level  $\omega=0$,  as
\begin{align}
\rho_{d\sigma}=\frac{\sin^2\delta_\sigma}{\pi\Delta}\,.  
\label{eq:rho_d_def}
\end{align}
Note that we are using the unit in which the Boltzmann constant 
is set as unity, i.e.,\  $k_B=1$.

The static linear susceptibilities $\chi_{\sigma\sigma'}$ 
are also  essential parameters that determine the  Fermi-liquid properties,  
\begin{align}
\chi_{\sigma\sigma'}\equiv&
-\frac{\partial^2 \Omega}{\partial \epsilon_{d\sigma}\partial\epsilon_{d\sigma'}}=\int_0^\frac{1}{T} \!  d\tau\,
\langle T_\tau\,\delta n_{d\sigma}(\tau)\,\delta n_{d\sigma'}\rangle.
\label{eq:linear_susceptibility}
\end{align}
Here,
$\delta n_{d\sigma}=n_{d\sigma}-\langle n_{d\sigma}\rangle$ is the fluctuation of the number of electrons in the dot from the average value, 
and  $T_\tau$ is an imaginary-time ordering operator.
The following Fermi-liquid relations hold  at $T=0$ 
between the susceptibilities and the parameters defined with respect 
to the self-energy \cite{Yamada1975II,Shiba1975,Yoshimori1976}, 
\begin{align}
\chi_{\sigma\sigma'}=\rho_{d\sigma}\widetilde{\chi}_{\sigma\sigma'},
\qquad \quad
\widetilde{\chi}_{\sigma\sigma'}\equiv\delta_{\sigma\sigma'}
+\frac{\partial \Sigma_\sigma^\mathrm{eq}(0)}{\partial \epsilon_{d\sigma'}},
\end{align}
and  $1/z_{\sigma} = \widetilde{\chi}_{\sigma\sigma}$. 
From this relation,  the $T$-linear specific heat of the impurity electrons 
$\mathcal{C}_\mathrm{imp}^{\mathrm{heat}}$ has been 
shown to be expressed in terms of  the diagonal part of the linear susceptibility 
\cite{Yamada1975II,Shiba1975,Yoshimori1976}, as 
\begin{align}
\mathcal{C}_\mathrm{imp}^{\mathrm{heat}} \,=\,   
\frac{\pi^2}{3}
\sum_\sigma \chi_{\sigma\sigma}^{} \, T \,.   
\end{align}

It has recently been shown that 
the static three-body correlation functions 
$\chi_{\sigma_1\sigma_2\sigma_3}^{[3]}$ 
also contribute to the next-leading terms of 
the transport coefficients when 
the system does not have particle-hole symmetry  
or time-reversal symmetry 
\cite{MMvDZ2015,FMvDM2018,AO2017_I,AO2017_II,AO2017_III}, 
\begin{align}
& \chi_{\sigma_1\sigma_2\sigma_3}^{[3]}\,\equiv
\, -\frac{\partial^3\Omega}{\partial\epsilon_{d\sigma_1}\partial\epsilon_{d\sigma_2}\partial\epsilon_{d\sigma_3}}
=\frac{\partial \chi_{\sigma_1\sigma_2}}{\partial \epsilon_{d\sigma_3}}
\nonumber
\\
=&-\int_0^\frac{1}{T} \! d\tau_1\int_0^\frac{1}{T} 
\! d\tau_2\, \langle T_\tau\delta n_{d\sigma_1}(\tau_1)\,
\delta n_{d\sigma_2}(\tau_2)\,\delta n_{d\sigma_3}\rangle\,.
\label{eq:3body_correlation}
\end{align}
This function has permutation symmetry for the indexes:   
$
\chi_{\sigma_1\sigma_2\sigma_3}^{[3]} 
 = 
\chi_{\sigma_2\sigma_1\sigma_3}^{[3]} 
 =  
\chi_{\sigma_3\sigma_2\sigma_1}^{[3]} 
 = 
\chi_{\sigma_1\sigma_3\sigma_2}^{[3]} = 
 \cdots$. 

The main subject of this paper is to clarify  how the three-body correlations contribute to the nonlinear transport 
when the additional asymmetries are present in the tunnel couplings and bias voltages. 
We also focus on the impurity level shift 
which is  induced by the bias voltages in the presence of these asymmetries 
 and demonstrate how it affects order $(eV)^2$ nonlinear current using the NRG.

\section{Current formula for junctions with 
tunnel and bias  asymmetries}

\label{Differential conductance}

In this section, we  derive the  formula for the nonlinear current 
in the low-energy Fermi-liquid regime, taking into 
account the asymmetries in the tunnel couplings 
and applied bias voltages.

\subsection{Steady-state average of nonlinear current}

We start with a quantum dot embedded in a tunnel junction as illustrated in  Fig.\ \ref{schematic}. 
The bias voltage $eV=\mu_L-\mu_R$ is applied through the chemical potential 
in the left lead  $\mu_L=\alpha_L\,eV$ and that 
in the right lead $\mu_R=-\alpha_R\,eV$.  
Here,  the parameters $\alpha_L$ and $\alpha_R$ 
are introduced such that   $\alpha_L+\alpha_R=1$ 
in order to specify the way 
in which the system has been driven from equilibrium.  
For $\alpha_L=\alpha_R =0.5$, it describes 
the bias as symmetric with respect to the Fermi level 
 $E_F=0$ at equilibrium. 
In an extreme $\alpha_L=1$ and $\alpha_R=0$,  
it describes the situation where 
the drain  is grounded at $\mu_R=E_F=0$ and  
the bias voltage is applied to the source side $\mu_L=eV$, 
corresponding to one natural experimental situation.

Specifically,  in this work, 
we examine  transport properties 
of the Anderson impurity for $N=2$ at zero field $b=0$, 
where  $\epsilon_{d\sigma}\equiv \epsilon_d$. 
In this case, the system has an SU(2) rotational symmetry in the spin space,
and thus the components of the correlation functions are 
related to each other, as      
 $\chi_{\uparrow\uparrow}=\chi_{\downarrow\downarrow}$, 
 $\chi_{\uparrow\uparrow\uparrow}^{[3]} = 
\chi_{\downarrow\downarrow\downarrow}^{[3]}$, 
and  $\chi_{\uparrow\downarrow\downarrow}^{[3]} = 
\chi_{\uparrow\uparrow\downarrow}^{[3]}$. 
Since the two spin components, $\uparrow$ and $\downarrow$, 
become equivalent in this case, we also use a simplified notation 
suppressing the  suffix for spin degrees of freedom 
from the phase shift $\delta$, the impurity  density of states
$\rho_d$ defined in Eq.\ \eqref{eq:rho_d_def}, and 
the self-energy. 
Furthermore,  the characteristic energy scale $T^\ast$ 
and the Wilson ratio $R$ can be expressed in the form 
\begin{align}
T^\ast \,\equiv\,
\frac{1}{4\chi_{\uparrow\uparrow}}\,,
\qquad \qquad R \,\equiv\, 
1-
\frac{\chi_{\uparrow\downarrow}}{\chi_{\uparrow\uparrow}}\,.
\label{Fermiparaorigin}
\end{align}

Nonequilibrium current through a quantum dot can be calculated using 
a Landauer-type formula \cite{Hershfield1992,M_W1992}, 
\begin{align}
I&=\frac{2e}{h}\,\frac{4\Gamma_L\Gamma_R}{(\Gamma_L+\Gamma_R)^2} \nonumber
\\
&\times\int_{-\infty}^\infty d\omega\,\Bigl[f(\omega-\mu_L)-f(\omega-\mu_R)\Bigr]\, \pi \Delta A(\omega,\,T,\,eV)\, .
\label{M-WFormula} 
\end{align}
Here, $f(\omega)=[e^{\beta\omega}+1]^{-1}$ is the Fermi distribution function,  
and $A(\omega,\,T,\,eV)$ is
the spectral function defined by Eq.\ \eqref{spectral_green}.
Thus at low energies, behavior  
of the spectral function $A(\omega,\,T,\,eV)$ 
for small $\omega$, $T$, and $eV$ determines the current. 
In particular, at zero temperature, 
 the spectral weight in the  bias window region $\mu_R\leq\omega\leq\mu_L$  
contributes to the current $I$ flowing through the quantum dot, 
\begin{align}
I &\xrightarrow{T\to 0}\frac{2e}{h}\,\frac{4\Gamma_L\Gamma_R}{(\Gamma_L+\Gamma_R)^2}\int_{\mu_R}^{\mu_L} d\omega\,\pi\Delta A(\omega,\,0,\,eV)\,. 
\label{M-Wabs0} 
\end{align}

In order to study the nonlinear current in the Fermi-liquid regime, 
 we use the exact low-energy asymptotic form of the self-energy,
 obtained recently up to terms of order $\omega^2$, $T^2\,$, and $(eV)^2$ \cite{FMvDM2018,AO2017_III}, 
specifically the extended one which is  applicable to 
 asymmetric junctions and bias voltages \cite{AO2017_III},
\begin{align}
&
\!\!\!  
\epsilon_{d}+{\rm Re} \, \Sigma_{}^r(\omega,T,eV)= \nonumber
\\
&\qquad \ \ 
\Delta \cot \delta \,+\, (1-\widetilde{\chi}_{\uparrow\uparrow})\,\omega 
\,+\,
\frac{1}{2}
\frac{\partial \widetilde{\chi}_{\uparrow\uparrow}}{\partial \epsilon_{d\uparrow}}
\omega^2 
\nonumber
\\
&\qquad 
+\frac{1}{6}\frac{1}{\rho_{d}}\frac{\partial \chi_{\uparrow\downarrow}}{\partial \epsilon_{d\downarrow}}\Biggl[ \, \frac{3\Gamma_L\Gamma_R}{(\Gamma_L+\Gamma_R)^2} (eV)^2+(\pi T)^2 \,\Biggr] 
\nonumber
\\ 
&\qquad 
- \widetilde{\chi}_{\uparrow\downarrow}\alpha \, eV
+
\frac{\partial \widetilde{\chi}_{\uparrow\downarrow}}{\partial \epsilon_{d\uparrow}} 
\alpha \, eV  \omega \,
+\frac{1}{2}
\frac{\partial \widetilde{\chi}_{\uparrow\downarrow}}{\partial \epsilon_{d\downarrow}} \alpha^2(eV)^2
+ \cdots,  
\label{ReSelf}
\\
&{\rm Im} \, \Sigma_{}^r (\omega,T,eV)=-\frac{\pi}{2} \frac{1}{\rho_{d}} 
\chi_{\uparrow\downarrow}^2 \Biggl[ \,(\omega-\alpha eV)^2 \nonumber
\\ 
&
\qquad \qquad \quad + \frac{3\Gamma_L\Gamma_R}{(\Gamma_L+\Gamma_R)^2} (eV)^2+(\pi T)^2 \, \Biggr]+\cdots\,. 
\label{ImSelf}
\end{align}

The parameter $\alpha$ plays a central role throughout this work,  
and it consists of the following two parts 
which represent the tunnel and the bias asymmetries, 
respectively, 
\begin{align}
\alpha\equiv\frac{\alpha_L\Gamma_L-\alpha_R\Gamma_R}{\Gamma_L+\Gamma_R}
\,=\,\frac{1}{2}(\alpha_L-\alpha_R)
+\frac{1}{2}\frac{\Gamma_L-\Gamma_R}{\Gamma_L+\Gamma_R}.
\label{alphadivided}
\end{align}
Note that  $\alpha_L+\alpha_R=1$. 
The bias asymmetry can also be described by 
a single parameter, as  $\alpha_L=(1+\alpha_{\mathrm{dif}}^{})/2$ and 
 $\alpha_R=(1-\alpha_{\mathrm{dif}}^{})/2$, with
\begin{align}
\alpha_{\mathrm{dif}}^{} \,\equiv\, \alpha_L-\alpha_R\,, 
\label{eq:alpha_diff}
\end{align}
which takes values in the 
range 
$-1 \leq \alpha_{\mathrm{dif}}^{}\leq 1$. 
The coefficients for low-energy expansion of the self-energy up to quadratic order terms, 
given in Eqs.\ \eqref{ReSelf} and \eqref{ImSelf}, are  
described by five  Fermi-liquid parameters, i.e.,\     
the phase shift $\delta$, the susceptibilities 
$\chi_{\uparrow\uparrow}$ and $\chi_{\uparrow\downarrow}$, 
and the two  three-body correlations 
 $\chi_{\uparrow\uparrow\uparrow}^{[3]}$ 
and $\chi_{\uparrow\downarrow\downarrow}^{[3]}$. 
In particular, the three-body correlations determine  the quadratic-order terms 
of the real part  $\mathrm{Re}\, \Sigma_{}^r(\omega,T,eV)$.

\begin{widetext}

\begin{table}[t]
\caption{
Low-energy expansion of $dI/dV$  for $N=2$ at  $T=0$. 
The tunnel and bias asymmetries enter through the parameters  
 $(\Gamma_L-\Gamma_R)/\Delta$ and 
$\alpha_\mathrm{dif}^{}\equiv \alpha_L -\alpha_R$, respectively,  
in addition to the factor 
${4\Gamma_L\Gamma_R}/{(\Gamma_L+\Gamma_R)^2}$ in front. 
The characteristic energy scale $T^*= 1/(4\chi_{\uparrow\uparrow})$ and  
the Wilson ratio $R-1= -\chi_{\uparrow\downarrow}/\chi_{\uparrow\uparrow}$ 
are determined by the linear susceptibilities  $\chi_{\uparrow\uparrow}$ and 
 $\chi_{\uparrow\downarrow}$. 
The three-body contributions 
 $\chi_{\uparrow\uparrow\uparrow}^{[3]}$
and  $\chi_{\uparrow\downarrow\downarrow}^{[3]}$
 enter through dimensionless parameters 
$\Theta_\mathrm{I}$ and $\Theta_\mathrm{II}$. 
}
\begin{tabular}{l} 
\hline \hline
$
\frac{dI}{dV}= 
\frac{2e^2}{h}
\frac{4\Gamma_L\Gamma_R}{(\Gamma_L+\Gamma_R)^2}
\left[\,\sin^2\delta 
+\,C_V^{(2)}\left(\frac{eV}{T^\ast}\right)
\,-\,C_V^{(3)}\left(\frac{eV}{T^\ast}\right)^2\,+\,\cdots\right]$\,,  
\rule{0.4cm}{0cm}
$\Theta_\mathrm{I}\equiv
\frac{\sin2\delta}{2\pi\chi_{\uparrow\uparrow}^2}\,
\chi_{\uparrow\uparrow\uparrow}^{[3]}$,
\rule{0.3cm}{0cm}
$\Theta_\mathrm{II}\equiv\frac{\sin2\delta}{2\pi\chi_{\uparrow\uparrow}^2}\,
\chi_{\uparrow\downarrow\downarrow}^{[3]}$\,, 
\rule{0cm}{0.70cm}
\\
\hline
$C_V^{(3)}=\frac{\pi^2}{64}\,(W_V\,+\,\Theta_V)$,
\rule{0cm}{0.60cm}
 \rule{3.4cm}{0cm}
$C_V^{(2)}=\frac{\pi}{4}\,
\Bigl[\,
\alpha_\mathrm{dif}^{}\,
- \left( \alpha_\mathrm{dif}^{}\,+\, 
\frac{\Gamma_L-\Gamma_R}{\Gamma_L+\Gamma_R}\right) (R-1)
\,\Bigr]\,\sin2\delta$, 
\rule{0cm}{0.60cm}
\\
$W_V=
 -\cos2\delta\,
\Biggl[ 
 1 + 3\alpha_\mathrm{dif}^{2} \,
-6 \alpha_\mathrm{dif}^{}  \left\{
\alpha_\mathrm{dif}^{} \,+
\,\frac{\Gamma_L-\Gamma_R}{\Gamma_L+\Gamma_R}
\right\} (R-1)  
+   \left\{
5 + 3\alpha_\mathrm{dif}^{2} 
+6\,\alpha_\mathrm{dif}^{}
\,\frac{\Gamma_L-\Gamma_R}{\Gamma_L+\Gamma_R}
\right\}
(R-1)^2
\,\Biggr]$,
\rule{0cm}{0.60cm}
\\
$\Theta_V=\Bigl[\, 1 + 3\alpha_\mathrm{dif}^{2} \,\Bigr]\,\Theta_\mathrm{I}\,
+\,
3\left[\, 1 + 3\alpha_\mathrm{dif}^{2} + 
4  \alpha_\mathrm{dif}^{}\,\frac{\Gamma_L-\Gamma_R}{\Gamma_L+\Gamma_R}
\,\right] 
\,\Theta_\mathrm{II}$. 
\rule{0cm}{0.60cm}
\\
\hline
\hline
\end{tabular}
\label{tab:C_W_SUN}
\end{table}

\end{widetext}

\subsection{Current formula for the local Fermi liquid}

 We obtain the generic expression for the differential conductance $dI/dV$ 
which is exact up to terms of order $T^2$ and $(eV)^2$, 
substituting the above asymptotic form of the self-energy 
 into the energy denominator of the Green's function given in   
 Eq.\ (\ref{eG}), and then carrying out the $\omega$ integral
 in the Meir-Wingreen formula (see Appendix A), 
\begin{align}
\!\!\! 
\frac{dI}{dV}\,=& \ 
\frac{2e^2}{h}\frac{4\Gamma_L\Gamma_R}{(\Gamma_L+\Gamma_R)^2}
\Biggl[\,\sin^2\delta\, 
-\,C_T^{}\left(\frac{\pi T}{T^\ast}\right)^2\,
\nonumber
\\
& \qquad 
+\,C_V^{(2)}\left(\frac{eV}{T^\ast}\right)\,-\,C_V^{(3)}\left(\frac{eV}{T^\ast}\right)^2\,+\,\cdots\Biggr]\,. 
\label{DiffCond}
\end{align}
The first  two terms on the right-hand side correspond to the linear conductance: 
the coefficient $C_T^{}$ for the $T^2$ term is given by  
\begin{align}
C_T^{}&=\frac{\pi^2}{48}\,\Bigl(\, W_T\,
+\,\Theta_\mathrm{I}\,+\,\Theta_\mathrm{II} \, \Bigr)\,, 
\\
W_T&=-\,\cos2\delta\, \Bigl[\, 1+2(R-1)^2 \,\Bigr] \,,  
\nonumber
\end{align}
 with   $\Theta_\mathrm{I}$ and $\Theta_\mathrm{II}$ the dimensionless 
three-body correlation functions, defined by 
\begin{align}
\Theta_\mathrm{I}&\equiv
\frac{\sin2\delta}{2\pi\chi_{\uparrow\uparrow}^2}\,
\chi_{\uparrow\uparrow\uparrow}^{[3]}\,,
\qquad \quad 
\Theta_\mathrm{II}\equiv\frac{\sin2\delta}{2\pi\chi_{\uparrow\uparrow}^2}\,
\chi_{\uparrow\downarrow\downarrow}^{[3]}\,.
\label{eq:ThetaI_II_def}
\end{align}
Note that  $\sin2\delta$ is proportional to the derivative of the spectral weight, as  
\begin{align}
\sin2\delta
 =\frac{\Delta}{\chi_{\uparrow\uparrow}^{}}
\frac{\partial \rho_{d}^{}(\omega)}{\partial \omega}
\Biggr|_{\omega=0}\,. 
\label{eq:sin2delta_def}
\end{align}
The linear-response part of the differential conductance 
depends on the tunnel asymmetry only through the prefactor 
$4\Gamma_L\Gamma_R/(\Gamma_L+\Gamma_R)^2$ 
on the right-hand side of Eq.\ \eqref{DiffCond},
and it does not
depend on the bias asymmetry $\alpha_\mathrm{dif}^{}$.

In contrast, the nonlinear part of $dI/dV$, 
i.e.,\  $C_V^{(2)}$ and $C_V^{(3)}$,  
 depends essentially on the tunnel and bias asymmetries:  
\begin{align}
C_V^{(2)}
&=
\frac{\pi}{4} \left[ 
\alpha_\mathrm{dif}^{}
- \left(\alpha_\mathrm{dif}^{}+\frac{\Gamma_L-\Gamma_R}{\Gamma_L+\Gamma_R}\right)(R-1)
\right] \sin2\delta,
\label{Cv2}
\\ 
C_V^{(3)}\,&=\,
\frac{\pi^2}{64}\,\Bigl(\, W_V\,+\,\Theta_V \,\Bigr)\,. 
\label{Cv3eq}
\end{align}
Here,  $W_V$ and $\Theta_V$ represent the two-body and three-body 
contributions, 
\begin{align}
&W_V\,=\,  -\cos2\delta 
\Biggl[ 
 1 + 3\alpha_\mathrm{dif}^{2} \,
\nonumber
 \\
&   \qquad \quad
-6 \alpha_\mathrm{dif}^{}  \left(
\alpha_\mathrm{dif}^{} \,+
\,\frac{\Gamma_L-\Gamma_R}{\Gamma_L+\Gamma_R}
\right) (R-1)  
\nonumber \\
& \qquad  \quad 
+   \left(
5 + 3\alpha_\mathrm{dif}^{2} 
+6\,\alpha_\mathrm{dif}^{}
\,\frac{\Gamma_L-\Gamma_R}{\Gamma_L+\Gamma_R}
\right)
(R-1)^2
\,\Biggr] , 
\label{Wv}
\\
&\Theta_V
\,=\,
 \Bigl[\, 1 + 3\alpha_\mathrm{dif}^{2} \,\Bigr]\, 
\Theta_\mathrm{I}\,\, \nonumber
\\
& \qquad \quad 
+3\left[\, 1 + 3\alpha_\mathrm{dif}^{2} + 
4 \alpha_\mathrm{dif}^{}\,\frac{\Gamma_L-\Gamma_R}{\Gamma_L+\Gamma_R}
\,\right] \Theta_\mathrm{II}\, .
\label{THv}
\end{align}
In order to express the coefficients in this form, 
we have used Eqs.\  \eqref{alphadivided} and \eqref{eq:alpha_diff} and 
the following relation between the asymmetry parameters  
 due to $\alpha_L+\alpha_R=1$, 
\begin{align}
\alpha^2 +  \frac{\Gamma_L \Gamma_R}{\Delta^2} 
\,= \, 
\frac{1}{4} \left[
1+ \alpha_\mathrm{dif}^{2}
+2 \alpha_\mathrm{dif}^{}\,\frac{\Gamma_L-\Gamma_R}{\Gamma_L+\Gamma_R}
\right] . 
\label{eq:alpha_relation}
\end{align}
Note that the dependence of 
$C_V^{(2)}$ and $C_V^{(3)}$ on 
  tunnel asymmetries $(\Gamma_L-\Gamma_R)/\Delta$ 
enters through the self-energy given in Eqs.\ \eqref{ReSelf} and \eqref{ImSelf}, 
and thus it emerges only for interacting-electron systems with finite $U$. 
While $C_V^{(2)}$  depends on the Coulomb interaction 
only through  the real part of the self-energy, 
the coefficient $C_V^{(3)}$ depends also 
on the  $\omega^2$  and $(eV)^2$ imaginary parts  
that destroy phase coherence \cite{Zaland2004,Kehrein2005}. 
In contrast, the  bias asymmetry 
$\alpha_\mathrm{dif}^{}$ 
affects these coefficients already at $U=0$.

The expression of $C_V^{(2)}$  given in Eq.\ \eqref{Cv2} agrees 
with the previous result derived by Aligia\ using the renormalized perturbation theory 
\cite{Aligia2011}. 
For the current of order  $(eV)^3$,
 the coefficients $\Theta_V$ and $W_V$ represent 
the three-body contributions and the two-body ones, respectively.
Our result for  $C_V^{(3)}$ extends the previous one  
which was derived  for  the particle-hole symmetric case 
by Sela and Malecki \cite{Sela-Malecki2009} 
to the particle-hole asymmetric case taking place away from half-filling.
These formulas for $dI/dV$ in the Fermi-liquid regime 
are also summarized in Table \ref{tab:C_W_SUN} for quick reference.

\subsubsection{Inversion of ($\alpha_L,\Gamma_L$) 
and ($\alpha_R,\Gamma_R$)}

One of the common features of the Landauer type conductance, 
defined in Eq.\ \eqref{M-WFormula}, is 
 a factor  ${4\Gamma_L\Gamma_R}/{(\Gamma_L+\Gamma_R)^2}$ 
in front that  suppresses the current  for asymmetric tunnel junctions. 
The explicit expressions of $dI/dV$ given in the above 
show explicitly that the nonlinear current components of the current 
depend further on the junction asymmetries.     
The coefficients  $C_V^{(2)}$ and $C_V^{(3)}$ vary as 
the tunnel and bias asymmetries enter also through 
the terms with the parameters  
$(\Gamma_L-\Gamma_R)/\Delta$ and $\alpha_\mathrm{dif}^{}$  
 in Eqs.\  \eqref{Cv2}--\eqref{THv}. 
Thus, these coefficients can be regarded as functions 
of the two asymmetry parameters 
$\alpha_{\mathrm{dif}}^{}$ and $\Gamma_L/\Gamma_R$ 
for a fixed value of $\Delta = \Gamma_L+\Gamma_R$.  

The order $(eV)^2$ component  has the following properties 
with respect to an inversion of 
 ($\alpha_L,\Gamma_L$) and ($\alpha_R,\Gamma_R$),   
\begin{align}
C_V^{(2)} (\alpha_{\mathrm{dif}}^{},\Gamma_L/\Gamma_R) 
\,=\, -\,
C_V^{(2)} (-\alpha_{\mathrm{dif}}^{},\Gamma_R/\Gamma_L) \;.
\label{eq:LR_inversion_C2}
\end{align}
It shows that the order $(eV)^2$ nonlinear current emerges 
when the system has the tunnel or the bias asymmetry, 
i.e.,\ $\alpha_\mathrm{dif}^{}\neq 0$ 
or $\Gamma_L \neq \Gamma_R$.  
In addition, Eq.\ \eqref{Cv2} also shows that 
this component of $I$ vanishes at half filling, where $\delta=\pi/2$. 

Similarly, the order $(eV)^3$ component  
 is invariant with respect to the same inversion of $L \Leftrightarrow R$,  
\begin{align}
C_V^{(3)} (\alpha_{\mathrm{dif}}^{},\Gamma_L/\Gamma_R) 
\,=\, 
C_V^{(3)} (-\alpha_{\mathrm{dif}}^{},\Gamma_R/\Gamma_L) \;.
\label{eq:LR_inversion_C3}
\end{align}
In the rest of this paper,
we demonstrate how the coefficients  $C_V^{(2)}$ and $C_V^{(3)}$ 
vary over a wide range of the asymmetries 
and of the gate voltage, by which the ground state evolves 
from the Kondo regime toward the valence fluctuation regime.

\section{NRG results for Fermi-liquid parameters}

\label{Gate voltage dependence of FL parameters}

We discuss in this section how the five Fermi-liquid (FL) parameters,
 $\delta$, $\chi_{\uparrow\uparrow}$, $\chi_{\uparrow\downarrow}$, 
 $\chi_{\uparrow\uparrow\uparrow}^{[3]}$, 
and $\chi_{\uparrow\downarrow\downarrow}^{[3]}$, 
 evolve as the impurity level  $\epsilon_d$ and the Coulomb 
interaction $U$ are varied \cite{Wilson1975,Hewson2004}.
These parameters are defined with respect to 
the equilibrium ground state,  
for which only a single  {\it bonding\/}-type channel constructed 
by a linear combination of the left and right conduction bands,
$\psi_{\sigma}^\mathrm{bond} 
\equiv (v_L \,\psi_{L,\sigma}+v_R\,\psi_{R,\sigma})/\sqrt{v_L^2+v_R^2}$,
is coupled to the impurity level and the other channel is separated. 
Thus,  the FL parameters themselves do not depend 
on the tunnel nor bias asymmetries, 
and numerical  results have been reported previously 
 \cite{MMvDZ2015,FMvDM2018,AO2017_I,AO2017_II,AO2017_III}. 
Nevertheless, behaviors of the FL parameter determine the coefficients 
 of the nonlinear current  $C_V^{(2)}$ and $C_V^{(3)}$, 
defined in Eqs.\  \eqref{Cv2}--\eqref{THv},  
with  the other parameters 
$(\Gamma_L-\Gamma_R)/\Delta$ and $\alpha_\mathrm{dif}^{}$,  
and play an essential role also for transport through junctions with
the tunnel and bias asymmetries.   

\begin{figure}[t]
	\begin{minipage}[r]{\linewidth}
	\centering
	\includegraphics[keepaspectratio,scale=0.31]{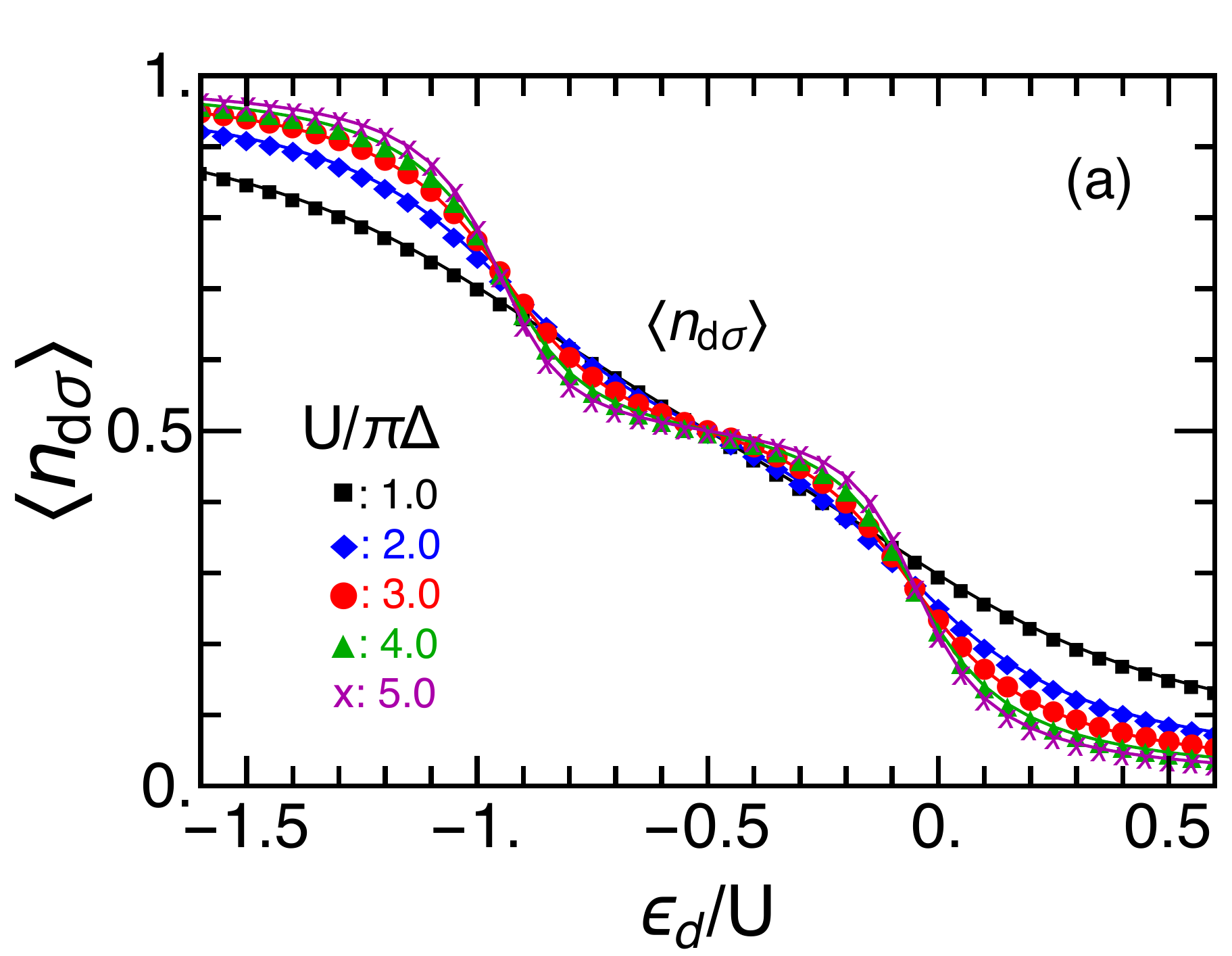}
	\end{minipage}
	\begin{minipage}[r]{\linewidth}
	\centering
	\includegraphics[keepaspectratio,scale=0.31]{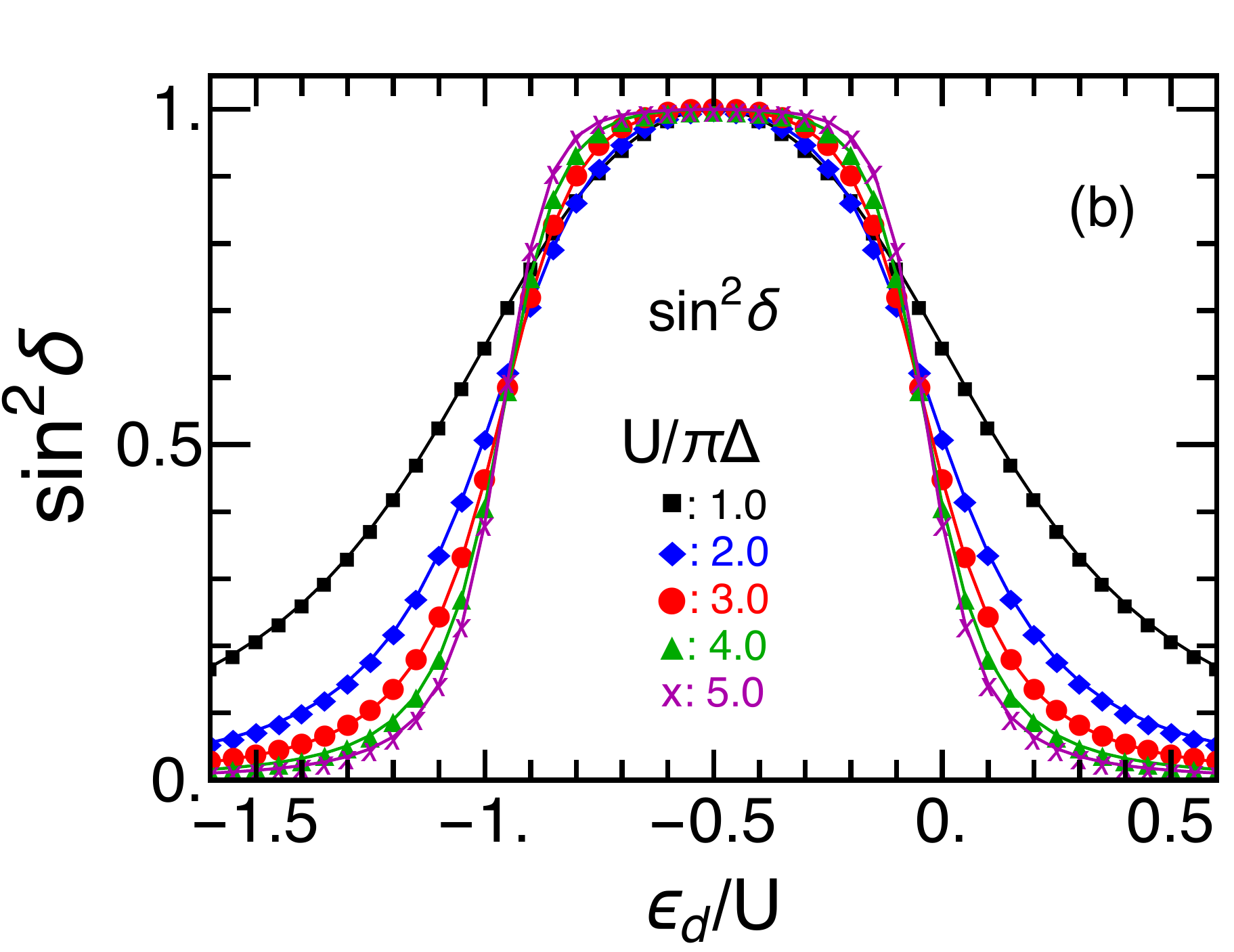}
	\end{minipage}
	\begin{minipage}[l]{\linewidth}
	\centering
	\includegraphics[keepaspectratio,scale=0.31]{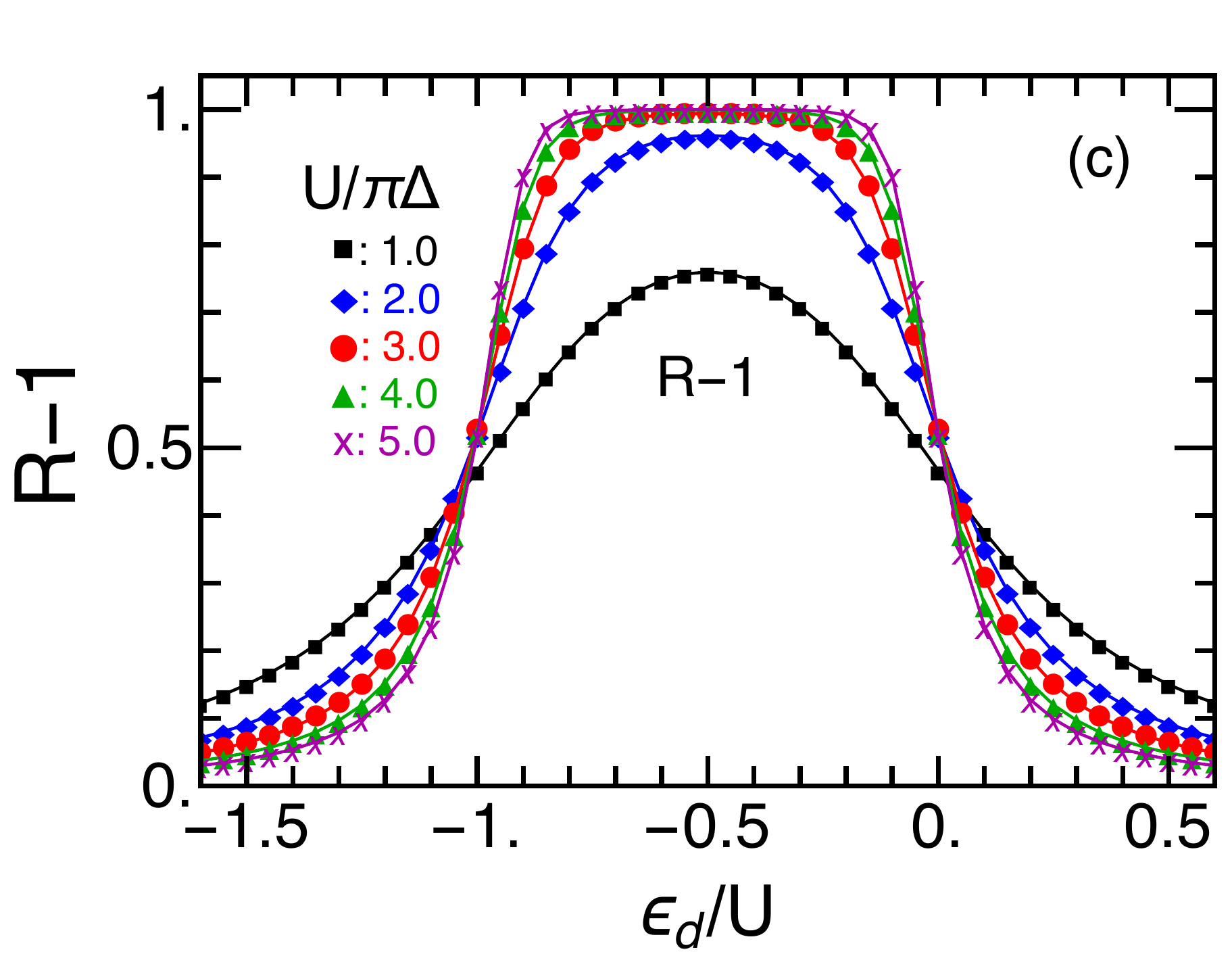}
	\end{minipage}
	\begin{minipage}[r]{\linewidth}
	\centering
	\includegraphics[keepaspectratio,scale=0.31]{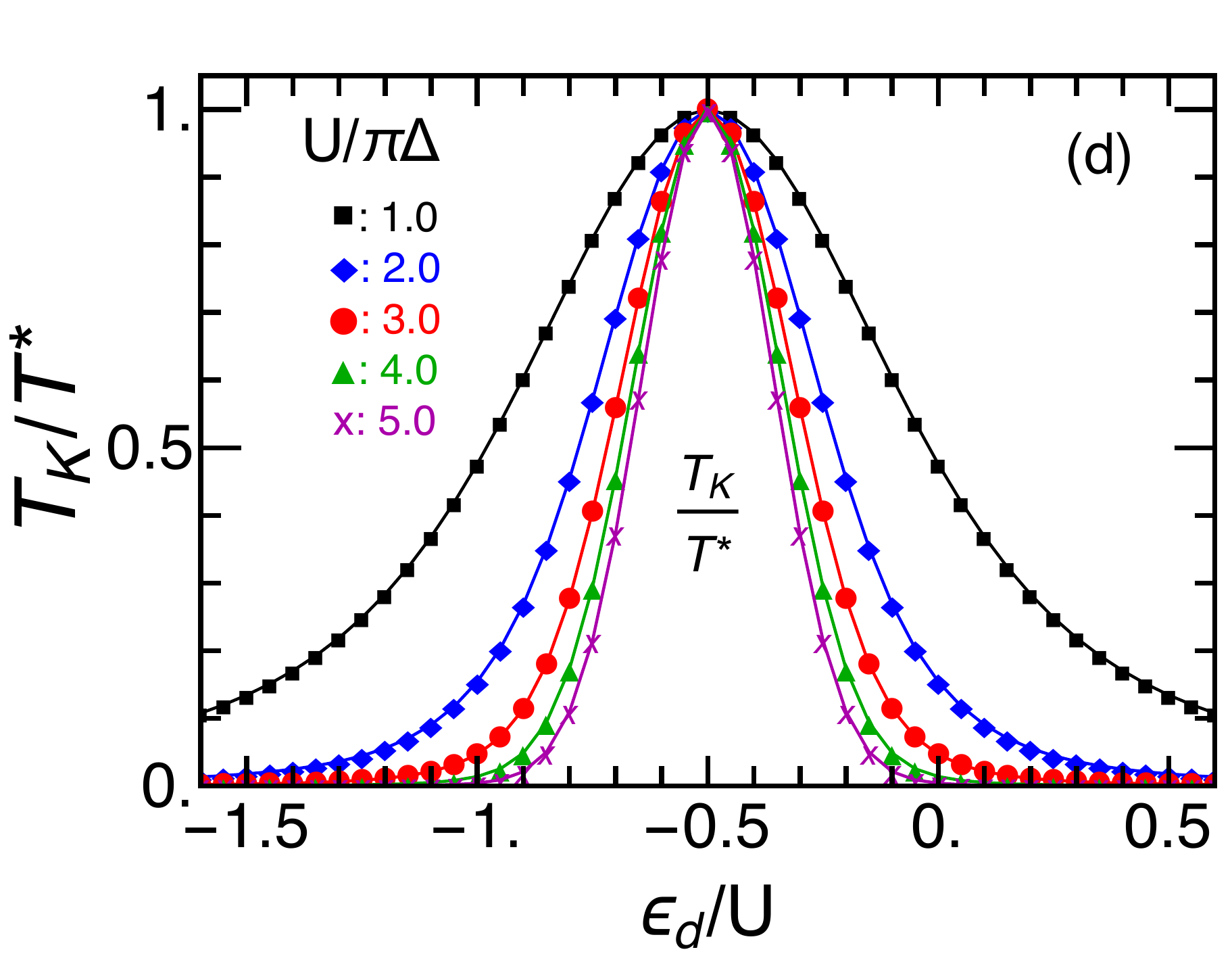}
	\end{minipage}
\caption{
NRG results  for the Fermi-liquid parameters are  
 plotted vs $\epsilon_d/U$, for   
 $U/(\pi\Delta) =1, 2, 3, 4, 5$.  
(a) $\langle n_{d\sigma}\rangle$, 
 (b) $\sin^2\delta$,   (c) 
$R-1=-\chi_{\uparrow\downarrow}/\chi_{\uparrow\uparrow}$, 
and  (d) $T_K/T^\ast$  
for which  $T^\ast  \equiv 1/(4\chi_{\uparrow\uparrow})$ depends on $\epsilon_d/U$    
and $T_K$ is defined as the value of  $T^\ast$ 
at half filling $\epsilon_d=-U/2$. 
}
  \label{Fermipara}
\end{figure}

\begin{figure}[t]
\begin{minipage}[t]{\linewidth}
	\centering
	\includegraphics[keepaspectratio,scale=0.4]{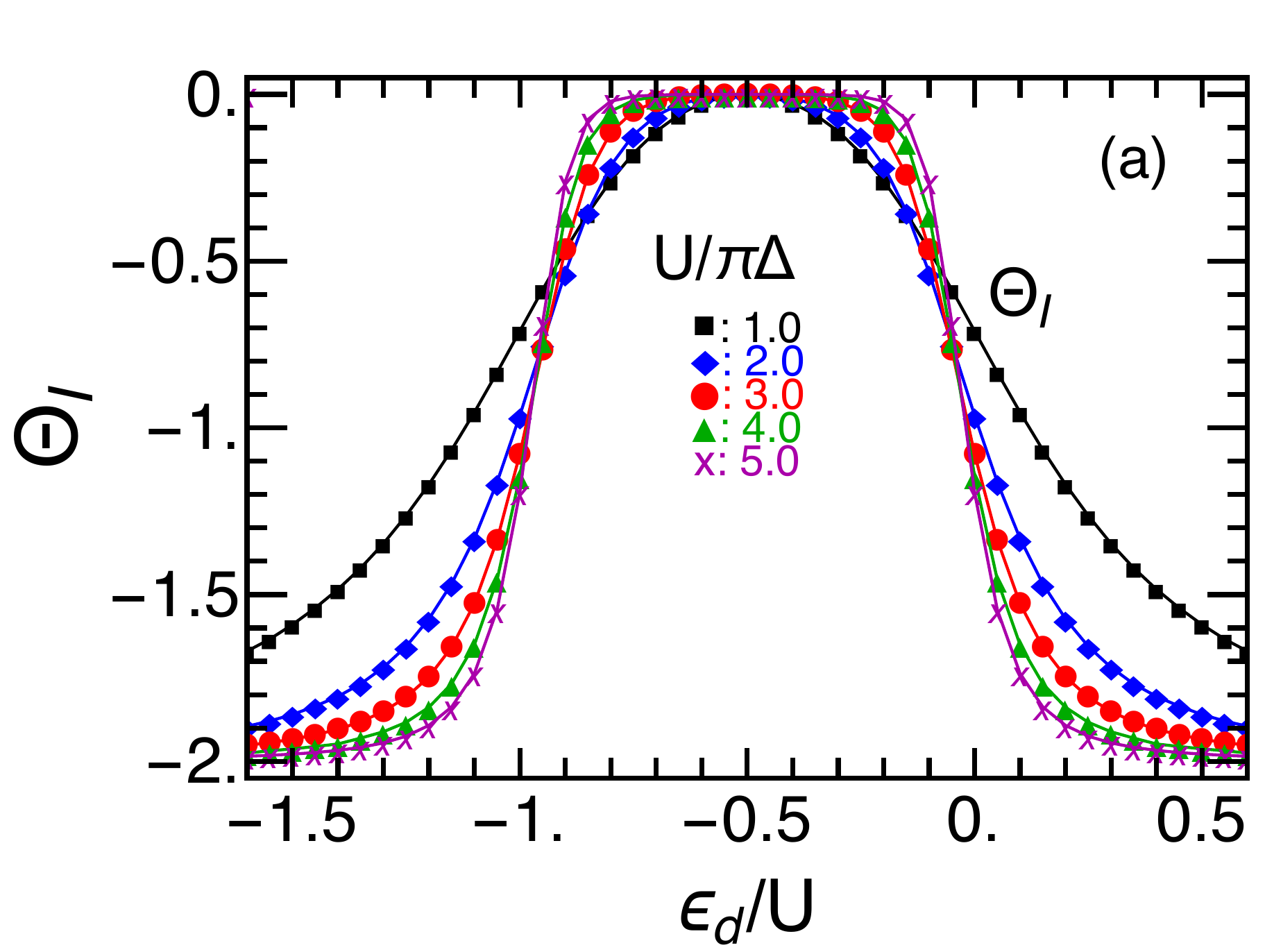}
\end{minipage}
\begin{minipage}[t]{\linewidth}
	\centering
	\includegraphics[keepaspectratio,scale=0.4]{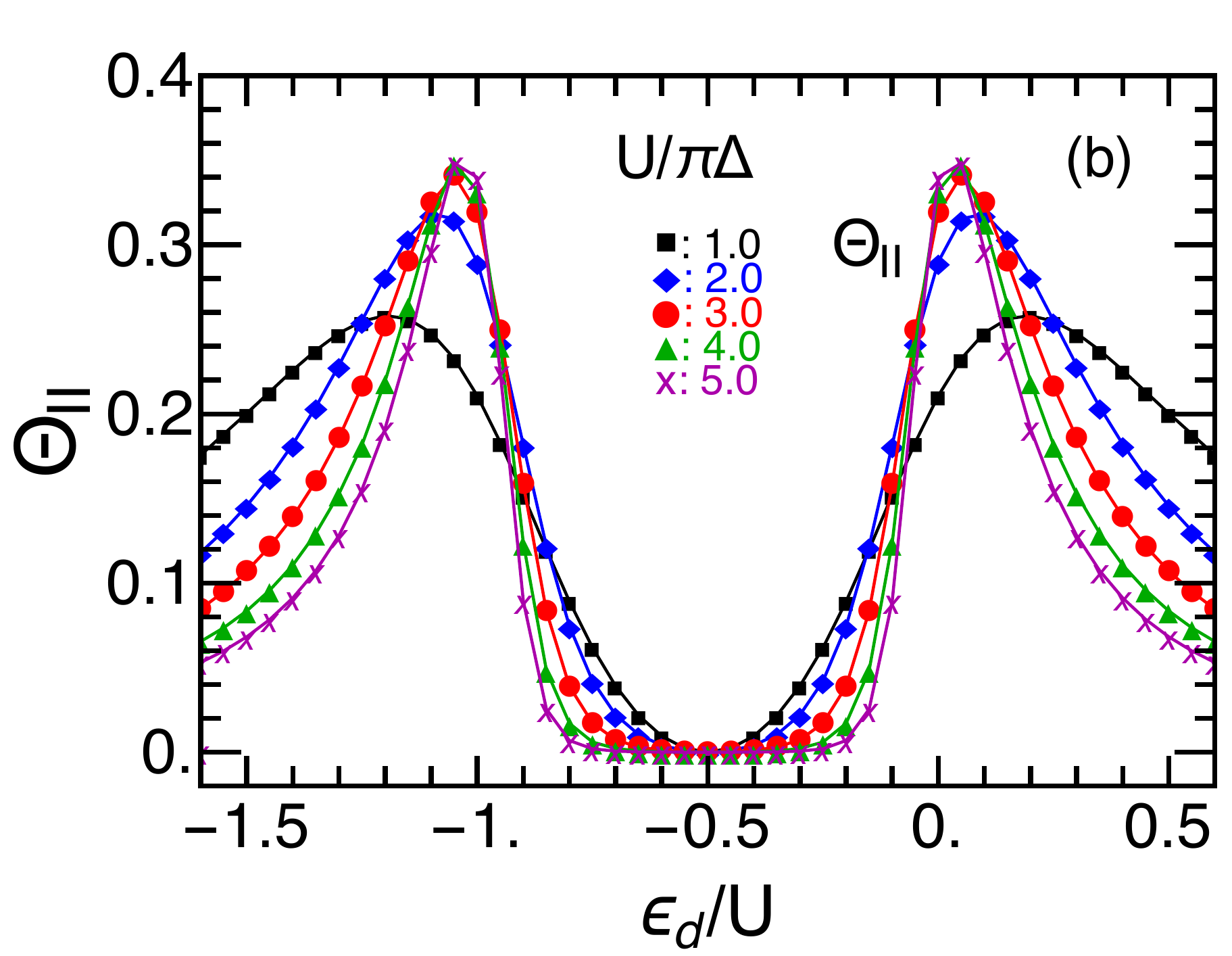}
\end{minipage}

\caption{
NRG results for the three-body correlation functions,   
 $\Theta_\mathrm{I}\equiv
\frac{\sin2\delta}{2\pi\chi_{\uparrow\uparrow}^2}\,
\chi_{\uparrow\uparrow\uparrow}^{[3]}$ and 
$\Theta_\mathrm{II}\equiv\frac{\sin2\delta}{2\pi\chi_{\uparrow\uparrow}^2}\,
\chi_{\uparrow\downarrow\downarrow}^{[3]}$, 
are plotted vs $\epsilon_d/U$ in (a) and (b), respectively, 
for  $U/(\pi\Delta) =1, 2, 3, 4, 5$.
}
\label{Thetaasym}
\end{figure}

We have calculated the FL parameters with the NRG approach,  
 choosing the discretization parameter to be  $\Lambda=2.0$ 
and keeping $N_\mathrm{trunc}=3600$ low-lying energy 
states at each step of the iterative procedure. 
Our NRG code uses the global  
$\mathrm{U}\left(1\right)\otimes\mathrm{SU}\left(2\right)$ 
symmetries and the method 
described in Refs.\ \onlinecite{Hewson2004,Teratani2020PRB}. 
The three-body correlation functions have been deduced  
from the linear susceptibilities  $\chi_{\uparrow\uparrow}$ 
and $\chi_{\uparrow\downarrow}$,  using the following relations, 
\begin{align}
\chi_{\uparrow\uparrow\uparrow}^{[3]}=\frac{\partial \chi_{\uparrow\uparrow}}{\partial \epsilon_d}-\frac{1}{2}\,\frac{\partial \chi_{\uparrow\downarrow}}{\partial \epsilon_d},\qquad 
\chi_{\uparrow\downarrow\downarrow}=\frac{1}{2}\,\frac{\partial \chi_{\uparrow\downarrow}}{\partial \epsilon_d}\,.
\end{align}

\noindent
Figure  \ref{Fermipara} shows the FL parameters  
 $\langle n_{d\sigma}\rangle$,  $\sin^2\delta$,  
the Wilson ratio $R-1$, and the inverse of $T^\ast$ 
as functions of  $\epsilon_d/U$, 
obtained with the NRG  
for several different values of the Coulomb interaction $U/(\pi\Delta)$ 
taking $\pi \Delta =D/100\,$ \cite{AO2017_III}. 
 Note that the system has particle-hole (PH) symmetry at  $\epsilon_d =-U/2$. 

We can see in panel (a) that 
the average number of  impurity electrons, 
 $\langle n_{d\sigma}\rangle = \delta/\pi$, 
shows the Coulomb staircase behavior, 
and the  impurity level is singly occupied  
 $\langle n_{d\uparrow} + n_{d\downarrow} \rangle \simeq 1.0$ 
in the Kondo regime $-U\lesssim\epsilon_d\lesssim0$.   
The second panel, Fig.\ \ref{Fermipara}(b), 
shows $\sin^2\delta$, which corresponds to the zero-bias conductance.  
Thus, the broad peak of the unitary-limit value  
 $\sin^2 \delta  \simeq  1.0$  at $ -U\lesssim\epsilon_d\lesssim0$ 
describes the Kondo ridge of the conductance at $T=0$. 
The Wilson ratio, plotted in Fig.\ \ref{Fermipara}(c),  
 also shows a broad ridge the height of which reaches 
the strong-coupling value 
$R-1 \simeq 1$ for $U \gtrsim 3.0 \pi \Delta $ in the Kondo regime.  
Panel (d) shows the $\epsilon_d$ dependence of  $1/T^\ast$,  
which is normalized by $T_K$ defined as the value of $T^\ast$ 
at half filling $\epsilon_d=-U/2$.  The characteristic temperature $T^\ast$ 
increases rapidly away from half filling as $\epsilon_d$ deviates 
from the particle-hole symmetric point.

These results also demonstrate clearly that the 
Fermi-liquid parameters approach the noninteracting values 
far away from half filling $|\epsilon_d+U/2| \gg U/2$ (see Appendix \ref{Cv3edinfty}).  
This is because in the limit $\epsilon_d\to\infty$   ($\epsilon_d\to -\infty$)
the impurity level is almost empty (fully filled) 
and the effects of the Coulomb interaction become less important.

The  results for the  three-body correlations 
 $\Theta_\mathrm{I}$ and $\Theta_\mathrm{II}$, 
defined in Eq.\ \eqref{eq:ThetaI_II_def}, 
are plotted in Fig.\ \ref{Thetaasym}
 as  functions of $\epsilon_d/U$ for several values  of $U/(\pi\Delta)$. 
These two dimensionless parameters, 
 $\Theta_{\mathrm{I}}$ and $\Theta_{\mathrm{II}}$,  
vanish in the particle-hole symmetric case  $\epsilon_{d}=-U/2$, 
and are suppressed 
 in the wide Kondo regime $-U \lesssim \epsilon_d \lesssim 0$ for large $U$.
At both ends of this region near $\epsilon_d \simeq 0$ and $-U$, 
 the three-body correlations evolve significantly and 
  play an important role in next-leading order terms of the transport   
coefficients, i.e.,\  $C_T^{}$, $C_V^{(2)}$, and $C_V^{(3)}$
 in Eq.\ \eqref{DiffCond}. 
In particular, $\Theta_{\mathrm{II}}$  has a peak in the valence fluctuation regime, 
which becomes sharper as $U$ increases. 
Farther away from half filling $|\epsilon_d| \to \infty$,  
the three-body correlations also approach the noninteracting values
$\Theta_{\mathrm{I}}\to -2$ and $\Theta_{\mathrm{II}}\to 0$ 
(see Appendix \ref{Cv3edinfty}).
Note that the three-body susceptibility $\chi_{\sigma\sigma'\sigma''}^{[3]}$ 
itself is an odd function of  $\xi_d \equiv \epsilon_{d}+U/2$. 
The dimensionless parameters $\Theta_{\mathrm{I}}$ and 
 $\Theta_{\mathrm{II}}$ become even functions of  $\xi_{d}$ 
since $\sin2\delta$ is also an odd function.

\section{Properties of order $(eV)^2$ nonlinear current}

\label{Cv2Results}

In this section, we examine how  the order  $(eV)^2$ term 
of the nonlinear current $I$ varies with gate voltage $\epsilon_d$ and Coulomb interaction $U$ 
in the presence of the tunnel and bias asymmetries.

\subsection{General properties of $C_V^{(2)}$}

The coefficient defined in  Eq.\ \eqref{Cv2}  consists of two parts: 
 $C_V^{(2)}=  (\pi/4)\left[\,
\overline{C}_{V}^{(2a)}+ \overline{C}_{V}^{(2b)} \,\right] \sin2\delta\,$ 
with 
\begin{align}
\overline{C}_{V}^{(2a)}& \equiv \,
\alpha_{\mathrm{dif}}^{}
\,,
\label{Cv21} 
\\
\overline{C}_{V}^{(2b)}& \equiv \,
-\left(\alpha_{\mathrm{dif}}^{}
\,+\, \frac{\Gamma_L-\Gamma_R}{\Gamma_L+\Gamma_R}
\right)(R-1)\,.
\label{Cv22}
\end{align}
These two parts represent contributions of  
 $\omega$-linear and $eV$-linear terms, respectively, 
 of  the low-energy expansion of the spectral function $A(\omega, T, eV)$, 
described in Appendix \ref{Differential conductanceAppendix}.  

As shown in Fig.\  \ref{Fermipara} (c),  
the Wilson ratio has a plateau of the height $R-1\simeq 1$ 
in the Kondo regime, 
i.e.,\  $-U\lesssim \epsilon_d \lesssim 0$ for $U/(\pi \Delta) \gtrsim 2$.  
Thus, in this region, the coefficient 
 becomes independent of bias asymmetry, 
\begin{align}
C_V^{(2)}\xrightarrow{R-1 \to 1\,}\, -\frac{\pi}{4}\,
\frac{\Gamma_L-\Gamma_R}{\Gamma_L+\Gamma_R}\,\sin2\delta\,. 
\label{Cv2Kondo}
\end{align}
Outside the Kondo regime, especially in the limit of $|\epsilon_d|\to\infty$, 
the Wilson ratio approaches the noninteracting value $R-1\to 0$, 
and the coefficient becomes independent of 
tunnel asymmetry,
\begin{align}
C_V^{(2)}\xrightarrow{R-1 \to 0\,}\frac{\pi}{4}\,\alpha_{\mathrm{dif}}^{}
\,\sin 2\delta\,.
\label{Cv2Empty}
\end{align}

\begin{figure}[t]
	\begin{minipage}[r]{\linewidth}
	\centering
	\includegraphics[keepaspectratio,scale=0.38]{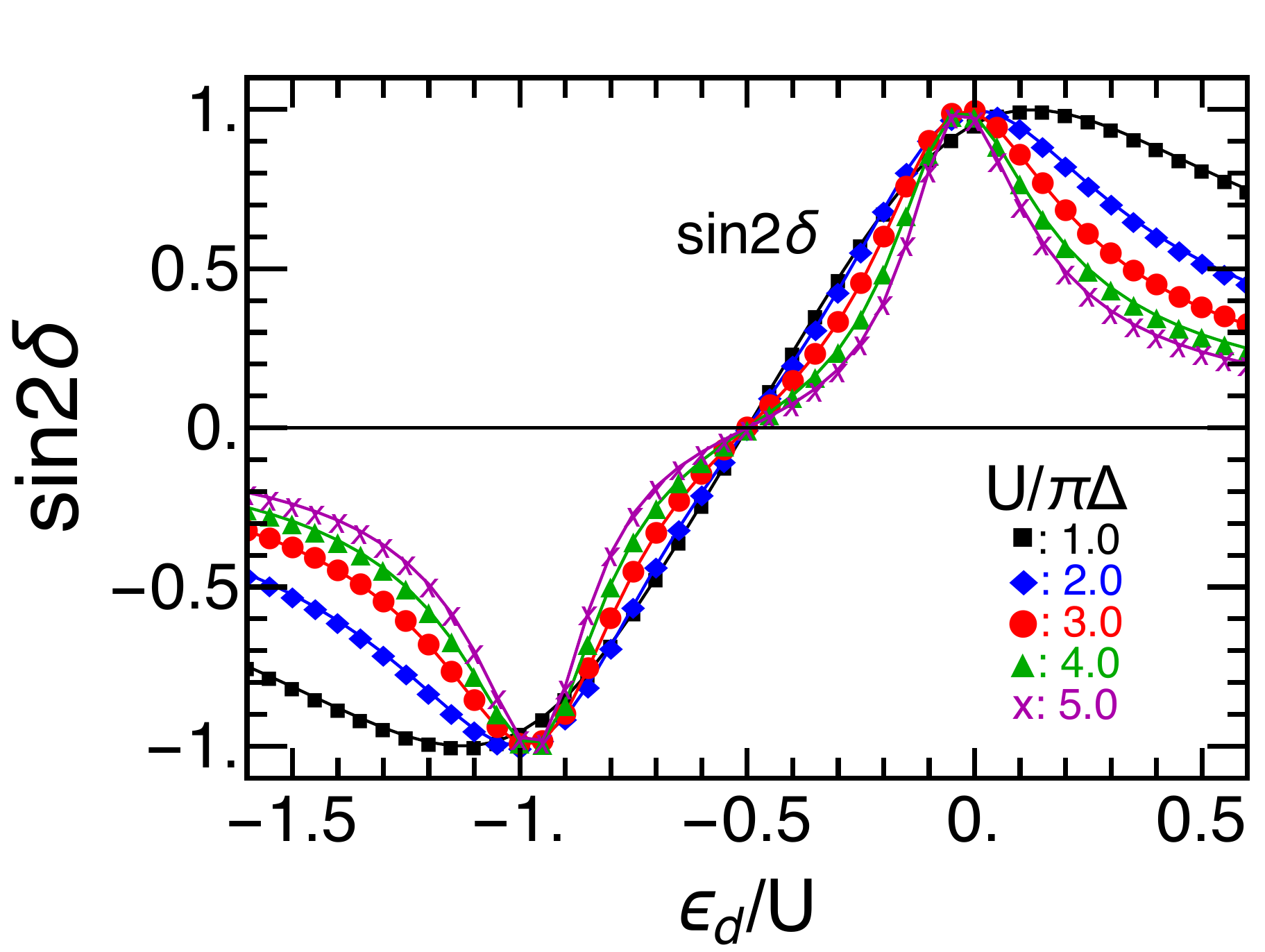}
	\end{minipage}
\caption{
NRG results  for  $\sin 2\delta$ are  plotted vs $\epsilon_d/U$, for   
 $U/(\pi\Delta) =1, 2, 3, 4, 5$.  
}
  \label{fig:sin_2_delta_odd}
\end{figure}

The dependence of $C_V^{(2)}$ on  impurity level $\epsilon_d$ is 
determined also by the other factor $\sin 2\delta$.   
It is related to the derivative of 
the spectral weight at equilibrium $\partial \rho_{d}^{}/\partial \omega$  
as shown in Eq.\ \eqref{eq:sin2delta_def}, 
and vanishes at half filling where $\delta=\pi/2$. 
In Fig.\  \ref{fig:sin_2_delta_odd}, the NRG results 
for $\sin 2\delta$ are plotted vs $\epsilon_d/U$ 
for several values of $U$. 
It has a peak at $\delta = \pi/4$ and a dip at $3\pi/4$, 
which correspond to the level positions 
for  $\epsilon_d \simeq 0$  and $-U$, 
respectively. 
For strong interactions $U\gg \Delta$,   
the peak and dip become sharper as 
the occupation number $\langle n_{d\sigma}^{}\rangle$ 
varies abruptly at these two valence fluctuation regions 
as shown in Fig.\  \ref{Fermipara} (a).

\subsection{Level shift due to bias voltage $eV$}
\label{Cv2Analytical}

We next examine how the impurity level evolves 
at finite bias voltage in the junction with tunnel and bias asymmetries, 
which affects significantly the nonlinear current of order $(eV)^2$.   
 As $C_V^{(2)}$ is determined by 
the linear-order terms of the self-energy $\mathrm{Re}\,\Sigma_{}^{r}(\omega,T,eV)$  
 with respect to the frequency and bias voltage, 
 we consider the low-energy form of the spectral function   
which is correct up to terms of order $\omega$ and $eV$,   
\begin{align}
&\pi\Delta A^{(1)}(\omega)\,\equiv  \, 
\frac{\widetilde{\Delta}^2}
{\Bigl(\omega-\widetilde{\epsilon}_d^{(1)}\Bigr)^2+\widetilde{\Delta}^2}\,, 
\label{Spectral1}
\\ 
&
\widetilde{\epsilon}_d^{(1)}\,\equiv \,  
\widetilde{\Delta}\cot\delta \,+\, \frac{R-1}{2}
\left(\alpha_{\mathrm{dif}}^{}
\,+\, \frac{\Gamma_L-\Gamma_R}{\Gamma_L+\Gamma_R}
\right)  eV\,.
\label{EdRen}
\end{align}
Here,  $\widetilde{\epsilon}_d^{(1)}$  is 
the renormalized impurity level 
that situates at $\widetilde{\Delta}\cot\delta$ for $eV=0$, 
with  $\widetilde{\Delta}=z\Delta$  the renormalized level width.
When a bias voltage $eV$ is applied,  
it moves upward or downward from the equilibrium position 
depending on sign of the coefficient $\alpha_{\mathrm{dif}}^{}
+ (\Gamma_L-\Gamma_R)/\Delta$. 
This shift of the impurity level is induced for interacting electrons, 
and is suppressed as the Wilson ratio approaches the noninteracting 
value $R-1 \to 0$. 
Note that the occupation number of the Anderson impurity 
is determined by the spectral function also at finite $eV$ \cite{DuttHan}, 
\begin{align}
\langle n_{d\sigma}^{} \rangle 
  = & \, \int_{-\infty}^{\infty}  \!\! 
  d\omega\, f_{\rm eff}^{}(\omega) \, A(\omega,T,eV) 
\,,\\
f_\mathrm{eff}(\omega) \,\equiv & \ 
\frac{\Gamma_L \,f(\omega-\mu_L) + \Gamma_R f(\omega-\mu_R) }{ 
 \Gamma_L +\Gamma_R } \,,
\label{eq:f_0}
\end{align}
and thus $\widetilde{\epsilon}_d^{(1)}$ directly affects charge distribution 
around quantum dots.

For $\Gamma_L>\Gamma_R$,  
the number of electrons entering the quantum dot from the source side
becomes larger than the number of electrons leaving from 
the dot toward the drain side. 
Therefore, the number of electrons in the dot increases 
and the repulsive interaction pushes effective level toward the high-energy side. 
In the opposite case,  for $\Gamma_L < \Gamma_R$,  
the effective level moves toward the low-energy side. 
Similarly, the bias symmetry affects the charge distribution. 
The number of electrons in the dot increases  
for $(\mu_L+\mu_R)/2 >0$,  namely for $\alpha_\mathrm{dif}^{}>0$, 
and the Coulomb repulsion pushes $\widetilde{\epsilon}_d^{(1)}$ upward. 
The number of electrons and $\widetilde{\epsilon}_d^{(1)}$  evolve 
in the opposite direction for $(\mu_L+\mu_R)/2 <0$.

 \begin{figure}[t]
	\begin{minipage}[t]{\linewidth}
	\centering
	\includegraphics[keepaspectratio,scale=0.4]{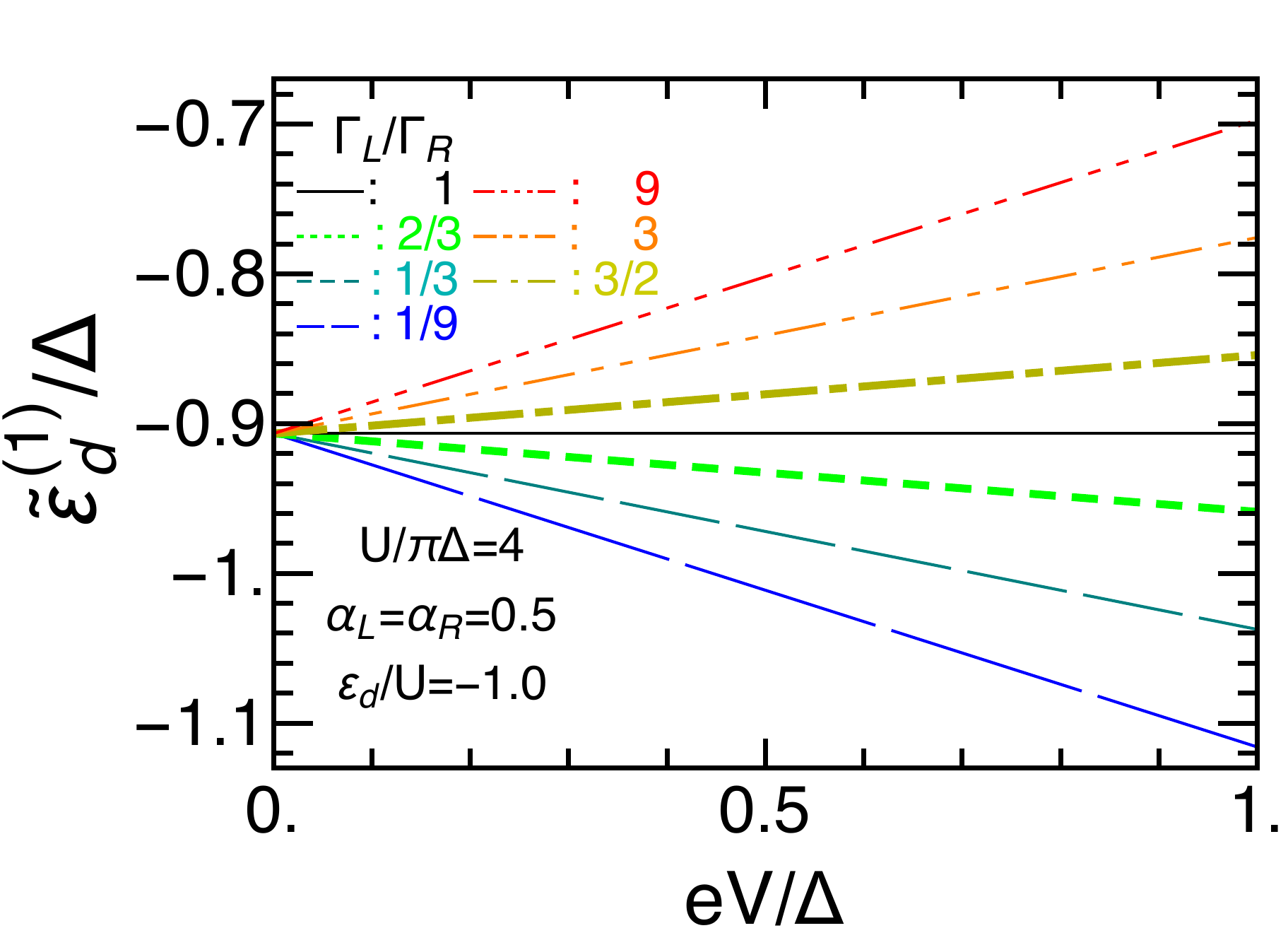}
\end{minipage}
\caption{
Effective impurity  level $\widetilde{\epsilon}_d^{(1)}$ vs $eV$ 
for a junction with tunnel asymmetries 
 $\Gamma_L/\Gamma_R=1/9,\,1/3,\,2/3,\,1,\,3/2,\,3,\,9$   
for chemical potentials $\mu_L=-\mu_R=eV/2\,$ 
 ($\alpha_\mathrm{dif}^{}=0$).  
The other parameters chosen to be $U/(\pi\Delta)=4.0$ and $\epsilon_d=-U$, 
at which the FL parameters take values 
 $z=0.75$, $R-1=0.52$, and 
 $\widetilde{\epsilon}_d^{(1)}=-0.91\Delta$ at  $eV=0$. 
}
\label{edren} 
\end{figure}

\begin{figure}[t]
	\centering
	\includegraphics[keepaspectratio,scale=0.4]{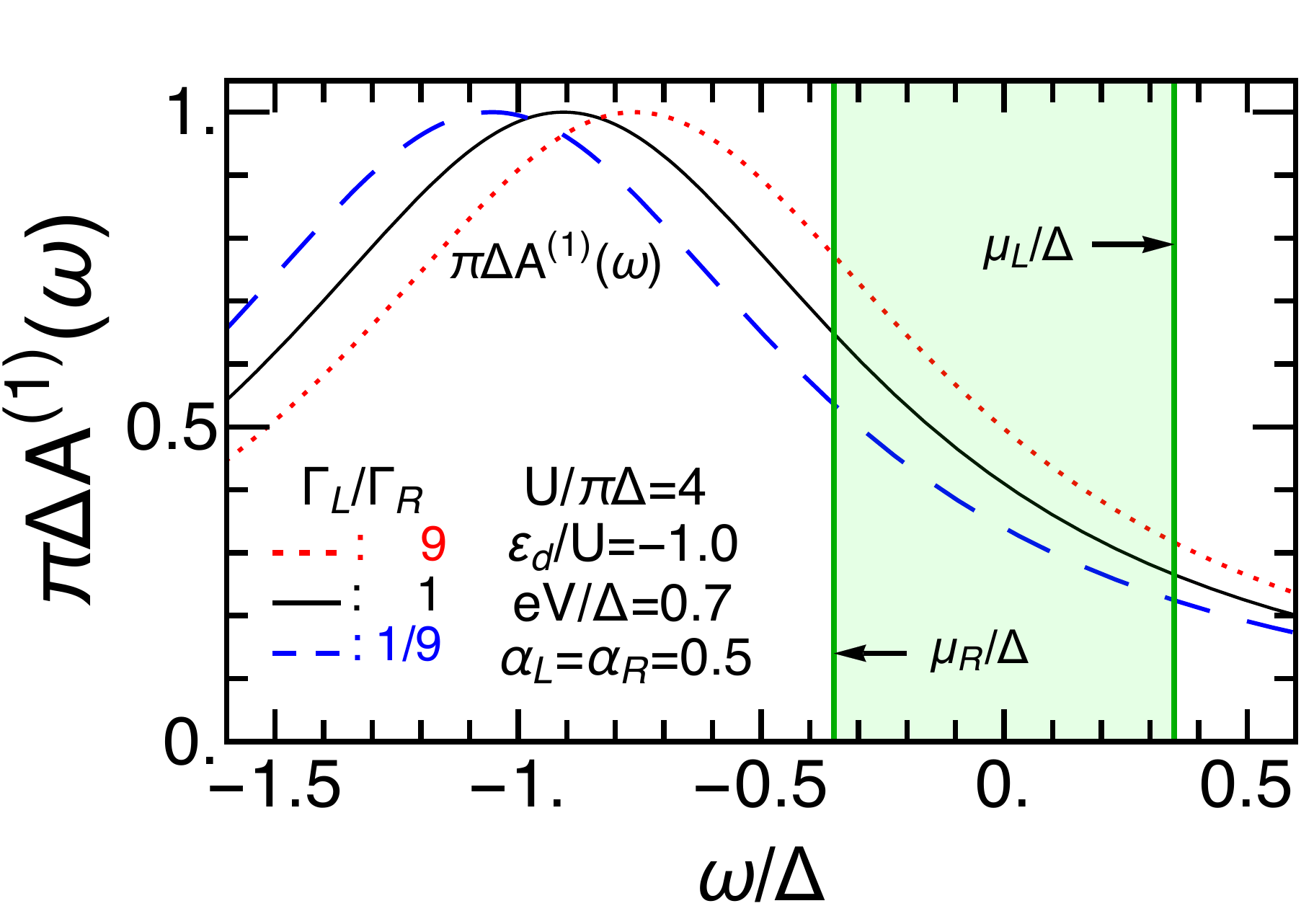}
\caption{
Behavior of  $A^{(1)}(\omega)$ defined in Eq. \eqref{Spectral1} 
near the Fermi level for a junction with tunnel asymmetries 
of $\Gamma_L/\Gamma_R=1/9$ (dashed line), $1$ (solid line), $9$ (dotted line). 
Bias voltage is applied such that  $\mu_L=-\mu_R=eV/2$ with $eV=0.7\Delta$, 
choosing $U/(\pi\Delta)=4$ and $\epsilon_d=-U$. 
The area between the curve for $A^{(1)}(\omega)$ 
 and horizontal axis in the shaded region, which represents the bias window, 
determines order $(eV)^2$ nonlinear current with the formula Eq.\ \eqref{M-Wabs0}. 
}
  \label{Cv2spectral}
\end{figure}

Figure \ref{edren} demonstrates how 
  $\widetilde{\epsilon}_d^{(1)}$ evolves with bias voltage $eV$ 
in a junction with tunnel asymmetries  $\Gamma_L \neq \Gamma_R$  
for a symmetric bias $\alpha_\mathrm{dif}^{}=0$, 
choosing $U/(\pi\Delta) =4.0$ and  $\epsilon_d=-U$. 
It describes the level shifts occurring 
 in the middle of the valence fluctuation region, and in this case 
 $\widetilde{\epsilon}_d^{(1)}=-0.91\Delta$ at  $eV=0$. 
The slope of the lines can be steeper for strong repulsions that make $R-1$ larger,  
and the coefficient  $C_V^{(2)}$ is enhanced significantly in such cases.
Figure \ref{Cv2spectral} also demonstrates the behavior of  
  $A^{(1)}(\omega)$  near the Fermi level 
for a junction with tunnel asymmetries $\Gamma_L/\Gamma_R=1/9, 1, 9$,  
 applying bias voltages such that  $\alpha_\mathrm{dif}^{}=0$ and $eV=0.7\Delta$  
for the same valence fluctuation state as that examined in Fig.\ \ref{edren}. 
In this example, the spectral peak corresponding to the effective level situates 
at $\widetilde{\epsilon}_d^{(1)}=-0.91\Delta$ for $\Gamma_L=\Gamma_R$, 
below the bias window $\mu_R\leq\omega\leq \mu_L$, which is 
described as a shaded region. 
When a bias voltage $eV$ is applied to an 
asymmetric tunnel junction with $\Gamma_L \neq \Gamma_R$,  
the peak shifts in such a way as 
the dashed or dotted line in Fig.\ \ref{Cv2spectral}.  
Simultaneously,  the spectral weight in the bias-window region varies,  
and it yields order $(eV)^2$ term of 
the nonlinear current $I$ defined in Eq.\ \eqref{M-Wabs0}. 
Especially, shifts of the effective level cause one of the two parts,  
i.e.,\  $\overline{C}_V^{(2b)}$ defined in Eq.\ \eqref{Cv22}, 
of the  order $(eV)^2$ term.

The other part $\overline{C}_V^{(2a)}$ is determined by 
the $\omega$-linear term of $A^{(1)}(\omega)$, which is renormalized by 
the differential coefficient  $\partial \Sigma_{}^\mathrm{eq}/\partial \omega$.    
 It contributes to order $(eV)^2$ nonlinear current 
for $\alpha_\mathrm{dif}^{}\neq 0$, i.e.,\      
when the bias voltage is applied asymmetrically. 
Thus, the $\overline{C}_V^{(2a)}$ part can also regarded 
as contributions caused by the shift of the bias window, 
 taking place in the low-frequency region. 
An alternative interpretation along this line is described in Appendix \ref{Cv2meaning}.

\subsection{
 $\epsilon_d^{}$ dependence of order $(eV)^2$ current 
through junctions with tunnel and bias asymmetries
}

\label{Cv2NumericalSymmetry}

We next examine the behavior of 
 $C_V^{(2)}$ over a wide range of the parameter space,  
especially its dependence on  junction asymmetries 
 $(\Gamma_L-\Gamma_R)/\Delta$ and  $\alpha_\mathrm{diff}$, 
and  also the dependence on $\epsilon_d$ and $U$
 by which the Fermi-liquid parameters $\sin 2 \delta$ and $R-1$ evolve.

\subsubsection{
Effects of tunnel asymmetries $\Gamma_L \neq \Gamma_R$, 
at $\,\alpha_\mathrm{dif}^{}=0$}

\begin{figure}[t]
	\begin{minipage}[t]{\linewidth}
	\centering
	\includegraphics[keepaspectratio,scale=0.4]{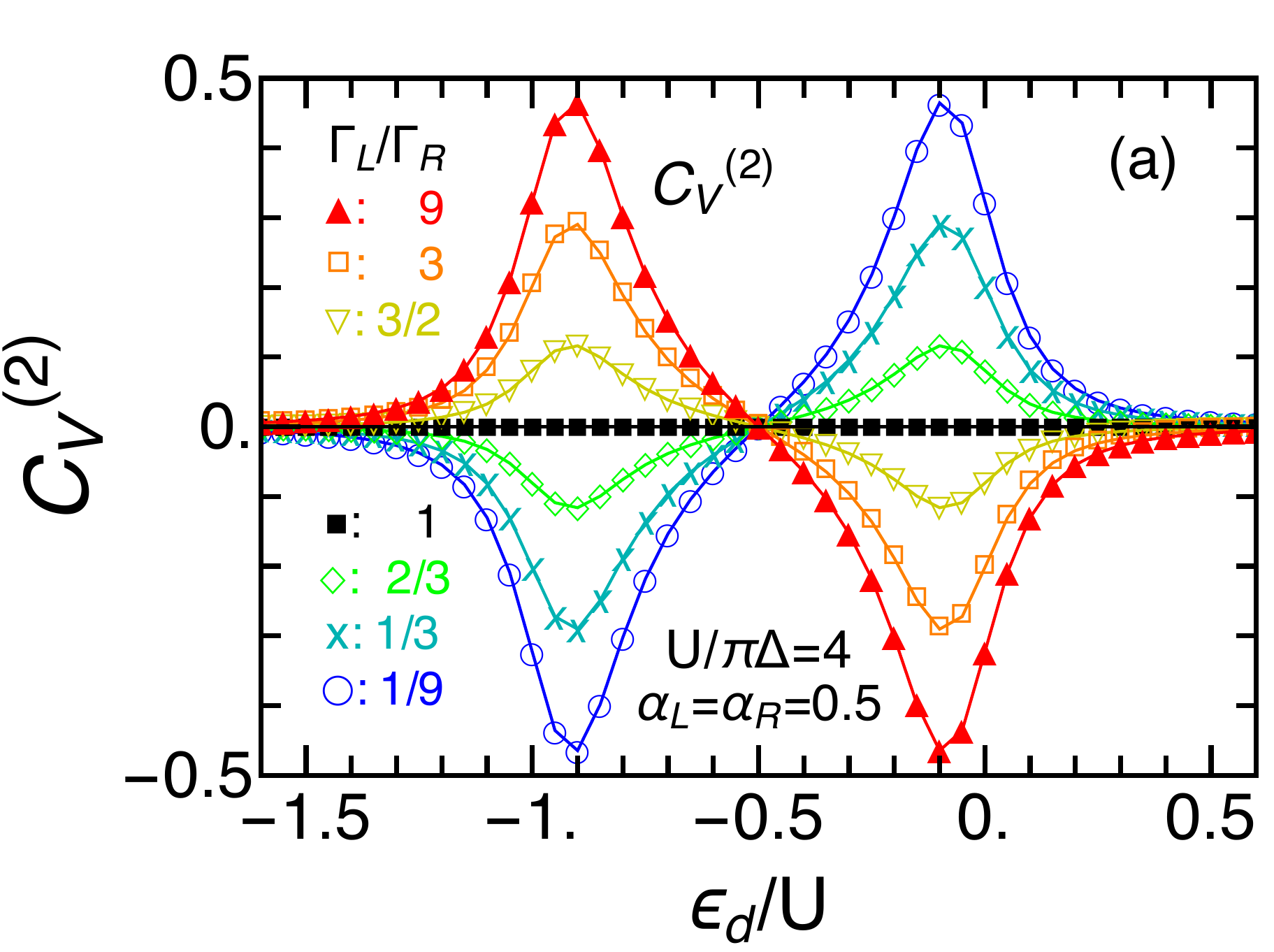}
	\end{minipage}
	\rule{1mm}{0mm}
	\begin{minipage}[t]{\linewidth}
	\centering
	\includegraphics[keepaspectratio,scale=0.4]{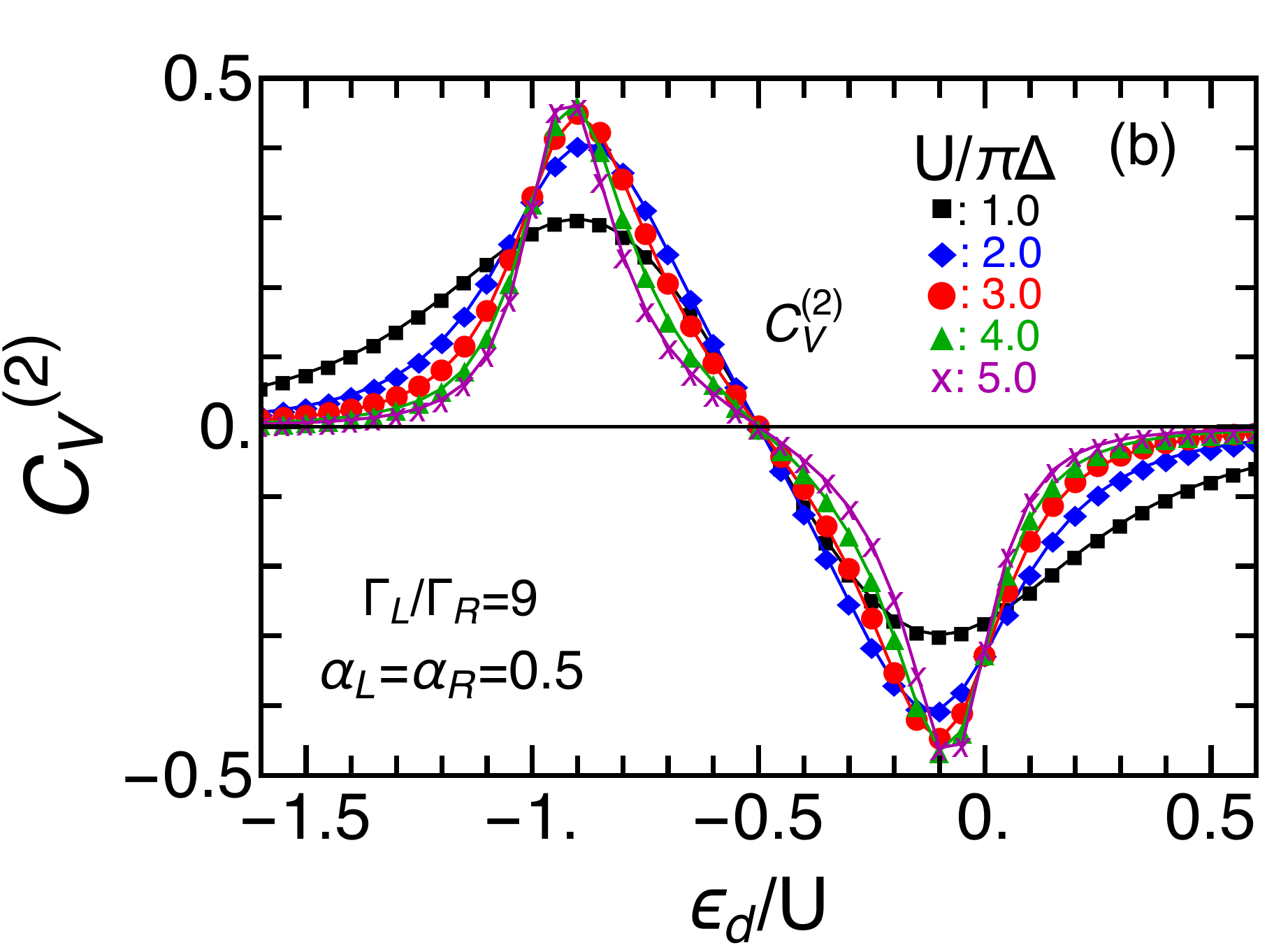}
	\end{minipage}
\caption{
(a) NRG results of $C_V^{(2)}$ vs $\epsilon_d/U$ for different 
tunnel asymmetries  $\Gamma_L/\Gamma_R=1/9(\circ)$,
$1/3(\times)$, $2/3(\diamond)$, $1(\blacksquare)$, $3/2(\triangledown)$,
$3(\square)$, $9(\blacktriangle)$  for $U/(\pi \Delta)=4.0$, 
keeping $\Delta=\Gamma_L+\Gamma_R$ unchanged. 
(b) $C_V^{(2)}$ 
for several values of interactions $U/(\pi\Delta)=1, 2, 3, 4, 5$, 
for  fixed tunnel couplings $\Gamma_L=9\Gamma_R$. 
For both (a) and (b), bias voltages are chosen to be symmetric   
 $\alpha_L=\alpha_R=1/2$, i.e.,\  $\alpha_\mathrm{dif}^{}=0$.
}
  \label{Cv2sym}
\end{figure}

We first of all  consider the effects of tunnel asymmetry, 
choosing bias voltages to be symmetric  $\alpha_\mathrm{dif}^{}=0$. 
In this case the coefficient takes the form 
$C_V^{(2)} 
 \xrightarrow{\alpha_\mathrm{dif}^{} = 0\,}
 -({\pi}/{4})\, [ (\Gamma_L-\Gamma_R)/\Delta]\,(R-1)\,\sin 2\delta$. 
Therefore, while the $\epsilon_d$ dependence enters through  
the product $(R-1)\,\sin 2\delta$, 
the sign and magnitude of $C_V^{(2)}$ are determined 
 also by tunnel asymmetries $(\Gamma_L-\Gamma_R)/\Delta$. 
In Fig.\ \ref{Cv2sym}(a),  the NRG results of this coefficient are plotted 
as a function of $\epsilon_d/U$ for $U/(\pi\Delta)=4$, 
keeping $\Delta =\Gamma_L+\Gamma_R$ unchanged. 
We see that all the curves for different 
$\Gamma_L/\Gamma_R=1/9,\,1/3,\,2/3,\,1,\,3/2,\,3,\,9$ 
pass through a common single zero point at $\epsilon_d=-U/2$, 
where the phase shift takes the value $\delta={\pi}/{2}$ 
due to the  particle-hole symmetry. 
Over a wide range of gate voltages  $ -U\lesssim \epsilon_d \lesssim 0$,  
the $\epsilon_d$ dependence of $C_V^{(2)}$ reflects that of 
$\sin 2 \delta$ since the Wilson ratio $R-1 \simeq 1$ for $U/(\pi \Delta) \gtrsim 2.0$ 
as shown in Fig.\ \ref{Fermipara}(c). 
Similarly,  the peak and dip structures 
appearing in the valence fluctuation regime 
near $\epsilon_d\simeq 0$ and $\epsilon_d\simeq -U$
reflect the behavior of $\sin 2 \delta$ 
at $\delta=\pi/4$ and $\delta=3\pi/4$, 
 shown in Fig.\ \ref{fig:sin_2_delta_odd}. 
These features of order $(eV)^2$ nonlinear current 
are pronounced as tunnel asymmetry $|\Gamma_L-\Gamma_R|$  
increases. 
For much larger  $|\epsilon_d|$, outside of the valence fluctuation regime,     
$C_V^{(2)}$ rapidly decreases 
much faster than $\sin 2 \delta$ 
since the effects of electron correlations are suppressed $R-1 \simeq 0$ 
as the impurity level is almost empty or doubly occupied.

In Fig.\ref{Cv2sym}(b),  the coefficient $C_V^{(2)}$ 
for largely imbalanced tunnel couplings  $\Gamma_L=9\Gamma_R$  
is plotted  for several different values of the interaction 
$U/(\pi\Delta)=1, 2, 3, 4, 5$. 
We can see that the peak and dip structures 
evolve with $U$ and become sharper 
as $U$ increases, 
reflecting the behavior of $\sin 2\delta$ shown in Fig.\ \ref{fig:sin_2_delta_odd}. 
Furthermore,  $C_V^{(2)}$ is suppressed in the Kondo regime 
 $-U\lesssim \epsilon_d \lesssim 0$ for strong interactions.

\begin{figure}[t]
	\begin{minipage}[r]{\linewidth}
	\centering
	\includegraphics[keepaspectratio,scale=0.4]{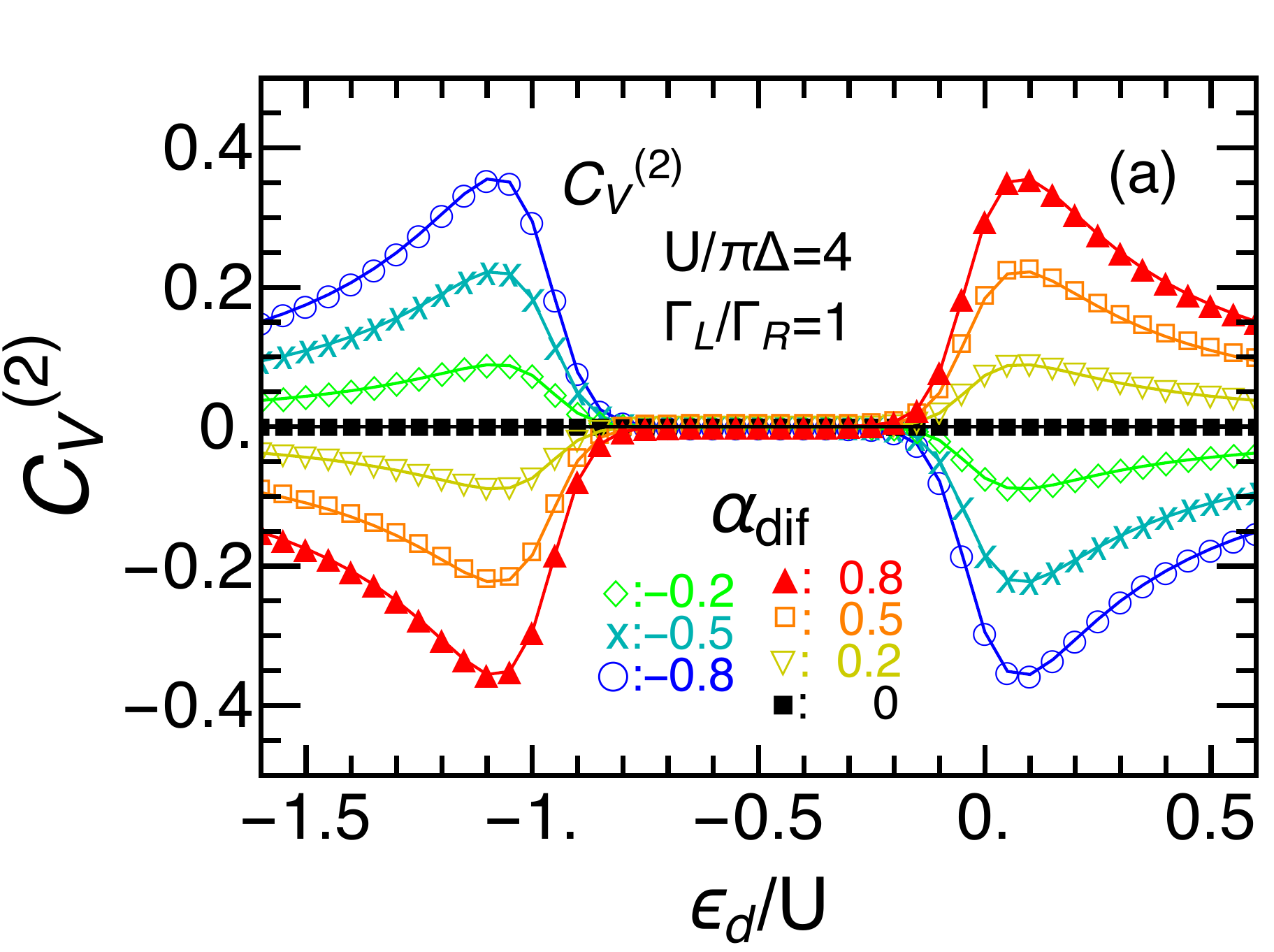}
	\end{minipage}
	\rule{1mm}{0mm}
	\begin{minipage}[r]{\linewidth}
	\centering
	\includegraphics[keepaspectratio,scale=0.4]{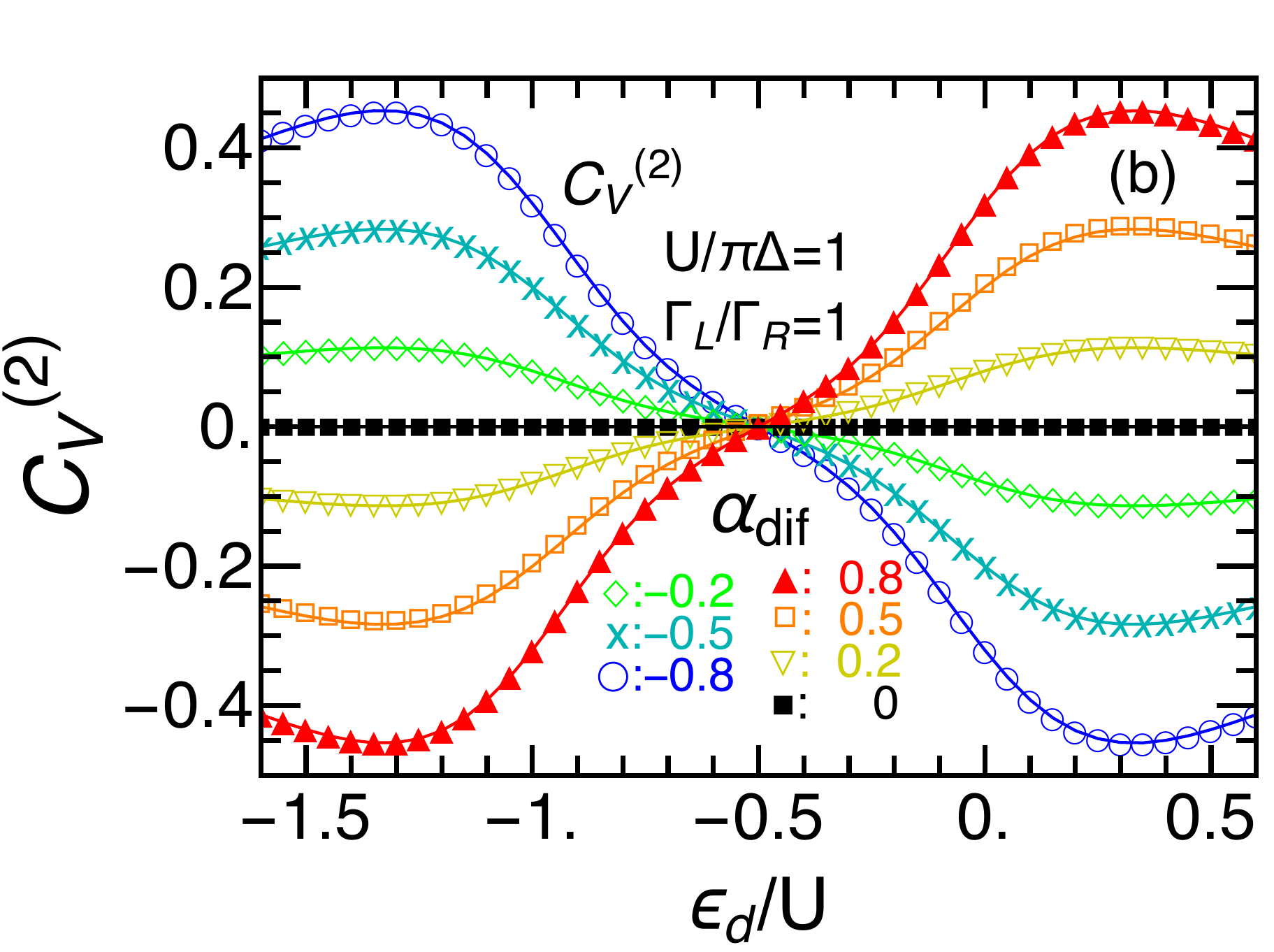}
	\end{minipage}
\caption{
NRG results of $C_V^{(2)}$ 
for symmetric tunnel couplings with $\Gamma_L=\Gamma_R$ 
are plotted for different degrees of bias asymmetries  
 $\alpha_{\mathrm{dif}}=-0.8(\circ)$, 
$-0.5(\times)$, $-0.2(\diamond)$, $0.0(\blacksquare)$, 
$0.2(\triangledown)$, $0.5(\square)$, $0.8(\blacktriangle)$.    
Coulomb interactions are chosen 
to be (a)  $U=4\pi\Delta$, and  (b)  $U=\pi\Delta$.
}
  \label{Cv2asymbias1}
\end{figure}

\subsubsection{
Effects of bias asymmetries  $\alpha_\mathrm{dif}^{} \neq 0$,  
at  $\,\Gamma_L = \Gamma_R$
}

We next consider effects caused by shift of the bias window 
which occurs for  $\alpha_L\neq\alpha_R$ 
($\alpha_\mathrm{dif}^{} \neq 0$). 
For this purpose, 
we  assume here the tunnel couplings to be symmetric $\Gamma_L=\Gamma_R$.
In this case, the coefficient takes the form 
$C_V^{(2)} \xrightarrow{\Gamma_L=\Gamma_R\,} 
 ({\pi}/{4})\,\alpha_\mathrm{dif}^{}\,
(2-R)\,\sin 2\delta$, 
and thus the  $\epsilon_d$ dependence is determined by the product 
$(2-R)\,\sin 2\delta$ while the sign and magnitude 
vary with bias asymmetries $\alpha_\mathrm{dif}^{}$. 
Note that the factor $2-R$ is proportional to the charge susceptibility 
$\chi_c$ defined at $T=0$,
 and  it can be written in the following form 
using Eq.\ \eqref{Fermiparaorigin}, 
\begin{align}
\chi_c \, \equiv \,
- \sum_{\sigma}
\frac{\partial \langle n_{d\sigma}^{}\rangle}{\partial \epsilon_d^{}}
\,=\, 
2 \left(\chi_{\uparrow\uparrow}^{}+\chi_{\uparrow\downarrow}^{} \right)
\ = \ \frac{2-R}{2T^\ast}\,.
\end{align}

In Fig.\ \ref{Cv2asymbias1}, the NRG results of $C_V^{(2)}$ 
for this case  are plotted vs $\epsilon_d$ 
for different degrees of bias asymmetries 
 $\alpha_\mathrm{dif}^{}
=-0.8,\,-0.5,\,-0.2,\,0,\,0.2,\,0.5,\,0.8$,  
choosing two different Coulomb repulsions  
 (a)  $U/(\pi\Delta) =4$ and  (b) $U/(\pi\Delta) =1$. 
For strong interactions, the order $(eV)^2$ component of the 
nonlinear current vanishes $C_V^{(2)}\simeq 0$ 
over a wide region  $-U \lesssim \epsilon_d \lesssim 0$,   
as seen in Fig.\ \ref{Cv2asymbias1}(a). 
This is because the charge fluctuations are suppressed and thus 
$\chi_c \propto 2-R \simeq 0$ in the Kondo regime. 
However, as demonstrated in Fig.\ \ref{Cv2asymbias1}(b), 
$C_V^{(2)}$ takes finite values for weak interactions
because the charge excitations occurring near the quantum dot  
are still active for  $U/(\pi \Delta) \lesssim 2.0$,  
as can be deduced from the behavior of $R-1$ shown in Fig.\ \ref{Fermipara}(c). 
The peak and dip structures seen in the valence fluctuation regime 
near $\epsilon_d\simeq 0$ and $\epsilon_d\simeq -U$
reflect the behavior of $\sin 2 \delta$ 
at $\delta=\pi/4$ and $\delta=3\pi/4$, 
similarly to those observed for the symmetric bias case 
in Fig.\ \ref{Cv2sym}. 
$C_V^{(2)}$ vanishes when 
both the tunnel and bias asymmetries are absent, i.e.,\ 
at  $\Gamma_L=\Gamma_R$ 
and  $\alpha_\mathrm{dif}^{}=0$, as mentioned.

\begin{figure}[t]
	\begin{minipage}[t]{\linewidth}
	\centering
	\includegraphics[keepaspectratio,scale=0.4]{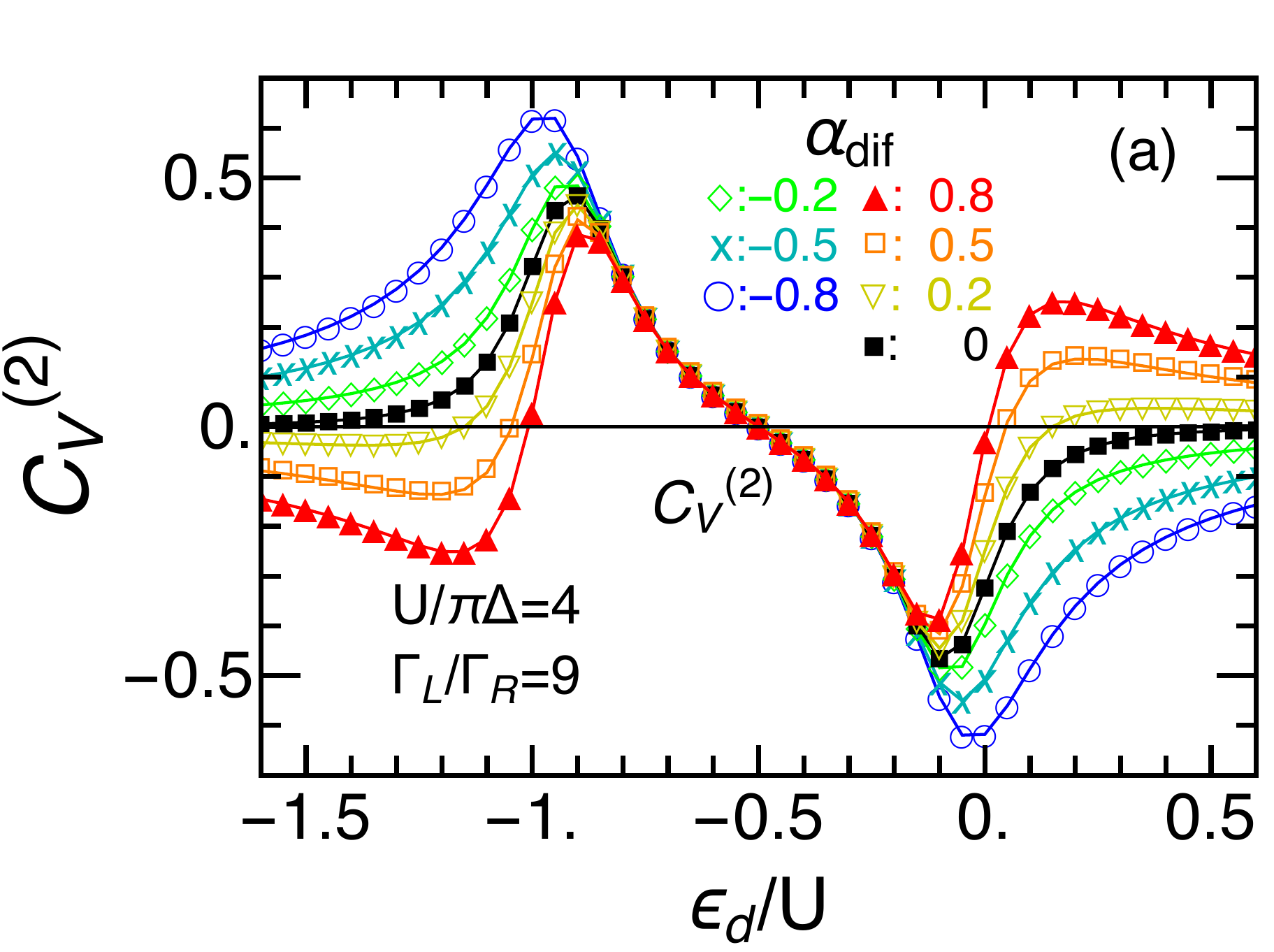}
	\end{minipage}
	\rule{1mm}{0mm}
	\begin{minipage}[t]{\linewidth}
	\centering
	\includegraphics[keepaspectratio,scale=0.4]{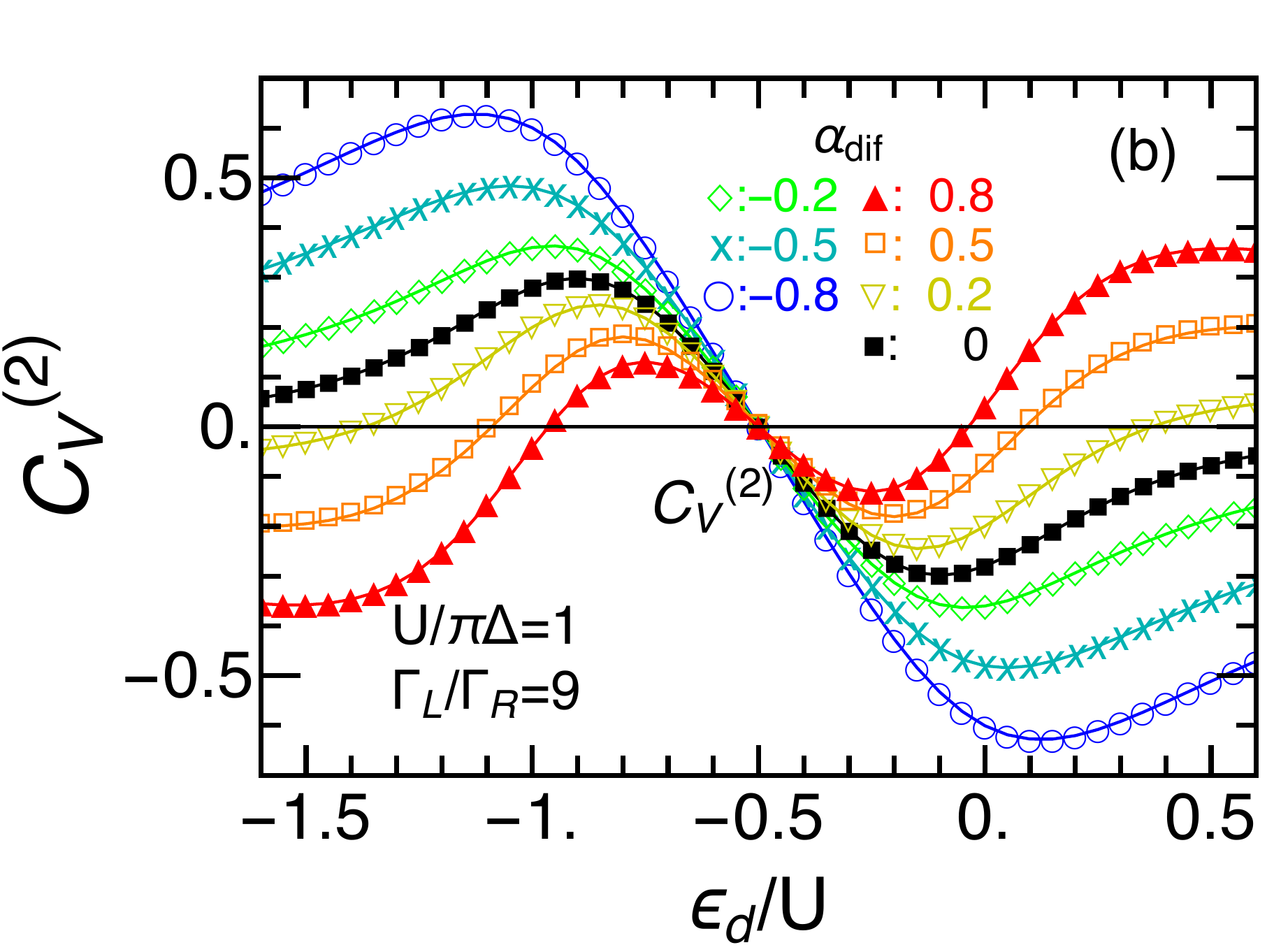}
	\end{minipage}
\caption{
NRG results of  $C_V^{(2)}$ 
for junctions with large imbalanced couplings $\Gamma_L=9\Gamma_R$ 
are plotted 
for different degrees of  bias asymmetries  
 $\alpha_{\mathrm{dif}}=-0.8(\circ)$, 
$-0.5(\times)$, $-0.2(\diamond)$, $0.0(\blacksquare)$, 
$0.2(\triangledown)$, $0.5(\square)$, $0.8(\blacktriangle)$.   
Coulomb interactions are chosen 
to be (a)  $U=4\pi\Delta$, and  (b)  $U=\pi\Delta$.
}
  \label{Cv2asymbias9}
\end{figure}

\subsubsection{
Effects of bias asymmetries  $\alpha_\mathrm{dif}^{}\neq 0$, 
 for junctions with largely imbalanced couplings $\Gamma_L \gg \Gamma_R$ 
}

So far, we have examined separately the effects of  tunnel asymmetry
and bias asymmetry 
which enter through the parameters  $\Gamma_L/\Gamma_R$ and 
 $\alpha_\mathrm{dif}^{}$, respectively, 
 assuming that the perturbations due to one of these two asymmetries is absent. 
We next consider the behavior of  the order $(eV)^2$ nonlinear response  
in the situation where both of these asymmetries are present 
in particular,  
 for junctions with largely 
imbalanced tunnel couplings $\Gamma_L \gg \Gamma_R$, 
specifically  for  $\Gamma_L= 9\Gamma_R$.

In Fig.\ \ref{Cv2asymbias9},  the coefficient  $C_V^{(2)}$ 
is plotted for different degrees of bias asymmetries 
$\alpha_\mathrm{dif}^{}
=-0.8$, $-0.5$, $-0.2$, $0.0$, $0.2$, $0.5$, $0.8$, 
choosing interactions to be  (a) $U= 4\pi\Delta$  and (b) $U= \pi\Delta$.  
One of the lines in Fig.\ \ref{Cv2asymbias9} (a) 
plotted for $\alpha_\mathrm{dif}^{}=0$
is identical to the line 
for $\Gamma_L= 9\Gamma_R$ in Fig.\ \ref{Cv2sym} (a). 
We can see that  $C_V^{(2)}$ for strong interactions $U= 4\pi\Delta$ 
is not affected by the bias asymmetries  $\alpha_\mathrm{dif}^{}$ 
over the wide region  $-U\lesssim \epsilon_d \lesssim 0$, 
where the Wilson ratio approaches $R-1 \simeq 1$ and  
 $C_V^{(2)}$ can be expressed in the form of Eq.\ \eqref{Cv2Kondo}. 
The coefficient  $C_V^{(2)}$ captures two extra zero points  
for $\alpha_\mathrm{dif}^{}>0$ 
in  the valence fluctuation region, 
 at  $\epsilon_d \simeq -U$ and  $\epsilon_d \simeq 0$. 
It happens at the points where 
 the two contributions defined in Eqs.\ \eqref{Cv21} and \eqref{Cv22} 
cancel each other out  $\overline{C}_V^{(2a)}+\overline{C}_V^{(2b)}=0$. 
This condition that determines 
the values of  $\epsilon_d$ at the extra zero points can also be expressed in the form
\begin{align}
R-1 \,= \, \frac{\alpha_\mathrm{dif}^{}}{\alpha_\mathrm{dif}^{} 
+ \frac{\Gamma_L-\Gamma_R}{\Gamma_L+\Gamma_R}  }\,, 
\label{eq:zero_point_C2}
\end{align}
and it has a solution if $(\Gamma_L-\Gamma_R)\,\alpha_\mathrm{dif}^{}>0$ 
as the Wilson ratio takes a value in the range $ 0 \leq R-1\leq 1$.
The extra zero points move 
toward the high-energy side as 
$\alpha_\mathrm{dif}^{}$ approaches zero.
Note that the contributions of the $\overline{C}_V^{(2a)}$ and $ \overline{C}_V^{(2b)}$ parts are caused by a shift of the bias window 
and that of the effective level $\widetilde{\epsilon}_d^{(1)}$, respectively, 
 as shown in Appendix \ref{Cv2meaning}. 
For $(\Gamma_L-\Gamma_R)\,\alpha_\mathrm{dif}^{}<0$,  
we can also see  in Fig.\ \ref{Cv2asymbias9} (a) that 
 these two contributions become cooperative 
and  $C_V^{(2)}$ is enhanced significantly in the valence fluctuation regime.

Outside of the Kondo and valence fluctuation regimes  $|\epsilon_d+U/2| \gg U/2$, 
the magnitude of  $C_V^{(2)}$ becomes independent of tunnel asymmetries 
and is proportional to $\alpha_\mathrm{dif}^{}$ obeying Eq.\ \eqref{Cv2Empty}. 
Thus,  as $|\epsilon_d|$ increases, 
 the behaviors of the curves in Fig.\ \ref{Cv2asymbias9} (a) 
approach those of the curves for the same $\alpha_\mathrm{dif}^{}$ 
shown  in Fig.\ \ref{Cv2asymbias1} (a) for the $\Gamma_L=\Gamma_R$ case. 
For weak interactions $U/(\pi\Delta) \lesssim 2.0$, 
charge fluctuations contribute to low-energy transport also 
near half-filling $\epsilon_d \simeq -U/2$,  as seen in Fig.\ \ref{Fermipara}.
It causes variations in $C_V^{(2)}$ 
as seen in whole regions of $\epsilon_d$ in Fig.\ \ref{Cv2asymbias9} (b)  
against $\alpha_\mathrm{dif}^{}$. 
The behaviors in the case of the opposite tunnel asymmetries,
for which $\Gamma_L$ is much smaller than $\Gamma_R$,  
can also be deduced from the results shown in Fig.\ \ref{Cv2asymbias9}, 
specifically for $\Gamma_L=\Gamma_R/9$, 
using the relation 
$
C_V^{(2)} (\alpha_{\mathrm{dif}}^{},\Gamma_L/\Gamma_R) 
= - C_V^{(2)} (-\alpha_{\mathrm{dif}}^{},\Gamma_R/\Gamma_L) 
$ given in Eq.\ \eqref{eq:LR_inversion_C2}.

\begin{figure}[t]
	\begin{minipage}[t]{\linewidth}
	\centering
	\includegraphics[keepaspectratio,scale=0.4]{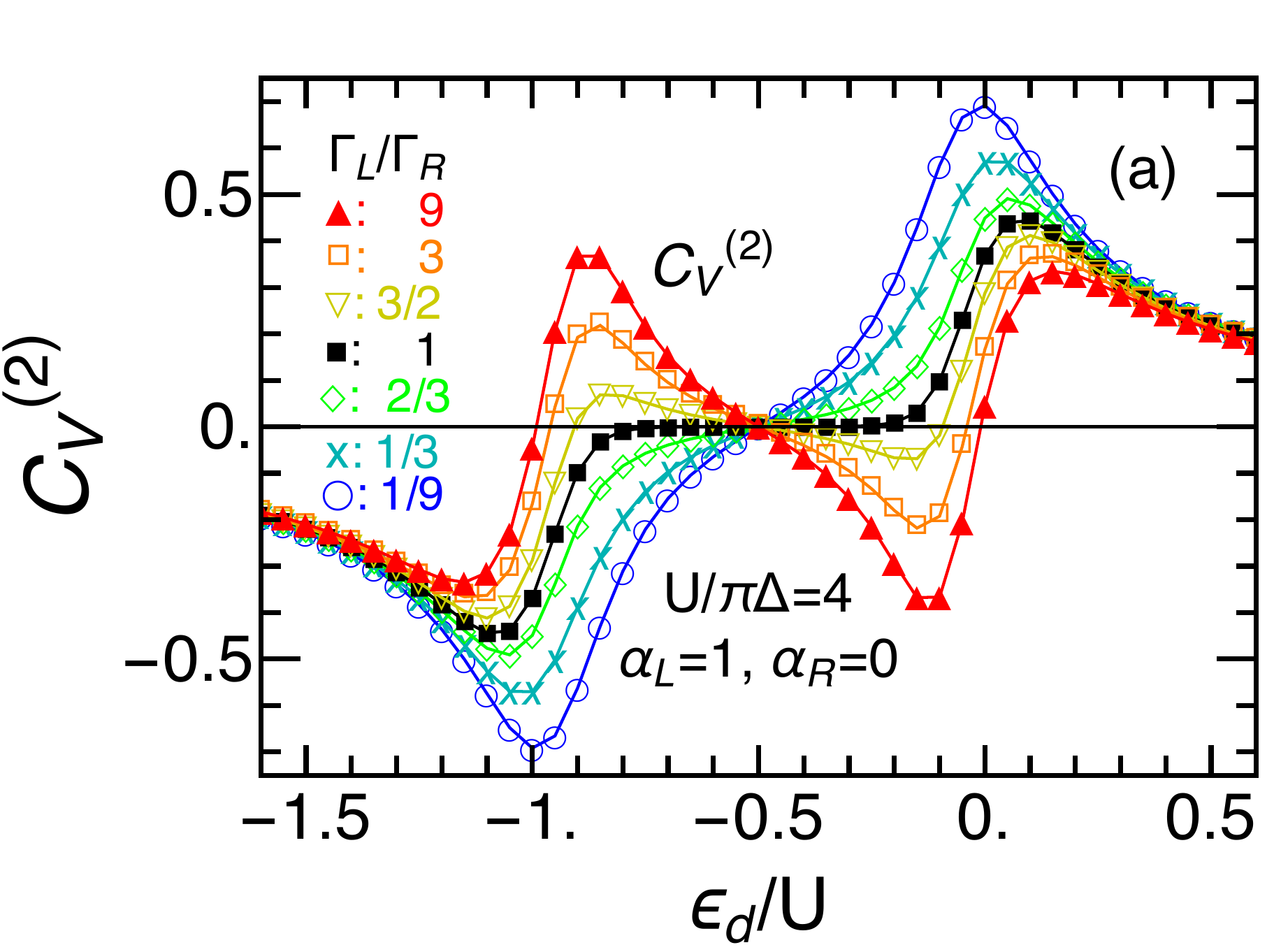}
	\end{minipage}
	\rule{1mm}{0mm}
	\begin{minipage}[t]{\linewidth}
	\centering
	\includegraphics[keepaspectratio,scale=0.4]{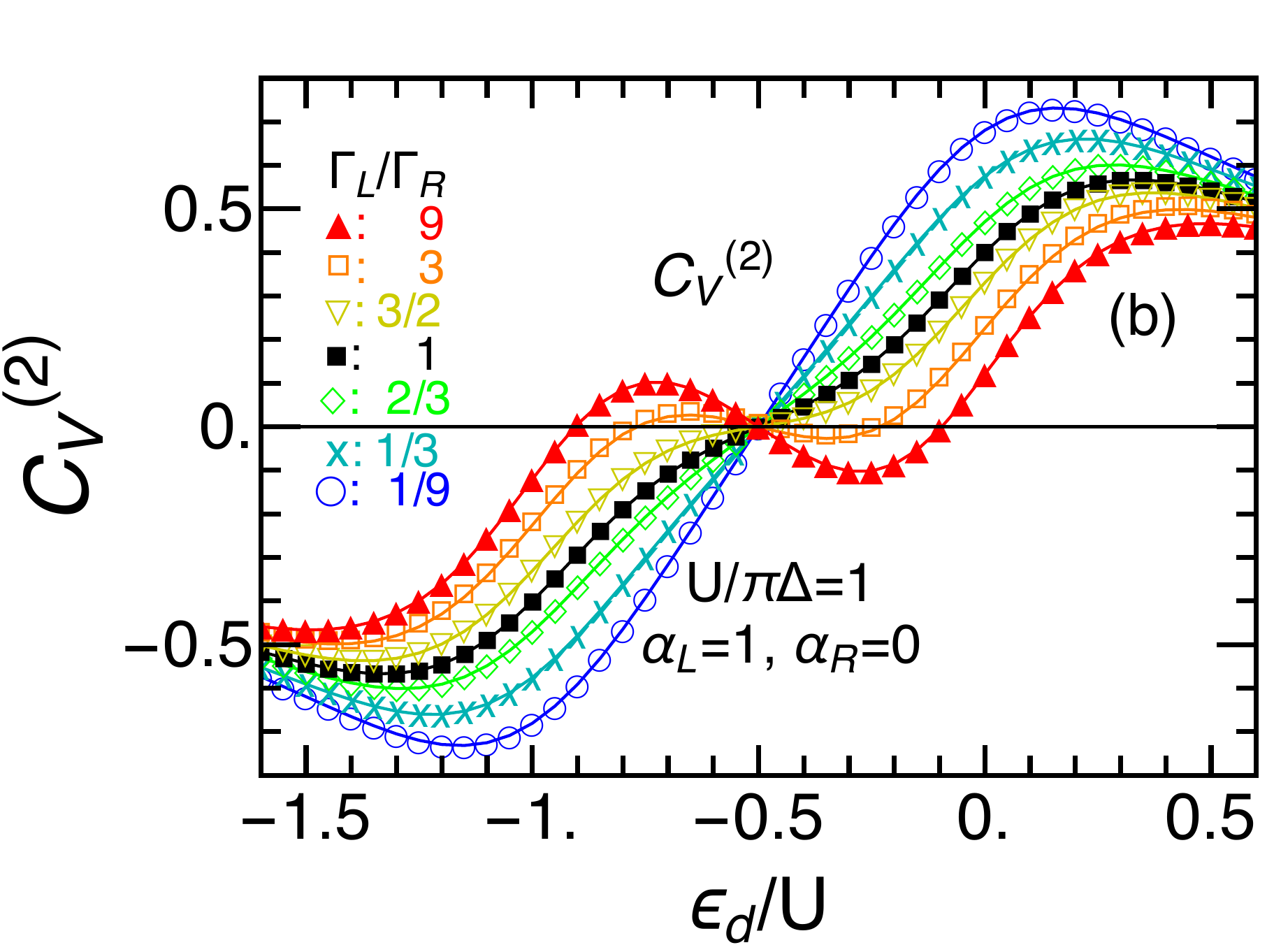}
	\end{minipage}
\caption{
NRG results of $C_V^{(2)}$ 
 for a maximized bias asymmetry 
 $\alpha_L=1$ and $\alpha_R=0$  
(i.e.,\  $\mu_L =eV$ and $\mu_R=0$)  
for different degrees of tunnel asymmetries 
  $\Gamma_L/\Gamma_R=1/9(\circ)$,
$1/3(\times)$, $2/3(\diamond)$, $1(\blacksquare)$, $3/2(\triangledown)$,
$3(\square)$, $9(\blacktriangle)$, for a fixed Coulomb 
interaction  $U/(\pi \Delta)=4$. 
Coulomb interactions are chosen 
to be (a)  $U=4\pi\Delta$, and  (b)  $U=\pi\Delta$.
}
  \label{Cv2asym}
\end{figure}

\subsubsection{
Effects of tunnel asymmetries  $\Gamma_L \neq \Gamma_R$,  
for junctions with a maximized bias asymmetry $\alpha_\mathrm{dif}^{}=1$
}

We also examine the situation where the electrode of the drain side is grounded, 
choosing the parameters such that  $\alpha_L=1$ and $\alpha_R=0$,  
i.e.,\ $\alpha_\mathrm{dif}^{} =1$.
This corresponds to one typical situation occurring in real measurements.
In this case, the coefficient takes the form  
$C_V^{(2)}\xrightarrow{\alpha_\mathrm{dif} =1}
({\pi}/{4})\left[1-\, 2\,(R-1)\,\Gamma_L/\Delta \right] \sin2\delta$. 
The NRG results are shown in Fig.\ \ref{Cv2asym}. 

In Fig.\ \ref{Cv2asym}(a), 
the coefficient  $C_V^{(2)}$ in this case $\alpha_\mathrm{dif}^{}=1$ 
is plotted for seven different tunnel asymmetries   
$\Gamma_L/\Gamma_R=1/9,\,1/3,\,2/3,\,1,\,3/2,\,3,\,9$, 
choosing $U=4\pi\Delta$ at which the Wilson ratio 
is almost saturated to the strong-coupling value  $R-1 \simeq 1$ 
in the Kondo regime, as mentioned.  
Therefore,  $C_V^{(2)}$ varies 
with $\Gamma_L/\Gamma_R$ for  $-U\lesssim \epsilon_d \lesssim 0$, 
obeying  Eq.\ \eqref{Cv2Kondo}.
It also has two extra zero points  
for  $(\Gamma_L-\Gamma_R)\,\alpha_\mathrm{dif}^{}>0$  
in the valence fluctuation region as a result of the cancellation 
of the  $\overline{C}_V^{(2a)}$ and  $\overline{C}_V^{(2b)}$ components,
described in Eq.\ \eqref{eq:zero_point_C2}. 
Similarly, these two components become cooperative for the opposite cases 
where $(\Gamma_L-\Gamma_R)\,\alpha_\mathrm{dif}^{}<0$,  
and they enhance largely the order $(eV)^2$ nonlinear current.
For much larger $|\epsilon_d|$,  all the data collapse into a single curve    
as the coefficient $C_V^{(2)}$ becomes independent of tunnel asymmetries, 
which can be explained with Eq.\ \eqref{Cv2Empty}.

The coefficient $C_V^{(2)}$ for a weak interaction $U=\pi\Delta$
 is in Fig.\ \ref{Cv2asym} (b), keeping the other parameters 
 the same as those used for (a).  
All the curves for different $\Gamma_L/\Gamma_R$ 
become moderate  as charge fluctuations of electrons in quantum dots
 contribute to low-energy transport  whole regions of $\epsilon_d$. 
We see that  $C_V^{(2)}$  for the cases 
 $(\Gamma_L-\Gamma_R)\,\alpha_\mathrm{dif}^{}<0$ 
where the tunnel and bias asymmetries are cooperative 
are enhanced significantly, especially for $\Gamma_L/\Gamma_R=1/9$.

\section{Properties of order  $(eV)^3$ nonlinear current}

\label{Cv3Results}

In this section, we study the effects of 
tunnel and bias asymmetries on the order  $(eV)^3$ nonlinear current. 
The coefficient for this component in $dI/dV$ consists of two parts: 
 $C_V^{(3)}= (\pi^2/64) (W_V+\Theta_V )$  with  
the two-body $W_V$ and three-body $\Theta_V$ contributions   
 defined in Eqs.\ \eqref{Cv3eq}--\eqref{THv}. 
One of the important properties deduced from these formulas, 
summarized also in Table \ref{tab:C_W_SUN},
is that effects of  tunnel asymmetries enter into  
 $W_V$ and  $\Theta_V$ through the cross term 
of $\Gamma_L-\Gamma_R$ and  $\alpha_\mathrm{dif}^{}$ 
of the form  
\begin{align}
\frac{\Gamma_L-\Gamma_R}{\Gamma_L+\Gamma_R}  
\  \alpha_\mathrm{dif}^{}   
\,.
\label{eq:cross_term_tunnel_bias}
\end{align}
Thus, for  symmetrical bias voltages $\alpha_\mathrm{dif}^{}=0$, 
the coefficient  $C_V^{(3)}$ is not affected by tunnel asymmetries.

From the formulas  in Table \ref{tab:C_W_SUN},  
some general  properties of $C_V^{(3)}$ can also be deduced 
 for the three characteristic regions 
of a wide parameter space for quantum dots, i.e.,\ 
the Kondo regime, valence fluctuation regime, 
and empty (fully occupied) orbital regime for $|\epsilon_d| \to \infty$.

In the Kondo regime at   
$-U\lesssim \epsilon_d \lesssim 0$ and $U/(\pi \Delta) \gtrsim 2$,  
most of the Fermi-liquid parameters take constant values 
as shown in Figs.\  \ref{Fermipara} and \ref{Thetaasym}: 
  $\delta \simeq {\pi}/{2}$,  $R-1 \simeq 1$, 
and three-body correlations almost vanish, 
$\Theta_\mathrm{I} \simeq 0$ and $\Theta_\mathrm{II} \simeq 0$. 
Therefore, over a wide parameter region of the Kondo regime, 
$C_V^{(3)}$ becomes almost  independent of  tunnel and bias asymmetries, as 
\begin{align}
W_V \,\xrightarrow{\mathrm{Kondo\ regime}\,}\, 6\,, 
\qquad \quad
\Theta_V\,\xrightarrow{\mathrm{Kondo\ regime}\,}\,  0\,.
\label{Whalffilling}
\end{align}

In the valence fluctuation regime 
at $\epsilon_d \simeq 0$ and $-U$,   
the cosine factor passes through the zero point as $\cos 2\delta = 0$
at  $\delta = \pi/4$ and $3\pi/4$, corresponding to $1/4$ and $3/4$ fillings.
Therefore,  near these zero points,
the two-body part $W_V $ is suppressed
and the three-body part $\Theta_V$ gives 
important contributions  in the valence fluctuation regime. 

Outside these regions, 
 in the limit of $|\epsilon_d|\to\infty$, 
 the impurity level approaches empty or fully occupied, 
and the FL parameters 
asymptotically take noninteracting values, i.e.,\ 
$\cos 2\delta \simeq 1 $, 
 $R-1\simeq 0$, 
 $\Theta_\mathrm{I} \simeq -2$, 
 $\Theta_\mathrm{II} \simeq 0$, as shown in Appendix B. 
Thus, the two-body $W_V$ and three-body $\Theta_V$
 parts are given by  
\begin{align}
&W_V \, \xrightarrow{|\epsilon_d|\to\infty\,}\,
-\Bigl (1+3 \,\alpha_\mathrm{dif}^2 \Bigr),
\nonumber \\
&\Theta_V \, \xrightarrow{|\epsilon_d|\to\infty\,} \,
-2\Bigl( 1+3\,\alpha_\mathrm{dif}^2\Bigr).
\label{Thetainfty}
\end{align}
Note that $C_V^{(3)}$ becomes independent of  tunnel asymmetries 
 in this limit.

\subsection{$C_V^{(3)}$ for  symmetric bias: $\alpha_L=\alpha_R=1/2$ 
($\alpha_\mathrm{dif}=0$)}
\label{Cv3NumericalSymmetry}

We next consider the case where bias voltages 
are applied symmetrically, i.e.,\  $\alpha_\mathrm{dif}^{}=0$. 
In this case, $C_V^{(3)}$  takes the following form and  
becomes independent of the tunnel asymmetry  as mentioned, 
\begin{align}
W_V&\,\xrightarrow{\alpha_\mathrm{dif}=0\,}\,
-\Bigl[\,1\,+\,5(R-1)^2 \,\Bigr]
\cos2\delta\,\,,
\\
\Theta_V
&\,\xrightarrow{\alpha_\mathrm{dif}=0\,}\ 
\Theta_\mathrm{I} 
\,+\, 3\,\Theta_\mathrm{II} 
\,.
\end{align}

\begin{figure}[t]
	\begin{minipage}[t]{\linewidth}
	\centering
	\includegraphics[keepaspectratio,scale=0.4]{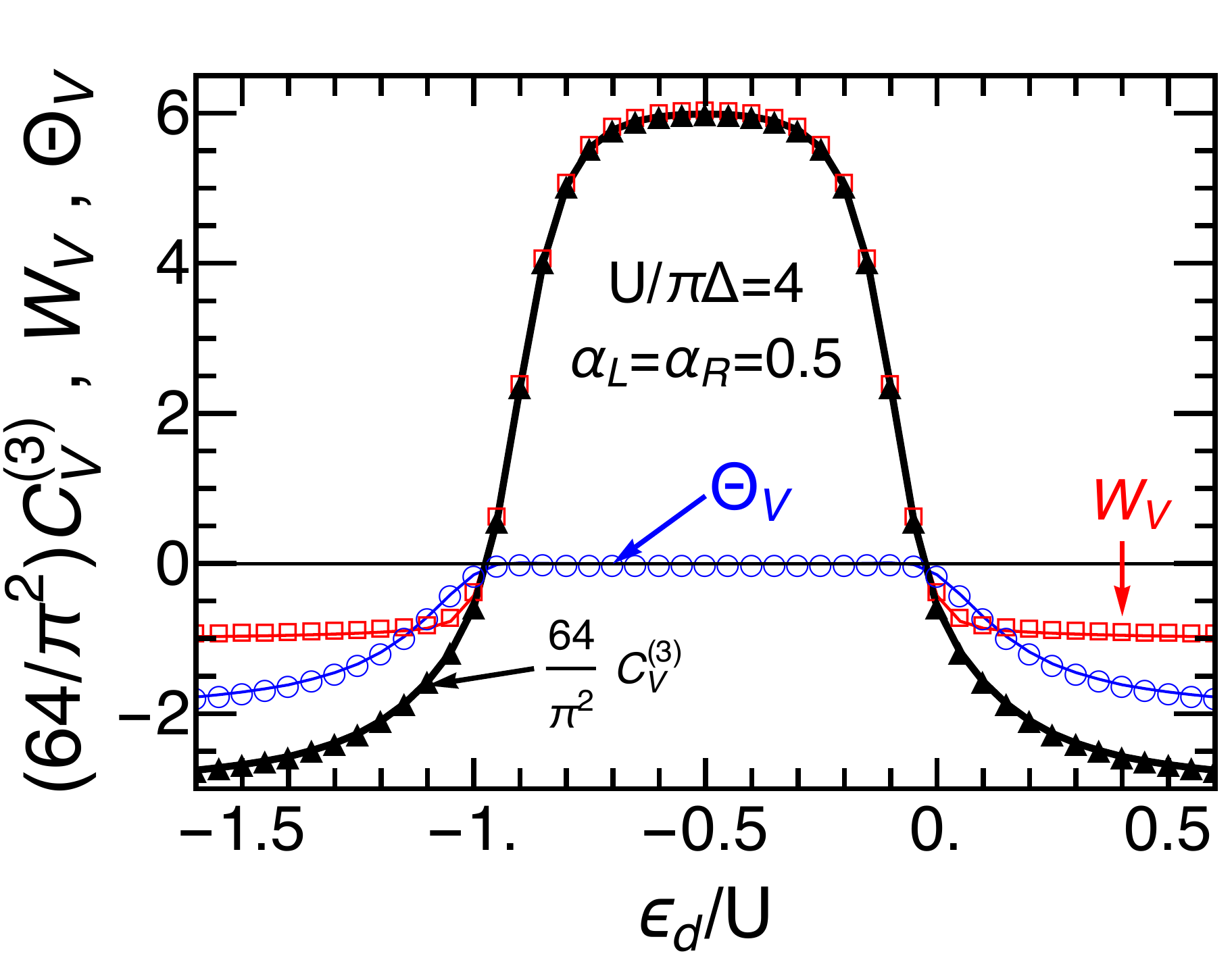}
\end{minipage}
\caption{
NRG results of the coefficient 
for order $(eV)^3$ nonlinear current 
 $C_V^{(3)}= ({\pi^2}/{64}) ( W_V+\Theta_V )$   
are plotted vs $\epsilon_d/U$ 
for a symmetric  bias $\alpha_L=\alpha_R=1/2$,   
choosing  $U/(\pi\Delta) =4$. 
In this case $\alpha_\mathrm{dif}^{}=0$; 
  both the two-body $W_V$ and three-body $\Theta_V$ contributions  
become independent of  tunnel asymmetries  $(\Gamma_L-\Gamma_R)/\Delta$   
from the definition summarized in Table \ref{tab:C_W_SUN}.
}
\label{Cv3symbias}
\end{figure}

Figure \ref{Cv3symbias} shows the  NRG results of 
$C_V^{(3)}$,   $W_V$, and $\Theta_V$ in this case 
$\alpha_\mathrm{dif}=0$. The Coulomb repulsion is 
chosen to be strong  $U/(\pi\Delta) = 4$. 
The coefficient $C_V^{(3)}$ has a broad peak in the Kondo regime, 
and it is determined by the two-body contributions $W_V$ 
since  $\Theta_V \simeq 0$ 
over the range of  $-U \lesssim \epsilon_d  \lesssim 0$
as both of the three-body components 
$\Theta_\mathrm{I}$ and $\Theta_\mathrm{II}$ 
almost vanish in this region,  as shown in Fig.\ \ref{Thetaasym}. 
The coefficient $C_V^{(3)}$ decreases rapidly  
in the valence 
fluctuation regime at $\epsilon_d \simeq 0$ and $\epsilon_d \simeq -U$. 
As $|\epsilon_d|$ increases further,
the two- and three-body components  
give comparable contributions, and 
approach $W_V\to -1$ and $\Theta_V\to -2$   
in the limit of $|\epsilon_d|\to\infty$.  
Thus,  $(64/\pi^2)\, C_V^{(3)} \xrightarrow{|\epsilon_d|\to\infty\,} -3$.

\begin{figure}[t]
	\begin{minipage}[t]{\linewidth}
	\centering
	\includegraphics[keepaspectratio,scale=0.4]{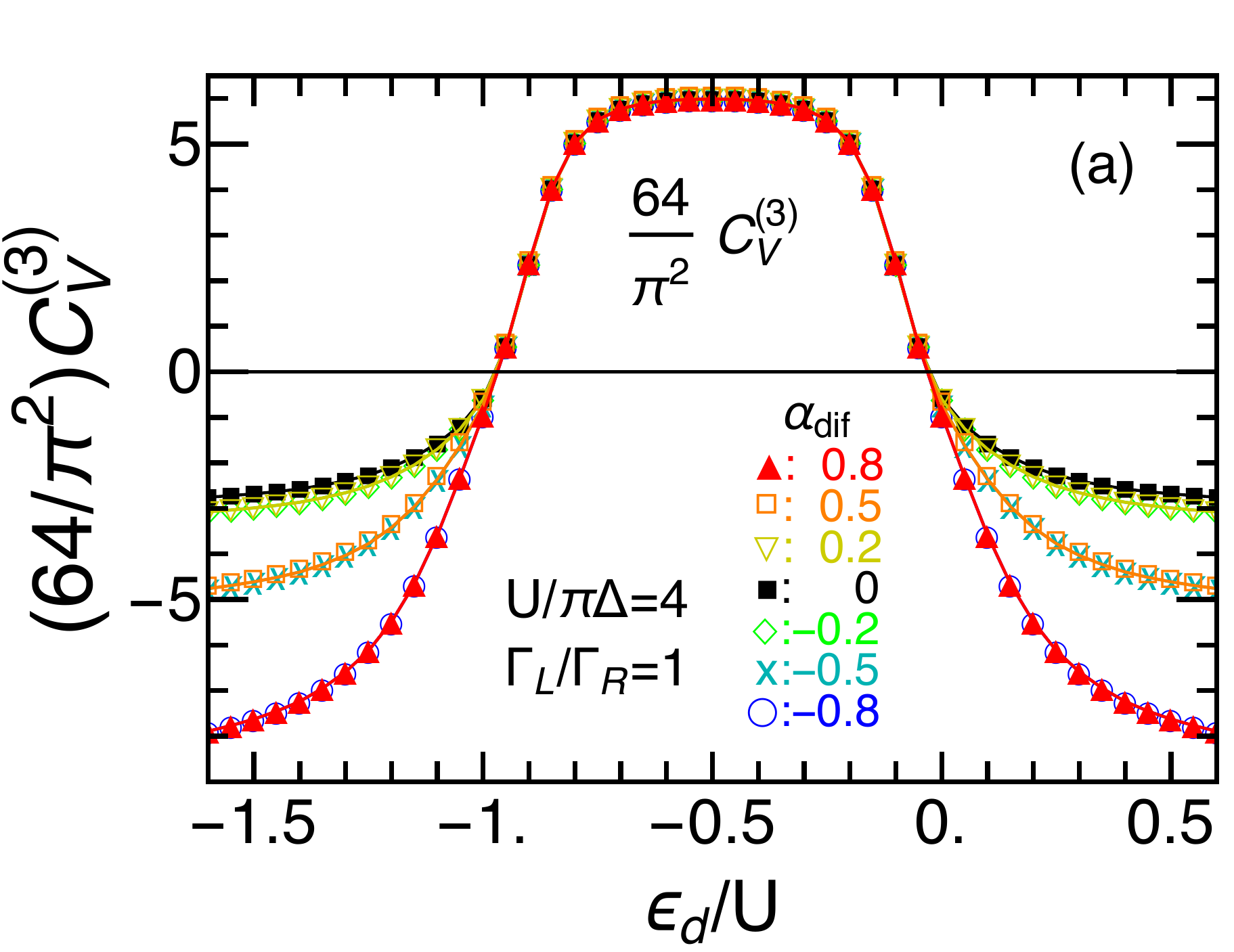}
\end{minipage}
\rule{1mm}{0mm}
	\begin{minipage}[t]{\linewidth}
	\centering
	\includegraphics[keepaspectratio,scale=0.4]{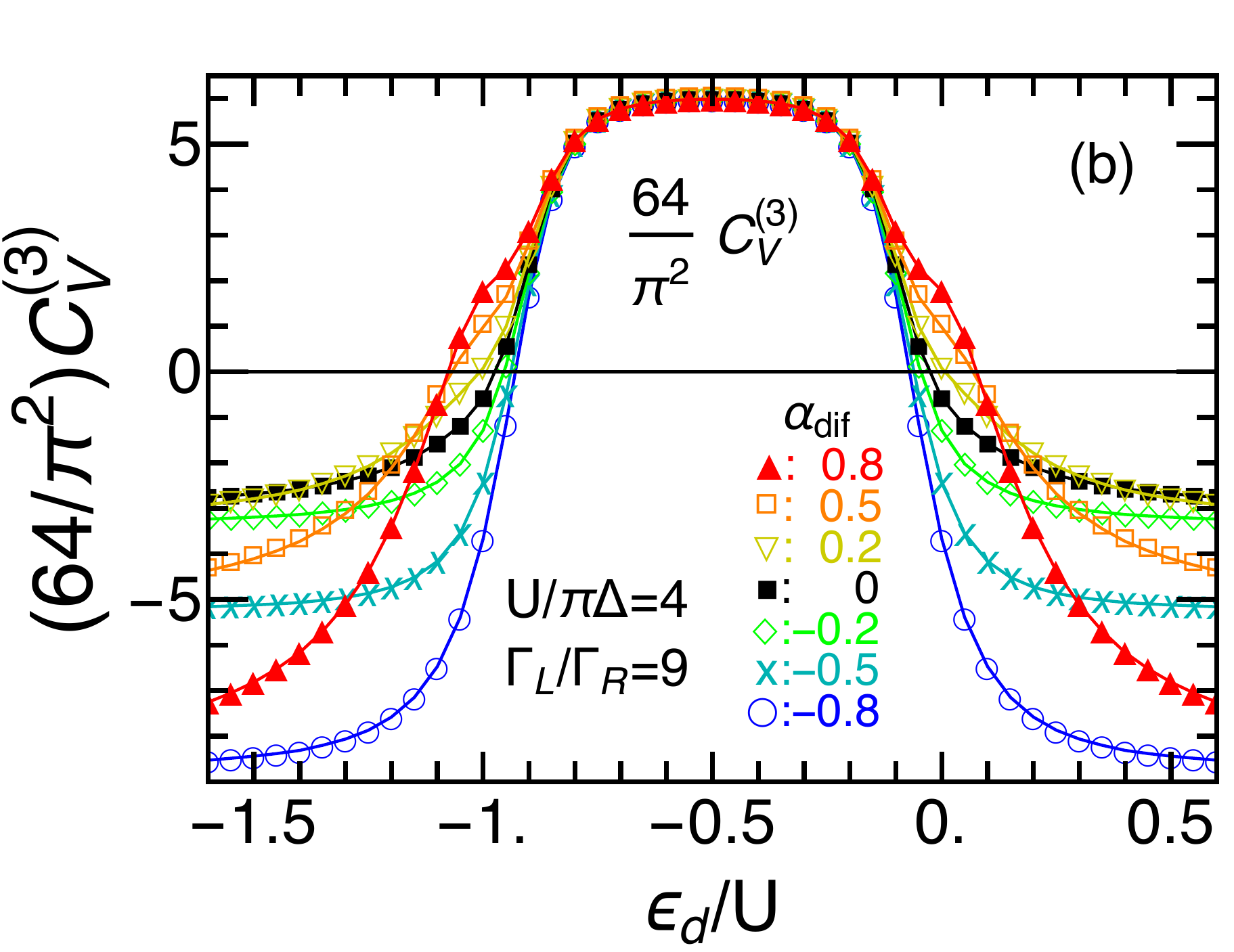}
\end{minipage}
\rule{1mm}{0mm}
\begin{minipage}[t]{\linewidth}
	\centering
	\includegraphics[keepaspectratio,scale=0.4]{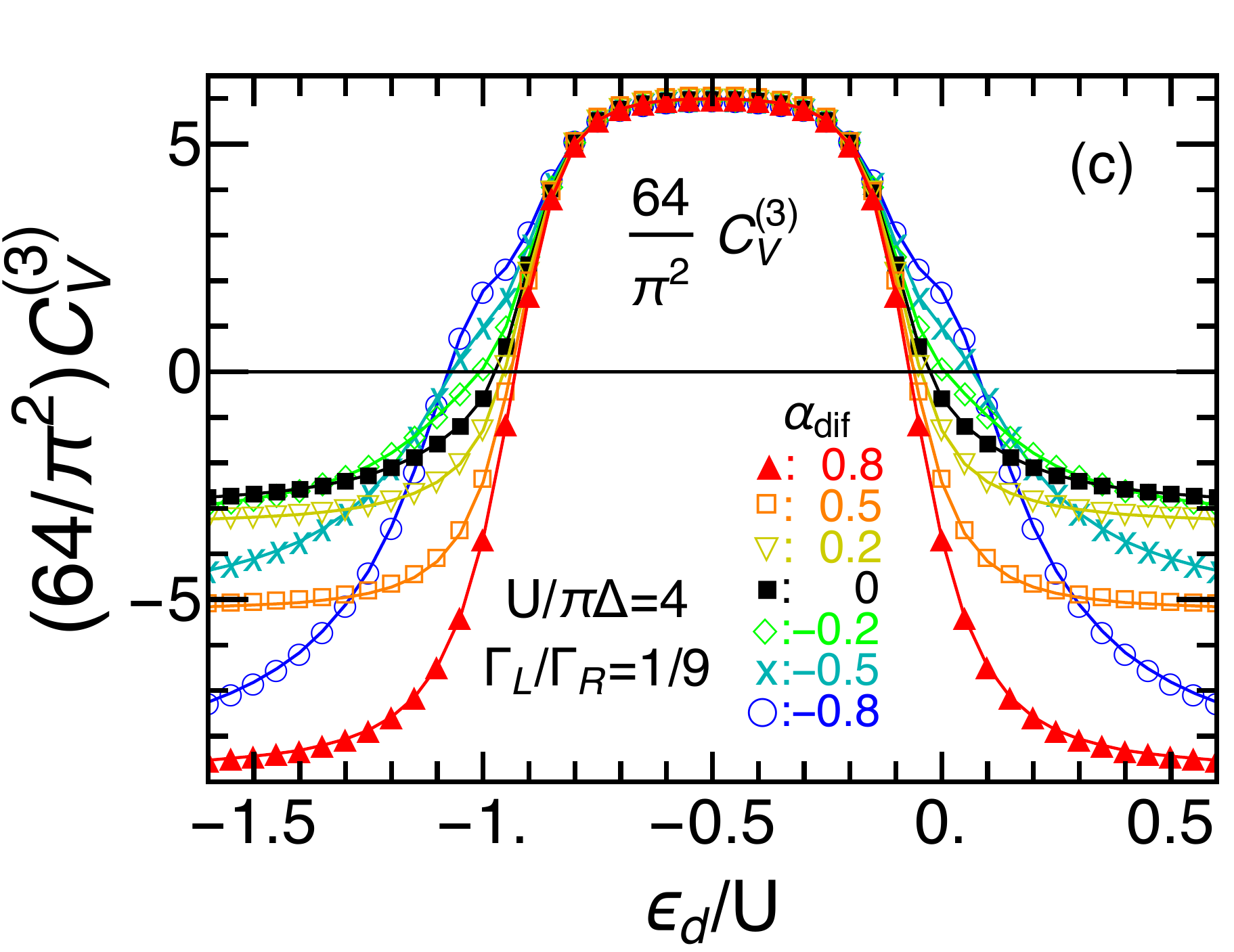}
	\end{minipage}
	\rule{1mm}{0mm}
\caption{
$C_V^{(3)}$  for $\alpha_{\mathrm{dif}}^{}=-0.8(\circ),$  $-0.5(\times)$, 
$-0.2(\diamond)$, $0.0(\blacksquare)$, $0.2(\triangledown)$, $0.5(\square)$, 
$ 0.8(\blacktriangle)$ are plotted for  three different tunnel couplings  
(a)  $\Gamma_L=\Gamma_R$, (b) $\Gamma_L/\Gamma_R=9$, 
and (c) $\Gamma_L/\Gamma_R=1/9$, choosing  
the interaction to be $U/(\pi \Delta)=4$.  
}
\label{Cv3asymbias}
\end{figure}

\subsection{$C_V^{(3)}$ for asymmetric bias: $\alpha_L\neq\alpha_R$ 
($\alpha_\mathrm{dif}\neq 0$)}
\label{Cv3NumericalAsymmetry}

We next examine the effects of bias asymmetries  $\alpha_L\neq\alpha_R$. 
In this case, the coefficient  $C_V^{(3)}$ for 
 order $(eV)^3$ nonlinear current depends on  
tunnel asymmetries 
through the cross term 
 $\alpha_\mathrm{dif}^{}\,(\Gamma_L-\Gamma_R)/\Delta$ 
described in Eq.\ \eqref{eq:cross_term_tunnel_bias}.

\subsubsection{
Effects of bias asymmetries ($\alpha_\mathrm{dif}\neq 0$)
on  $C_V^{(3)}$ for different degrees of tunnel asymmetries
}

In Fig.\ \ref{Cv3asymbias}, 
the coefficient  $C_V^{(3)}$ 
is plotted for different degrees of bias asymmetries 
$\alpha_{\mathrm{dif}}^{}=-0.8$, $-0.5$, $-0.2$, $0.0$, $0.2$, $0.5$, $0.8$, 
for strong Coulomb repulsion $U=4\pi\Delta$. 
The three panels  (a), (b), and (c) correspond to the results obtained for 
different tunnel couplings:  (a) the symmetric one $\Gamma_L=\Gamma_R$, 
and two largely imbalanced tunnel couplings  (b) $\Gamma_L=9\Gamma_R$ 
and (c) $\Gamma_L=\Gamma_R/9$. 
Note that there is a relation between these curves 
against the inversion of 
 $(\alpha_L,\Gamma_L)$  and  $(\alpha_R,\Gamma_R)$:  
$
C_V^{(3)} (\alpha_{\mathrm{dif}}^{},\Gamma_L/\Gamma_R) 
\,=\, 
C_V^{(3)} (-\alpha_{\mathrm{dif}}^{},\Gamma_R/\Gamma_L)$,  
as shown in Eq.\ \eqref{eq:LR_inversion_C3}.

In the Kondo regime  $-U \lesssim \epsilon_d  \lesssim 0$,  
the coefficient  $C_V^{(3)}$ is not affected by tunnel  
or bias asymmetries 
and has a wide plateau of the height $(64/\pi^2) C_V^{(3)} \simeq 6.0$ 
given in Eq.\ \eqref{Whalffilling}.
Outside the plateau region,
the coefficient $C_V^{(3)}$ varies with  $\alpha_\mathrm{dif}^{}$ 
and asymptotically approaches the saturation value 
$(64/\pi^2) C_V^{(3)} 
\xrightarrow{|\epsilon_d|\to\infty\,} 
 -3\bigl( 1+3\alpha_\mathrm{dif}^{2} \bigr)$ 
given in  Eq.\ (\ref{Thetainfty}) 
for the empty and fully occupied orbital regimes.  
However, for instance in Fig.\ \ref{Cv3asymbias} (b), 
while the curve for $\alpha_\mathrm{dif}^{}=-0.8$ 
approaches very closely to the saturation value $-8.76$ 
at both ends of the horizontal axes 
 $\epsilon_d=-1.6$ and  $0.6$, 
the curve for $\alpha_\mathrm{dif}^{}=0.8$ 
still deviates from the saturation value at these points. 
This is caused by a shoulder structure seen for  
the curve for  $\alpha_\mathrm{dif}^{}=0.8$ 
at $\epsilon_d \simeq -U $ and  $\epsilon_d \simeq 0$ 
in the valence fluctuation regime.  
The same shoulder structure emerges in the situation where 
the parameters  $(\alpha_L,\Gamma_L)$  and  $(\alpha_R,\Gamma_R)$ 
are inverted, i.e.,\ the curve for 
$\alpha_\mathrm{dif}^{}=-0.8$ in Fig.\ \ref{Cv3asymbias} (c). 
This structure is caused by an enhanced contribution  
of the three-body component  $\Theta_\mathrm{II}$ 
which has the characteristic peaks that evolve with $U$ 
in the valence fluctuation regions as seen in Fig.\ \ref{Thetaasym} (b). 
The shoulder structure emerges in the case where 
the cross term of tunnel and bias asymmetries, 
 $12\alpha_\mathrm{dif}^{}\,(\Gamma_L-\Gamma_R)/\Delta$,  
 in the formula for $\Theta_V$  given in Eq.\ \eqref{THv} 
becomes positive and large. 
We will discuss contributions of $\Theta_V$ more precisely below.

\subsubsection{
Effects of tunnel asymmetries on $\,C_V^{(3)}$ 
for junctions 
with a maximized bias asymmetry $\alpha_\mathrm{dif}^{} = 1$
}

\label{Cv3NumericalAsymmetryTypical}

We next examine the effects of tunnel asymmetries 
for junctions with a maximized bias asymmetry, 
i.e.,\ the case where bias voltage is applied to 
the source side keeping the drain grounded   
 $\mu_L=eV$ and $\mu_R=0$. 
In this case $\alpha_\mathrm{dif}^{}=1$, 
and the coefficient  $C_V^{(3)}$  takes the following form,
\begin{align}
W_V\xrightarrow{\alpha_\mathrm{dif}^{}=1\,}
& \ 
-2\cos2\delta\Biggl[\, 2
\,-3\left( 1+ \frac{\Gamma_L-\Gamma_R}{\Gamma_L+\Gamma_R}\right)(R-1)\,
\nonumber
\\
& \qquad \qquad \quad  
+ \left( 4+3\, 
\frac{\Gamma_L-\Gamma_R}{\Gamma_L+\Gamma_R}\right) (R-1)^2 \Biggr], 
\nonumber 
\\
\Theta_V\xrightarrow{\alpha_\mathrm{dif}^{}=1\,}& \ \, 
4\,\Theta_\mathrm{I}
\,+\,
12 \left( 1+ \frac{\Gamma_L-\Gamma_R}{\Gamma_L+\Gamma_R}\right)
\Theta_\mathrm{II}\,.
\label{THasym}
\end{align}
The last line shows that the contributions of  $\Theta_\mathrm{II}$  
increase with tunnel asymmetries $(\Gamma_L-\Gamma_R)/\Delta$. 
The two-body part  $W_V$ also 
depends on $(\Gamma_L-\Gamma_R)/\Delta$. 
However, since another factor $\cos 2\delta$ has a zero point 
in the valence fluctuation regions, 
 $W_V$ becomes less sensitive to tunnel asymmetries.

\begin{figure}[t]
	\begin{minipage}[t]{\linewidth}
	\centering
	\includegraphics[keepaspectratio,scale=0.39]{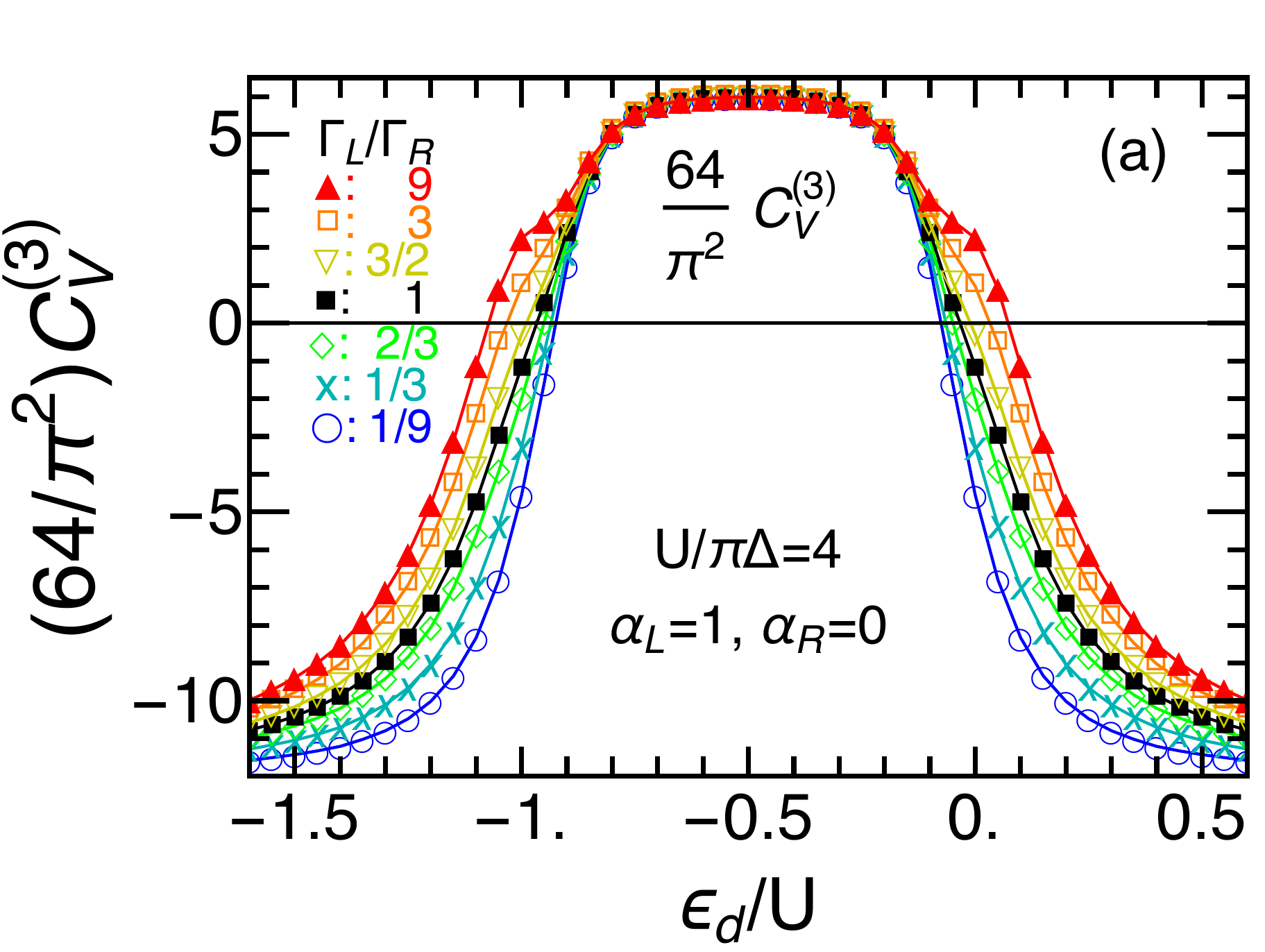}
\end{minipage}
\rule{1mm}{0mm}
\begin{minipage}[t]{\linewidth}
	\centering
	\includegraphics[keepaspectratio,scale=0.39]{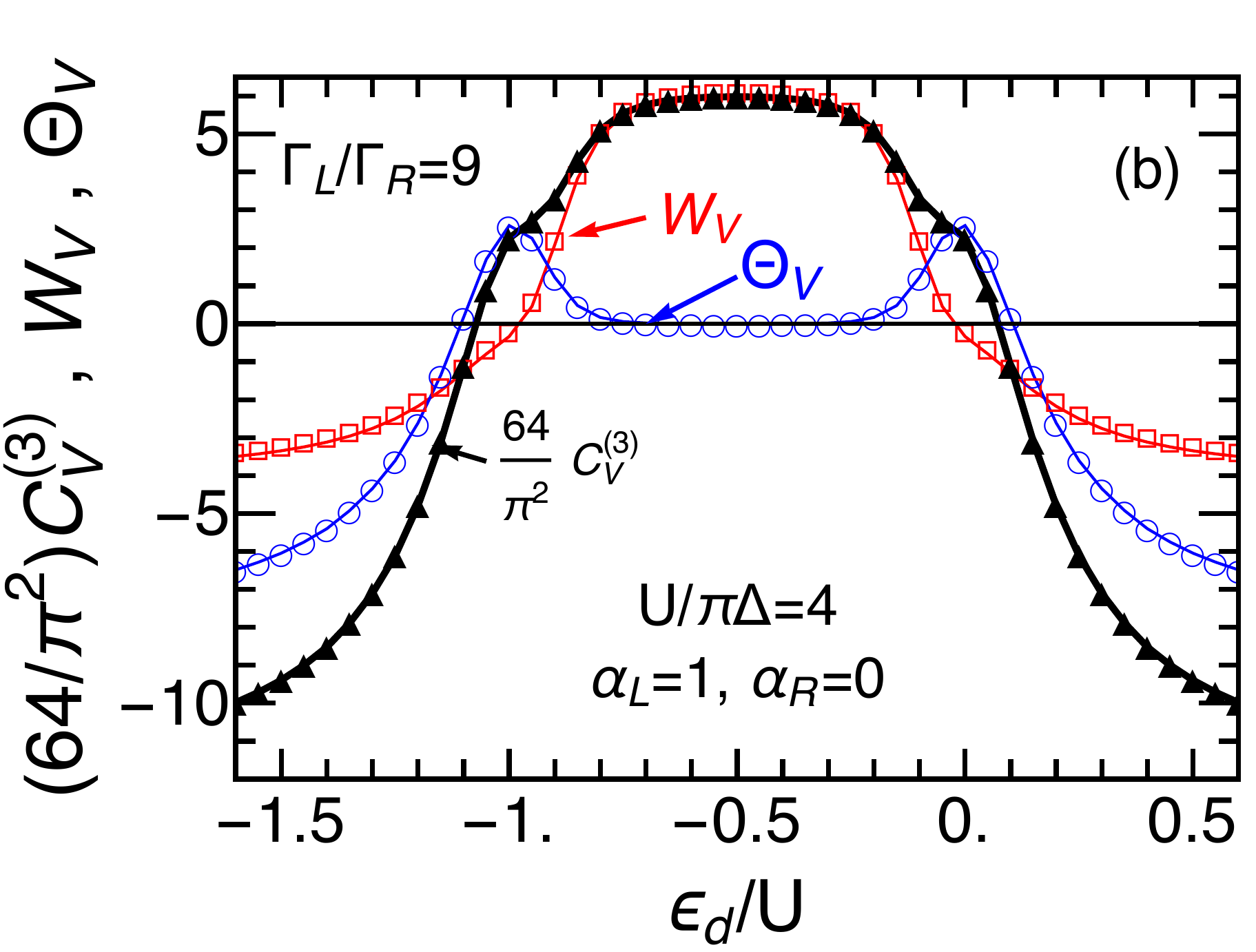}
	\end{minipage}
	\rule{1mm}{0mm}
	\begin{minipage}[t]{\linewidth}
	\centering
	\includegraphics[keepaspectratio,scale=0.39]{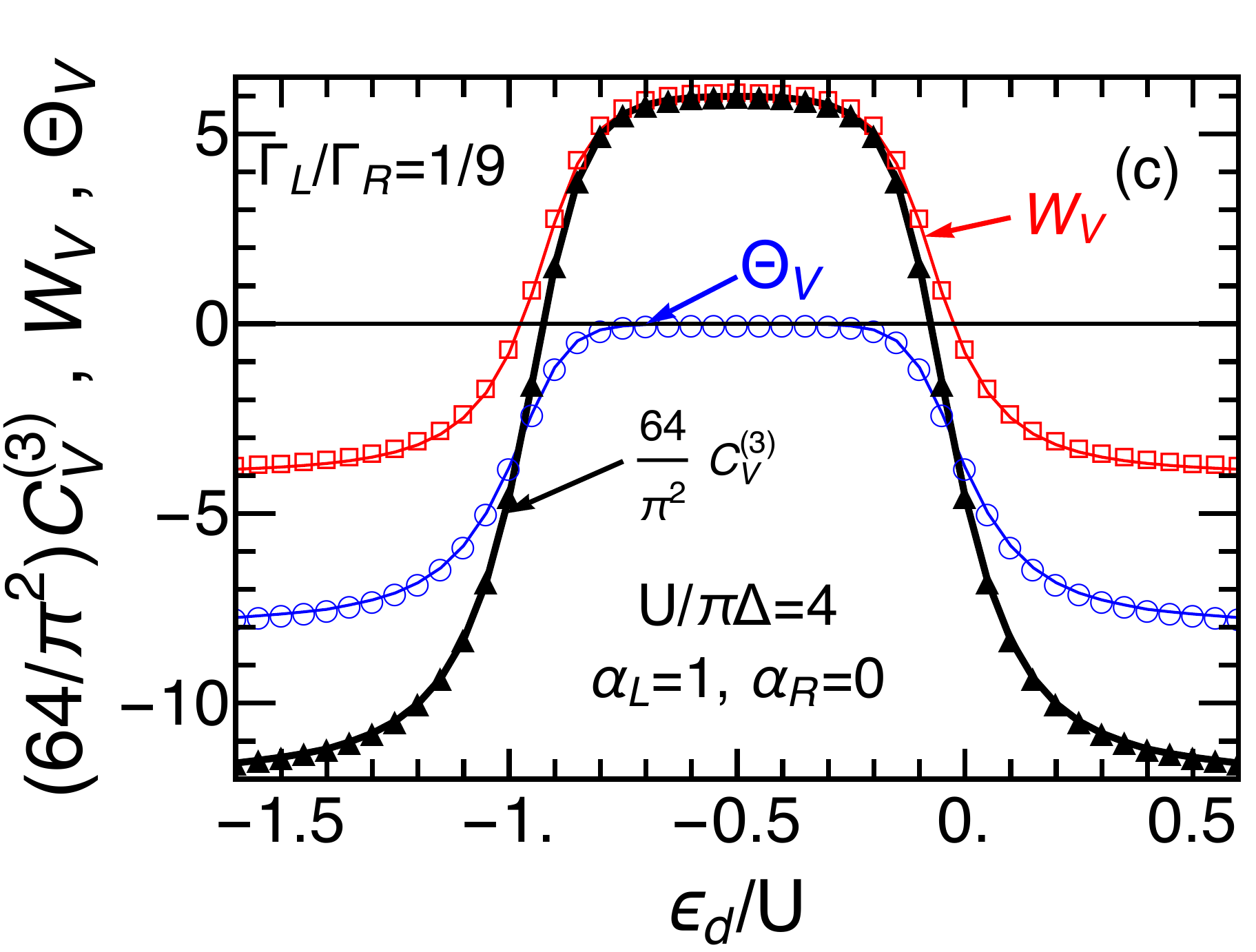}
	\end{minipage}
\caption{
 $C_V^{(3)}$ for a maximized bias asymmetry $\alpha_\mathrm{dif}^{}=1$ 
($\mu_L=eV$ and $\mu_R=0$). 
(a) Different degrees of tunnel asymmetries are examined, 
taking $\Gamma_L/\Gamma_R=1/9(\circ)$, 
$1/3(\times)$, $2/3(\diamond)$, $1(\blacksquare)$, $3/2(\triangledown)$,
$3(\square)$, $9(\blacktriangle)$. 
(b) and (c):  
$W_V$ and $\Theta_V$ are plotted together with $C_V^{(3)}$ 
for  (b) $\Gamma_L =9\Gamma_R$  and (c) $\Gamma_L=\Gamma_R/9$.  
The Coulomb interaction is chosen to be $U/(\pi\Delta)=4$. 
}
\label{Cv3}
\end{figure}

In Fig.\ \ref{Cv3} (a), the coefficient $C_V^{(3)}$ 
in this case  $\alpha_\mathrm{dif}^{}=1$  is plotted 
for seven different degrees of tunnel couplings 
  $\Gamma_L/\Gamma_R=1/9$, $1/3$, $2/3$, $1$, $3/2$, $3$, $9$,  
choosing interactions to be $U/(\pi\Delta)=4$,      
keeping $\Delta=\Gamma_L+\Gamma_R$ unchanged. 
It clearly shows that the shoulder structure evolves 
in the valence fluctuation regime at $\epsilon_d\simeq-U$ and $\epsilon_d\simeq 0$ 
as $\Gamma_L/\Gamma_R$, 
or equivalently $(\Gamma_L-\Gamma_R)/\Delta$, increases.
The shoulder structure makes the convergence 
of $C_V^{(3)}$ at  $|\epsilon_d|\to\infty$ protracted,
which in this case approaches  $C_V^{(3)} \to -12$.
As mentioned for this structure appearing in Fig.\ \ref{Cv3asymbias} (b), 
it occurs when the cross term 
described in Eq.\ \eqref{eq:cross_term_tunnel_bias},
or the corresponding coefficient for $\Theta_\mathrm{II}$
in  Eq.\ \eqref{THasym}, 
is positive  $\alpha_\mathrm{dif}^{}\,(\Gamma_L-\Gamma_R)/\Delta>0$ 
 and large.

The contributions of  $W_V$ and  $\Theta_V$ are also shown separately  
in Fig.\ \ref{Cv3},  for  (b) $\Gamma_L=9\Gamma_R$,  
 and  (c)  $\Gamma_L=\Gamma_R/9$, 
i.e.,\ for two opposite largely imbalanced tunnel asymmetries.  
The two-body contributions $W_V$ dominate $C_V^{(3)}$ 
in the Kondo regime as both three-body components  
$\Theta_\mathrm{I}$ and $\Theta_\mathrm{II}$ almost vanish 
for  $-U\lesssim\epsilon_d\lesssim 0$, 
 similarly to that in the symmetric bias case  Fig,\ \ref{Cv3symbias}. 
 In the valence fluctuation regime, the shoulder structure emerges 
when  $\alpha_\mathrm{dif}^{}\,(\Gamma_L-\Gamma_R)/\Delta$ 
is large and positive. 
These two examples show that  the two-body part  $W_V$ is less sensitive 
to tunnel asymmetries, i.e.,\ to  whether $\Gamma_L=9\Gamma_R$ 
or $\Gamma_L=\Gamma_R/9$.
In contrast, 
the three-body part $\Theta_V$ has a peak that causes the shoulder structure 
of $C_V^{(3)}$  in Fig.\ \ref{Cv3} (b)    
for which the cross term $\alpha_\mathrm{dif}^{}(\Gamma_L-\Gamma_R)/\Delta$ 
is positive  and large, 
whereas it is negative in (c) and 
$\Theta_\mathrm{V}$ decreases monotonically as  $\epsilon_d$ 
deviates from the particle-hole symmetric point $\epsilon_d=-U/2$.

\begin{figure}[t]
	\begin{minipage}[t]{\linewidth}
	\centering
	\includegraphics[keepaspectratio,scale=0.4]{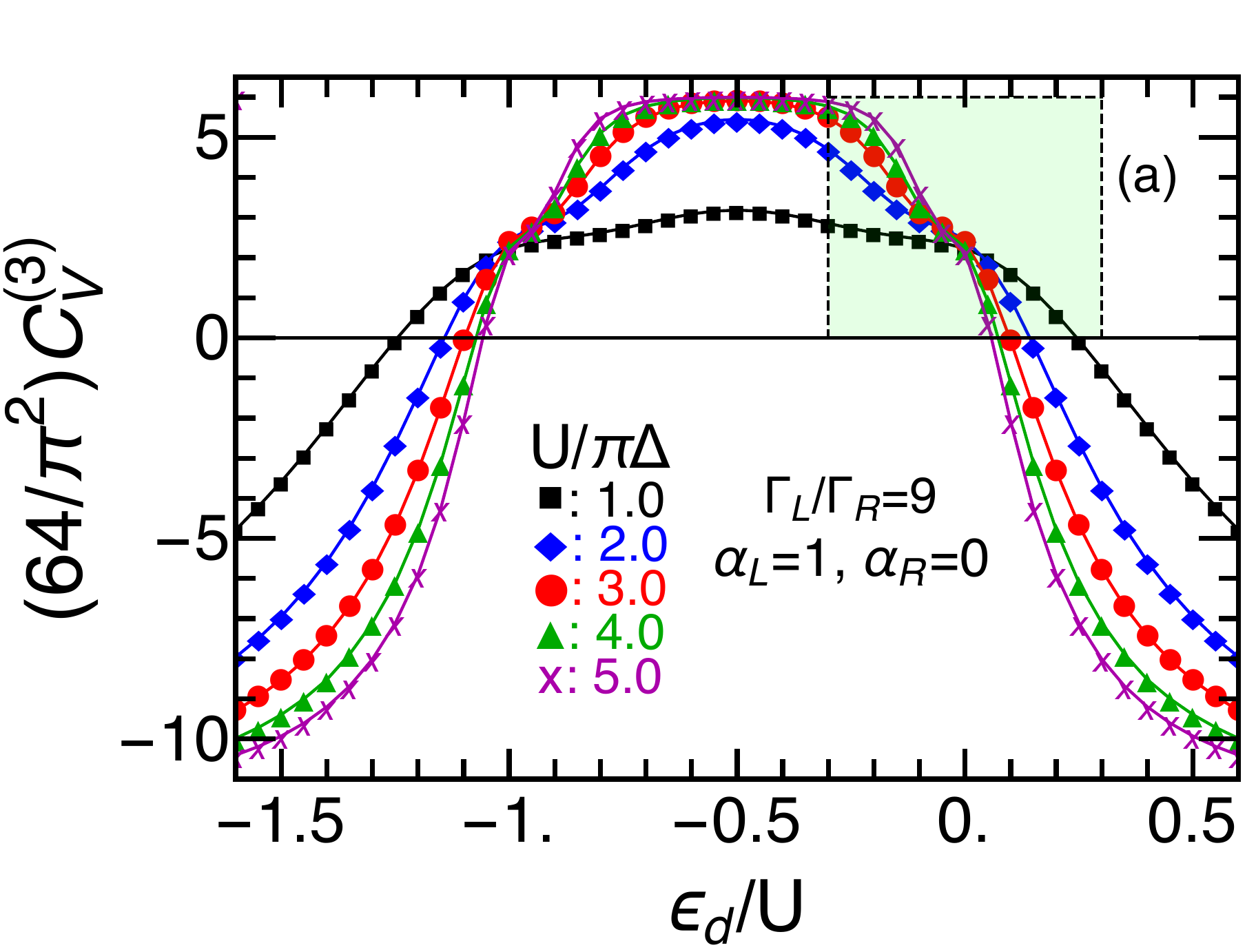}
	\end{minipage}
	\rule{1mm}{0mm}
	\begin{minipage}[t]{\linewidth}
	\centering
	\includegraphics[keepaspectratio,scale=0.375]{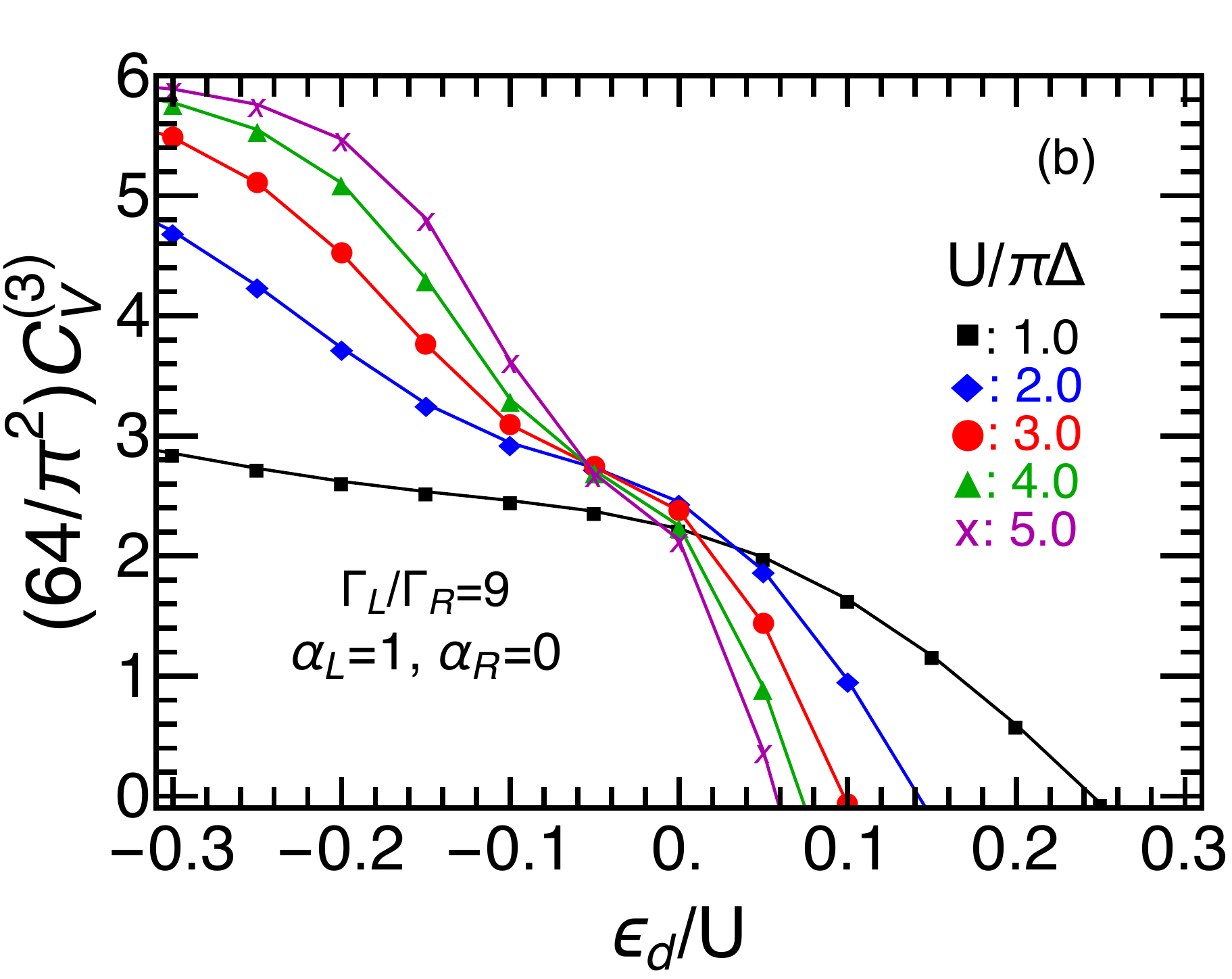}
	\end{minipage}
\caption{
 $C_V^{(3)}$ 
for  $\alpha_\mathrm{dif}^{}=1$ and  $\Gamma_L/\Gamma_R=9$ 
is plotted for different values of Coulomb 
interactions $U/(\pi\Delta)=1, 2, 3, 4, 5$. 
Panel (b) shows an enlarged view of the shaded region in (a),   
describing an evolution of the shoulder structure in the valence fluctuation regime. 
 }
\label{Cv3U}
\end{figure}

In Fig.\ \ref{Cv3U}, the coefficient $C_V^{(3)}$ 
is plotted for different values of interactions  $U/(\pi\Delta)=1, 2, 3, 4, 5$, 
for the same  bias asymmetries  $\alpha_\mathrm{dif}^{}=1$,  
choosing tunnel couplings to be $\Gamma_L=9\Gamma_R$. 
The lower panel  (b) shows an enlarged view of the shaded region in (a).   
The Kondo ridge of $C_V^{(3)}$ becomes flat for strong interactions
  $U/(\pi\Delta) \gtrsim 3.0$. 
Simultaneously, the shoulder structure evolves in the valence fluctuation regime 
at  $\epsilon_d \simeq -U$ and $\epsilon_d \simeq 0$ as  $U$ increases. 
It reflects the peak of the three-body correlation  $\Theta_\mathrm{II}$, 
the height of which increases with $U$ as shown in Fig.\ \ref{Thetaasym} (b). 
These observations show that contributions of $\Theta_\mathrm{II}$ 
can be enhanced selectively by tuning tunnel and bias asymmetries 
to make the cross term  $\alpha_\mathrm{dif}^{}(\Gamma_L-\Gamma_R)/\Delta$ 
positive and large. 
Conversely,  it also indicates that from measurements of the shoulder structure  
information about the nonlinear susceptibilities between 
different spins  $\chi_{\uparrow\downarrow\downarrow}^{[3]} 
\propto \Theta_\mathrm{II}$ can be extracted.

\section{Summary}
\label{Conclusion}

In summary, we have studied the effects of tunnel and bias asymmetries 
on the nonlinear current through quantum dots in the low-energy Fermi-liquid regime. 
We have derived the generic formula for the differential conductance 
through the Anderson impurity at $T=0$ up to terms of order $(eV)^3$, 
 using the exact results for the low-energy asymptotic form of the retarded self-energy. 

The formula, which  is summarized in Table \ref{tab:C_W_SUN}, 
 explicitly reveals the fact that the Coulomb interaction induces 
an extra dependence for $dI/dV$ on tunnel asymmetries
other than that which enters through the prefactor 
$g_0^{}=(2e^2/h)\, 4\Gamma_L\Gamma_R/(\Gamma_L+\Gamma_R)^2$. 
We have also examined the behavior of  the coefficients $C_V^{(2)}$ and $C_V^{(3)}$ 
for the nonlinear components of $dI/dV$ using the NRG over 
a wide range of the parameter space, varying degrees of tunnel and bias asymmetries, 
and relative position of the impurity level  $\epsilon_d$ from the Fermi level,    
for weak and strong  interactions $U$.

The order $(eV)^2$ component of the nonlinear current appears 
away from half filling $\epsilon_d \neq -U/2$ 
when the system has tunnel or bias asymmetries, i.e.,\ $\Gamma_L\neq \Gamma_R$ 
or $\alpha_\mathrm{dif}^{}\neq 0$.  
It reflects the shift of the impurity level of order $eV$ 
and  position of  the bias window relative to the Fermi level at equilibrium.
In the Kondo regime at  $-U \lesssim \epsilon_d \lesssim 0$ 
for strong interactions  $U/(\pi\Delta) \gtrsim 2.0$,  
the coefficient  $C_V^{(2)}$ 
does  not depend on bias asymmetries $\alpha_\mathrm{dif}^{}$ 
but on tunnel asymmetries $(\Gamma_L-\Gamma_R)/\Delta$.
The coefficient $C_V^{(2)}$ is enhanced especially 
in the valence fluctuation regime at $\epsilon_d \simeq 0$ and $\epsilon_d \simeq -U$. 
In the case where the sign of  $\alpha_\mathrm{dif}^{}$  
and that of $(\Gamma_L-\Gamma_R)/\Delta$ are the same, 
the tunnel and bias asymmetries tend to cancel their effects 
in the valence fluctuation regime, 
and it yields extra zero points in $C_V^{(2)}$.

The order $(eV)^3$ component of the nonlinear current is determined by 
the sum of the two-body contributions $W_V$ and three-body contributions $\Theta_V$. 
The three-body contributions are described by two independent 
nonlinear static susceptibilities  
$\chi_{\uparrow\uparrow\uparrow}^{[3]}$ 
and $\chi_{\uparrow\downarrow\downarrow}^{[3]}$. 
Although these nonlinear susceptibilities almost vanish in the Kondo regime,
they contribute to the transport outside of the Kondo regime, i.e.,\ 
in the valence fluctuation,  empty orbital, and fully occupied orbital regimes. 
The formula  shown in Table \ref{tab:C_W_SUN} reveals the fact that  
tunnel asymmetries affect $C_V^{(3)}$ only through 
the cross term  $\alpha_\mathrm{dif}^{} (\Gamma_L-\Gamma_R)/\Delta$, 
whereas bias asymmetries $\alpha_\mathrm{dif}^{}$ enter also 
through the other terms. Thus, the order $(eV)^3$  nonlinear current 
is not affected by tunnel asymmetries when the bias voltage is applied 
symmetrically, i.e.,\ $\mu_L=-\mu_R=eV/2$.  
Conversely, under asymmetrical bias voltages, 
the coefficient  $C_V^{(3)}$  depends on both  $\alpha_\mathrm{dif}^{}$ 
and $(\Gamma_L-\Gamma_R)/\Delta$. 
We find that $C_V^{(3)}$ shows  a shoulder structure 
in the valence fluctuation regime in the case where the cross 
term  $\alpha_\mathrm{dif}^{} (\Gamma_L-\Gamma_R)/\Delta$ 
is positive and large.
It is caused by the peaks that 
emerge at $\epsilon_d \simeq -U$ and $\epsilon_d \simeq 0$ 
in the three-body contributions $\Theta_V$.
Specifically,  these peaks come from one of the two three-body components  
 $\Theta_\mathrm{II}$ which is enhanced significantly 
by tunnel and bias asymmetries in this case.

Recently, the three-body contributions 
 $\Theta_V$ on the order $(eV)^3$ nonlinear current 
have been extracted experimentally for quantum dots in a magnetic field \cite{Hata2021}. 
To our knowledge, 
this is the first observation of three-body effects in the Fermi liquid 
of the Kondo systems.  
Our results show that the contributions of $\Theta_\mathrm{II}$ 
can be enhanced selectively by tuning tunnel asymmetries 
keeping the other contributions from $\Theta_\mathrm{I}$ unchanged,  
and it leads a characteristic shoulder structure in the gate-voltage dependence 
of  $C_V^{(3)}$. 
Therefore,  it can be used for future analysis to deduce separately  
the two different components of the three-body correlations, 
 $\chi_{\sigma\sigma\sigma}^{[3]}$ and $\chi_{\sigma\sigma'\sigma'}^{[3]}$ 
for $\sigma\neq \sigma'$ from measurements.
Furthermore,  for a multilevel quantum dot such as the SU($N$) dot 
with internal degrees of freedom $\sigma=1,2,\ldots,N$, 
there is another independent three-body component  
 $\chi_{\sigma\sigma'\sigma''}^{[3]}$ 
for three different flavors,  i.e.,\ $\sigma$'s \cite{Teratani2020PRL}. 
This will give interesting varieties to 
nonlinear conductance through junctions with  tunnel and bias asymmetries, 
and studies along this line are also in progress.

\section{Acknowledgements}
\label{Acknowledgements}

This work was supported by JSPS KAKENHI
 Grants No.\ JP18K03495, No. JP18J10205, No. JP21K03415, and No. JP26220711, 
  JST CREST Grant No.\ JPMJCR1876, and the Sasakawa Scientific Research Grant from the Japan Science Society, No.\ 2021-2009.

\appendix

\section{Low-energy asymptotic form of spectral function}

\label{Differential conductanceAppendix}

We show here the low-energy asymptotic form of 
the spectral function $A(\omega,\,T,\,eV)$ 
in the presence of the tunnel and bias asymmetries 
which are parametrized by 
$(\Gamma_L-\Gamma_R)/(\Gamma_L+\Gamma_R)$
and $\alpha_{\mathrm{dif}}^{} =\alpha_L-\alpha_R$, respectively.

In order to deduce the coefficients  $C_V^{(2)}$ and $C_V^{(3)}$ given
 in Eqs.\ \eqref{Cv2} and \eqref{Cv3eq} 
from the Landauer-type formula Eq.\ \eqref{M-Wabs0},  
we have  expanded  $A(\omega,\,T,\,eV)$ 
up to terms of order $\omega^2$, $T^2$, and $(eV)^2$.  
Substituting the low-energy asymptotic form of self-energy given in 
Eqs.\ \eqref{ReSelf} and \eqref{ImSelf} into the retarded
Green's function in Eq.\ \eqref{eG}, 
we obtain the asymptotic form of the spectral function, 
which is exact  up to terms of order $\omega^2$, $T^2$, and $(eV)^2$:  
\begin{widetext}
\begin{align}
\pi\Delta A(\omega,\,T,\,eV)&\simeq\sin^2\delta+\pi\sin2\delta\,\biggl[\chi_{\uparrow\uparrow}\,\omega+\frac{1}{2}\biggl(\alpha_{\mathrm{dif}}+\frac{\Gamma_L-\Gamma_R}{\Gamma_L+\Gamma_R}\biggr)\chi_{\uparrow\downarrow}\,\,eV\biggr] 
\nonumber
\\
& \ +\pi^2\biggl[\,\cos2\delta\biggl(\chi_{\uparrow\uparrow}^2+\frac{1}{2}\chi_{\uparrow\downarrow}^2\biggr)-\frac{\sin2\delta}{2\pi}\chi_{\uparrow\uparrow\uparrow}^{[3]}\biggr]\omega^2 
\nonumber
\\
& \ +\pi^2\biggl[\cos2\delta\biggl(\chi_{\uparrow\uparrow}\chi_{\uparrow\downarrow}-\frac{1}{2}\chi_{\uparrow\downarrow}^2\biggr)-\frac{\sin2\delta}{2\pi}\chi_{\uparrow\downarrow\downarrow}^{[3]}\biggr]\biggl(\alpha_{\mathrm{dif}}+\frac{\Gamma_L-\Gamma_R}{\Gamma_L+\Gamma_R}\biggr)\,\omega\,eV
\nonumber
\\
& \ +\frac{\pi^2}{3}\biggl(\frac{3}{2}\cos2\delta\chi_{\uparrow\downarrow}^2-\frac{\sin2\delta}{2\pi}\chi_{\uparrow\downarrow\downarrow}^{[3]}\biggr)
\biggl[\biggl(1+2\alpha_\mathrm{dif}\frac{\Gamma_L-\Gamma_R}{\Gamma_L+\Gamma_R}+\alpha_{\mathrm{dif}}^2\biggr)\frac{3}{4}(eV)^2+(\pi T)^2\biggr]+\cdots.
\label{spectral_expansion}
\end{align}
\end{widetext}

\section{Fermi-liquid parameters for  $|\epsilon_d|\to\infty$}

\label{Cv3edinfty}

In the limit of $|\epsilon_d|\to\infty$, 
 the Fermi-liquid parameters converge to the values 
for the noninteracting case $U=0$ \cite{MMvDZ2015}, 
in which the phase shift $\delta^0$ and the diagonal element of 
the linear susceptibility $\chi_{\uparrow\uparrow}^{(0)}$
and that of the three-body correlations 
$\chi_{\uparrow\uparrow\uparrow}^{[3](0)}$  are given by 
\begin{align}
\sin 2\delta^{(0)}\, = & \  
\frac{2\Delta \epsilon_{d}}{\epsilon_d^2+\Delta^2}\ 
\, \xrightarrow{ |\epsilon_d| \to \infty\,} \  
\frac{2\Delta}{\epsilon_d}\,,
\\
\chi_{\uparrow\uparrow}^{(0)}\, =  & \ 
\frac{1}{\pi}\,\frac{\Delta}{\epsilon_d^2+\Delta^2}
\  \xrightarrow{ |\epsilon_d| \to \infty\,} \  
\frac{\Delta}{\pi \epsilon_d^2}
\,,
\\
 \chi_{\uparrow\uparrow\uparrow}^{[3](0)}\,  
= & \ 
\frac{-1}{\pi}\,\frac{2\Delta \epsilon_{d}}{(\epsilon_{d}^2+\Delta^2)^2}
\ \xrightarrow{ |\epsilon_d| \to \infty\,} \ 
\frac{-2\Delta}{\pi\epsilon_d^3} \,.
\end{align}
Furthermore,  $\chi_{\uparrow\downarrow}^{(0)}= 0$ and 
$\chi_{\uparrow\downarrow\downarrow}^{[3](0)} = 0$ for $U = 0$;  
 i.e., the off-diagonal elements  of linear and nonlinear 
 susceptibilities for noninteracting electrons vanish and $R-1 \to 0$.   

 Therefore,   in the limit of $|\epsilon_d|\to\infty$, 
the three-body correlations tend to 
$\Theta_\mathrm{I} \to -2 $ and $\Theta_\mathrm{II} \to 0$,
and thus two-body and three-body parts of  $C_V^{(3)}$ 
approach the following form, 
\begin{align}
W_V
\,  \xrightarrow{\,|\epsilon_d|\to\infty\,} \,
 -\Bigl(\,  1 + 3\,\alpha_\mathrm{dif}^{2} \,\Bigr), 
\\
\Theta_V
\,  \xrightarrow{\,|\epsilon_d|\to\infty\,} \,
-2 \Bigl(\, 1 + 3\,\alpha_\mathrm{dif}^{2} \,\Bigr)\, 
\end{align}
Thus,  $C_V^{(3)}$ 
 depends only on the bias asymmetry and not on the tunnel asymmetry, 
as discussed in Eq.\  (\ref{Thetainfty}). 
Furthermore, the nonlinear current of order $(eV)^2$ vanishes in this limit:  
$C_V^{(2)}\,  \xrightarrow{\,|\epsilon_d|\to\infty\,} \,0$.

\section{Alternative interpretation for $C_V^{(2)}$}
\label{Cv2meaning}

We describe here an alternative interpretation 
for the two components of  $C_V^{(2)}$ 
defined in Eqs.\ (\ref{Cv21}) and  (\ref{Cv22}).  
In particular, we show that  $\overline{C}_{V}^{(2a)}$ 
and   $\overline{C}_{V}^{(2b)}$ 
 can also be interpreted as the contributions 
 caused by the shift of the bias window 
and the shift of the spectral weight, respectively.

For this purpose, we introduce the conductance  $g^{(2)}$ 
that is determined by the spectral function  $A^{(1)}$ 
given  in  Eq.\ \eqref{Spectral1} 
which is deduced from order  $\omega$ and $eV$ terms of the self-energy, 
\begin{align}
g^{(2)} \,\equiv & \ 
g_{0}^{}\,  \frac{\partial}{\partial eV}
\int_{-\alpha_ReV}^{\alpha_LeV}d\omega\,
\pi\Delta A^{(1)}(\omega,eV) \,,
\end{align}
where $g_{0}^{} =
({2e^2}/{h})
4\Gamma_L\Gamma_R/(\Gamma_L+\Gamma_R)^2$. 
This conductance  $g^{(2)}$ 
describes the nonlinear current exactly up to terms of order $(eV)^2$.

We also consider  the following contributions, using 
the spectral function $A^{(1)}(\omega,eV)$ for $\alpha=0$, 
\begin{align}
g_{}^{(1)}\,\equiv & \ 
g_{0}^{}\   \frac{\partial}{\partial eV}
\int_{-\frac{1}{2}eV}^{\frac{1}{2}eV}d\omega\,
\left. \pi\Delta  A^{(1)}(\omega,eV)\right|_{\alpha=0}^{}\, 
\nonumber\\
= & \ 
g_{0}^{}\,\sin^2\delta\, +\, O\left((eV)^2\right) \,.
\label{linearint} 
\end{align}
The last line shows that this term $g_{}^{(1)}$ describes the linear conductance. 
We now decompose the conductance  
  $g^{(2)}$   into  the following form,
by introducing another term $g_\mathrm{ref}^{(2)}$ as a reference,  
\begin{align}
g^{(2)} \,=& \ 
g_{}^{(1)}
\,+ \, \left[\,g_\mathrm{ref}^{(2)}-g_{}^{(1)} \,\right] 
\,+ \, \left[\,g^{(2)}-g_\mathrm{ref}^{(2)}\right]\,, 
\\
g_\mathrm{ref}^{(2)}\,=& \ 
g_{0}^{}\,  \frac{\partial}{\partial eV}
\int_{-\alpha_ReV}^{\alpha_LeV}d\omega\,
\left. \pi\Delta A^{(1)}(\omega,eV)\right|_{\alpha=0}^{}\,.
\end{align}
Here, 
$g_\mathrm{ref}^{(2)}-g_{}^{(1)}$ represents 
 variation of the current caused by  
the shift of the  bias window from the symmetric position 
because the integrand 
 $\left.A^{(1)}(\omega,eV)\right|_{\alpha=0}^{}$ 
is common for $g_\mathrm{ref}^{(2)}$ and $g_{}^{(1)}$,  
\begin{align}
g_\mathrm{ref}^{(2)}-g_{}^{(1)}\,
=\,g_{0}^{}\, 
\frac{\pi\sin2\delta}{4}\,\alpha_\mathrm{dif}^{} 
\left(\frac{eV}{T^\ast}\right)+ \cdots \,. 
\label{Cv21int} 
\end{align}
This term is identical to the one with the coefficient $\overline{C}_{V}^{(2a)}$, 
 defined in Eq.\ (\ref{Cv21}). 
The remaining term $g^{(2)}-g_\mathrm{ref}^{(2)}$  
represents variation of the  current caused by the shift of the spectral weight  
as  the integrand for this term can be rewritten into the difference  
$A^{(1)}(\omega,eV) -
\left.A^{(1)}(\omega,eV)\right|_{\alpha=0}^{}$,
\begin{align}
g^{(2)}-g_\mathrm{ref}^{(2)}\,
=\,-g_{0}^{}\, 
\frac{\pi \sin2\delta}{4}\,2\alpha\,(R-1)\left(\frac{eV}{T^\ast}\right) + \cdots\,. 
\label{Cv22int}
\end{align}
This term also corresponds to the one described with the coefficient 
 $\overline{C}_{V}^{(2b)}$.

\end{document}